# Packing, Scheduling and Covering Problems in a Game-Theoretic Perspective

Elena Kleiman

A THESIS SUBMITTED FOR THE DEGREE
”DOCTOR OF PHILOSOPHY”

University of Haifa
Faculty of Natural Sciences
Department of Mathematics

# Packing, Scheduling and Covering Problems in a Game-Theoretic Perspective

By: Elena Kleiman
Supervised by: Dr. Leah Epstein

A THESIS SUBMITTED FOR THE DEGREE
”DOCTOR OF PHILOSOPHY”

University of Haifa
Faculty of Natural Sciences
Department of Mathematics

Recommended by: ________________________ Date:______________
(Advisor)

Approved by: ________________________ Date:______________
(Chairman of Ph.D Committee)

# Contents

# Packing, Scheduling and Covering Problems in a Game-Theoretic Perspective

By: Elena Kleiman

## Abstract

The fairly recent appearance of modern distributed network systems such as the Internet and P2P applications has raised many challenging and principally new questions, both of algorithmic and socio-economic nature. The rapidly gained popularity of such systems as well as their extensive usage and the undeniable impact that they have on almost every aspect of our everyday life, has created a need to address the raised issues systematically, and analyze the underlying computational and economic processes occurring in these systems.

This required of us to make a major shift in our traditional view of system design, as besides their enormous scale and novel nature, the formation as well as various computation and maintenance tasks in such systems are executed by a multitude of uncoordinated and economically interested autonomous agents (such as Internet Service Providers, end-users, etc.), as opposed to having a single central unit that controls and regulates the entire network and all of its participants.

These agents have diverse personal goals, each aiming at optimizing his own objective without taking into consideration the global welfare of the system. They compete each other for expensive and scarce network resources, such as bandwidth and service time. This prompted establishing the Algorithmic Game Theory discipline, that uses concepts borrowed from the classic Game Theory and applies newly developed techniques to study various algorithmic problems, in particular, networking problems that involve interactions of rational strategic agents.

A topical issue for this line of study considers the effect of the "anarchistic" and uncoordinated behavior of the system participants on the overall performance of the system. The discrepancy between the personal goals of the agents and the global social goal often has a negative effect on the system. The inefficiency suffered by the system is expressed by a measure called the Price of Anarchy. Other directions of research include considering different quality measures and various computational aspects concerning the algorithmic and economic concepts in the setting in question.

To explore the different algorithmic aspects of modern computing systems, we study models that are simplification of problems arising in real networks, which seem appropriate for describing basic network problems.

Many packing, scheduling and covering problems that were previously considered by computer science literature in the context of various transportation and production problems, appear also suitable for describing and modeling various fundamental aspects in networks optimization such as routing, resource allocation, congestion control e.t.c..

Various combinatorial problems were already studied from the game theoretic standpoint, and we attempt to complement to this body of research.

Specifically, we consider the bin packing problem both in the classic and parametric versions, the job scheduling problem and the machine covering problem in various machine models. We suggest new interpretations of such problems in the context of modern networks and study these problems from a game theoretic perspective by modeling them as games, and then concerning various game theoretic concepts in these games by combining tools from game theory and the traditional combinatorial optimization. In the framework of this research we introduce and study models that were not considered before, and also improve upon previously known results.

We believe that the study of these combinatorial problems under the game theoretic framework contributes to enhancing our understanding of various processes occurring in modern decentralized systems and providing us with new insights on their effect on the functionality of these systems. This knowledge will allow us to design more efficient network algorithms and protocols, that are robust against the inherent constraints in the modern networks which are now taken into account.

# List of Tables

# List of Figures

# Chapter 1

The natural state of man is a state of pure selfishness

CHARLES G. FINNEY (1792-1875)

# Introduction

## 1.1 Context and motivations

From the dawn of computing era, the prevalent paradigm in system design assumed existence of a central authority which constructed and managed the computational system and its participants, with a purpose of optimizing a well-defined global objective. System designers developed algorithms for network problems under the implicit assumption that the algorithm can make definitive decisions which are always carried out by the entities (or agents) participating in the system.

Emergence of large-scale distributed information and communication systems about two decades ago, with the Internet being the most prominent example, put these traditional assumptions into question, creating a major shift in our view of computational networking systems.

The Internet is built, managed and used by an enormous number of autonomous and self-interested entities (such as network operators, service providers, designers, end-users, etc.), in different levels of competition and cooperation relationships with one another. These entities have diverse sets of interests and they compete each other over expensive network resources such as link bandwidth, storage space or processing time, while aiming at achieving their individual goals (maximize their payoffs or interchangeably, minimize their costs), as opposed to obtaining a global optimum of the system.

Such entities may have no incentive to cooperate and follow a predefined protocol and may prefer to selfishly deviate from protocols if it is beneficial for their personal interests.

For example, Internet service providers (ISPs) who often have competing interests need to cooperate in order to provide global connectivity. In presence of selfish considerations each ISP selects paths that are optimal within his own network. The result is that paths spanning across multiple networks can be poor from a global perspective.

On the other hand, the tremendous size and complexity of such networks, as well as their socio-economic nature, make it impossible to introduce a single centralized authority that can enforce a protocol or regulation on all participants of the system.

Obviously, the outcomes of such unregulated self-interested behavior often have properties that are very different from the centrally designed or managed networks which traditional system design have focused on. Therefore, traditional algorithmic and distributed systems approaches are insufficient to understand and solve various important design problems that arise in the modern networks. This requires quite different set of methods and considerations.

From the economics side, the appearance of large scale distributed computing systems such as the Internet and electronic commerce applications raised many principally new and challenging problems related to computational aspects in resource management, multi agent markets, auction theory and many other fields of economics, which traditional networking research could answer only partially.

Complex problems of strategic interaction and confrontation between multiple rational participants are usually considered by the classical Game Theory discipline, which offers formal framework for studying the results of such interactions. Network optimization problems that describe situations with selfish strategic agents in an environment which lacks a central control authority can be often modeled as non-cooperative games, which are the primary objects studied in Game Theory, by considering these agents to be the players in the game.

However, because of the novel nature of the problems encountered in modern computer network not all the issues could be handled by simply combining tools from computer science and game theory. While pure economic and game-theoretic literature concerned itself mainly with incentives and economic qualities that the proposed solution should satisfy without concerning the computational efficiency of the attained methods, traditional computer science puts a special emphasis on efficient identification and finding of solutions satisfying the required qualities.

This had motivated the development of an appropriate joint CS and economic framework that could embody the new paradigm in which distributed computation is performed by self-interested agents, which is called Algorithmic Game Theory. This new discipline which was initiated in [92, 95] studies various algorithmic issues in games, while consid-ering the selfishness of the players and the resulting lack of central control as an inevitable constraint, similar to the lack of unbounded computing resources in design of approximation algorithms or the lack of information about the future when designing online algorithms. For an exhaustive survey of the research in this area, we refer the reader to the textbook by Nisan et al. [91].

Among the topical questions investigated in this context are quantifying the effect of the selfish behavior on the performance of the system, measuring the quality of stable states, i.e. equilibria of the induced game, considering the complexity of computing these equilibria and designing algorithms to compute them. Another important issues concern designing of systems with equilibria of good quality, and causing players to act in a desirable manner by introducing incentives.

This thesis is located within the field of Algorithmic Game Theory and considers many of the questions that were raised above, in various settings.

## 1.2 Objectives and contributions

The rise of the Internet as a major economic, cultural and computational platform as well as its broad spectrum of applications that includes among others peer-to-peer, content delivery and social networks, had significantly influenced many aspects of our everyday life. This makes it extremely important to understand the processes that surround the development, structure and operation of the Internet and Internet-like systems in which selfishly-oriented users interact.

For this purpose we seek to build and study appropriate mathematical models of the problems that naturally occur in the multi-agent modern networks from a game-theoretic point of view. Game-theoretic concepts and techniques were applied to study various network issues like routing [95, 84, 35], network formation [4, 50], Web access [93], Quality of Service [82, 1], bandwidth allocation [107] and congestion control [77], to name only a few.

A commonly accepted approach is to study combinatorial problems that are abstractions of the problems that arise in real networks and are suitable for describing basic issues in networks. The motivation behind this approach is that analyzing the simplified scenarios modeled by these problems will help to shed light and provide useful insights into effects of interactions between selfish agents on the effectiveness of network protocols. By considering the tradeoffs and proving precise guarantees regarding the performance achievable in these models we hope to understand how to positively influence these interactions which will enable the algorithmic design of efficient network protocols that are and robust in presence of selfish and uncoordinated behavior of the agents.

Combinatorial problems that were considered in this context include job scheduling and load balancing (see e.g. [79, 36, 84, 13, 61, 5, 55, 52, 25, 70, 14, 16, 20] for a partial list), web caching [27, 85], facility location [104, 66, 21], set covering [60] etc..

It appears that a lot of packing, scheduling and covering problems are suitable for describing and modeling various fundamental aspects in networks optimization such as network routing, allocation of network resources, network dynamics, performance of queuing systems, service provision and so on.

Machine scheduling and packing problems have their origin in the optimization of production systems and their formal mathematical treatment dates back to the 1950s (see [35, 28] for surveys). These combinatorial problems are challenging by themselves. In many cases they are NP-hard [62], thus it is unlikely that there are algorithms that find optimal solutions to these problems in a computationally efficient manner (unless P=NP). In such cases, the goal is to design approximation algorithms which are relatively simple algorithms that are required to run in polynomial time and find feasible solutions that have cost within a constant factor from the exact optimal solution.

This thesis attempts to contribute to the body of knowledge regarding combinatorial optimization problems that are studied under under game-theoretic perspective in the context of modern networks.

We study naturally induced game theoretic versions of packing, scheduling and covering problems from both combinatorial as well as game theoretic standpoints. We define the models in a way that allows us to directly incorporate the structure of the relevant combinatorial problems into the description of a game, which promises applicability of game-theoretic reasoning, and consider different issues concerning the game-theoretic solution concepts of these games. In order to investigate the various game-theoretic solution concepts we use advanced tools and methodologies borrowed from the fields of combinatorial optimization and approximation algorithms, such as linear and integer programming, probabilistic analysis, primal-dual approach and the weighting functions technique.

As scheduling and packing problems naturally describe many scenarios encountered in networks, we believe that the study of these problems under game-theoretic framework is substantial for enhancing our understanding of various processes occurring in modern decentralized systems and providing us with new insights on their effect on the functionality of these systems.

In fact, this line of research has incentives that stretch far beyond the Internet and its applications, as such problems can be also used to model analogous issues that arise in any given traffic network which involves selfish agents. These include transportation and telecommunication networks, telephone and cellular networks, electric power supply systems and other economic applications. Hence, sound analysis of the combinatorial problems mentioned above is of practical importance for many settings.

In the rest of the thesis, however, we stick to the terminology of a communication network.

## 1.3 Preliminaries

### 1.3.1 Definitions and notations

To establish notation that is used throughout this thesis, we will briefly introduce the basic concepts from Game Theory. A non-cooperative strategic game is a tuple $G = \langle N, (S_i)_{i\in N}, (u_i)_{i\in N}\rangle$, where $N$ is a non-empty, finite set of rational players, each player $i \in N$ has a non-empty, finite set $S_i$ of *strategies* (actions) and a payoff function $u_i$. Each player chooses a strategy independently of the choices of the other players. The choices of all players can thus be thought to be made simultaneously. It is assumed that each player has a full knowledge over all possible strategies of all the players. In a setting of *pure* strategies, each player chooses exactly one strategy (with probability one); in a setting of *mixed* strategies, each player uses a probability distribution over the strategies. A combination of strategies chosen by the players $s = (x_j)_{j\in N} \in \times_{j\in N} S_j$, is called a *strategy profile* or a *configuration*. $X = \times_{j\in N} S_j$ denotes the set of the strategy profiles. Let $i \in N$. $X_{-i} = \times_{j\in N\setminus\{i\}} S_j$ denotes the strategy profiles of all players except for player $i$. Let $A \subseteq N$. $X_A = \times_{j\in A} S_j$ denotes the set of strategy profiles of players in $A$. Strategy profiles $s = (x_j)_{j\in N} \in X$ will be denoted by $(x_i, x_{-i})$ or $(x_A, x_{N\setminus A})$ if the strategy choice of player $i$ or of the set $A$ of players needs stressing. The payoff function $u_i : X \to \mathbb{R}$ specifies for each strategy profile $s \in X$ player *i*'s payoff $u_i(x) \in \mathbb{R}$. The payoff of each player depends not only on his own strategy but also on the strategies chosen by all other players. In some games it is more suitable to think of the payoffs as the costs incurred by players. In this case we have a cost function $c_i : X \to \mathbb{R}$ such that $u_i(s) = -c_i(s)$ for $s \in X$ instead of the payoff function. Each strategic player $i \in N$ would like to choose a strategy that maximizes his payoff, or interchangeably, minimizes his cost.

The most prominent stability (solution) concept in a strategic game is the *Nash equilibrium* (NE) [90] which was introduced by Nash in 1951. It is a stable state of the game where no player can increase his payoff by unilaterally changing his strategy, while the strategies of all other players remain unchanged. Formally,

**Definition 1.** *A strategy profile $s \in X$ is called a pure Nash equilibrium if for every $i$ and for all strategies $x'_i \in S_i$, $u_i(x_i, x_{-i}) \geq u_i(x'_i, x_{-i})$ holds.*

It is often more realistic to assume that players choose pure strategies rather than randomizing over many strategies, which is not appropriate for many settings. Thus, when we mention Nash equilibria throughout the chapter we consider the setting of pure strategies, unless it is specifically stated otherwise.

Nash equilibrium (perhaps in mixed strategies) exists in every finite game [90]. A game can have several Nash equilibria, with different social cost values. However, if only pure strategies are allowed, there may exist no Nash equilibrium at all. The set of pure Nash equilibria of a game $G$ is denoted by $NE(G)$.

Games as defined above assume that players are independent, and do not negotiate and cooperate with each other. Coalitional Game Theory considers cooperative games, where the notion of players is replaced by the set of possible coalitions (i.e., groups of players) rather than individuals. Of course, the players remain strategic, and agree to participate, if at all, in a coalition that ensures them a benefit from participation in that coalition. Each coalition can achieve a particular value (the best possible sum of payoffs among players in the coalition, against worst-case behavior of players outside the coalition).

While Nash equilibria are resilient to unilateral deviations of individuals, they are usually not stable against joint deviations of two or more cooperating players.

In 1959 Aumann proposed the concept of Strong Nash equilibrium (SNE) [10], which is resilient to coalitional deviations. This is a stable state where no subset of players can jointly deviate by simultaneously changing strategies in a manner that increases the payoffs of all its members, given that nonmembers stick to their original strategies. Formally,

**Definition 2.** *A strategy profile $s \in X$ is called a Strong Nash equilibrium if for every subgroup of players $S \subseteq N$ and for all strategy profiles $y_S \in X_S$, there is at least one player $i \in S$ such that $u_i(x_S, x_{-S}) \geq u_i(y_S, x_{-S})$.*

Since each player can either participate or decline to participate in a coalition, given the strategy he will be obligated to choose in case he does, and the payoff he will receive as a result, the Strong Nash equilibrium is typically studied only for settings that involve no randomization, that is, only pure strategies are considered. The set of Strong Nash equilibria of a game $G$ is denoted by $SNE(G)$.

Every Strong Nash equilibrium is a Nash equilibrium (as it is resilient to deviation by coalitions with a single player). Hence, $SNE(G) \subseteq NE(G)$. The opposite does not usually holds. A game can have no Strong Nash equilibrium at all. Several specific classes of congestion games (where players compete for network resources, and the cost of using each resource is a function of the number of players using it) were shown in [69, 102, 67] to possess Strong Nash equilibria. For any other game, the existence of Strong equilibria should be checked specifically in each case. Other variants of Strong equilibria that were studied consider static predefined coalitions [68, 59, 53] and coalitions that are not subject to deviations by subsets of their own members [106].

The *social cost* of a game $G$, is an objective function $SC(s) : X \to \mathbb{R}$ that numerically expresses the aggregated "social value" or "social cost" (depends whether we use a payoff or a cost interpretation) of an outcome of the game for a strategy profile $s \in X$. The *social optimum* of a game $G$, is the game outcome that optimizes the social cost function. It is denoted by OPT, and its value is defined by $OPT(G) = \min_{s \in X} \setminus \max_{s \in X} SC(s)$ (depending on the model in question).

### 1.3.2 Measuring the efficiency of equilibria

As there is often a discrepancy between the private goals of the selfishly motivated players and the global social goal, it may result in a negative effect on the system performance. Even in very simple settings, selfish behavior can lead to highly inefficient outcomes (see e.g. the well known Prisoners Dilemma).

An important matter concerns quantifying the efficiency loss incurred to the system by the selfish and uncoordinated behavior of system participants. This is achieved by considering the quality or "efficiency" of the game-theoretic stability concepts. The analysis of the equilibria can lead to conclusions on whether a system can survive even without a centralized protocol. There is an expansive body of work on quantifying this "efficiency" of various types of equilibria (for a comprehensive survey, go to Part III of the text by Nisan et al. [91]). We will concentrate herein on the measures that are relevant for the scope of our contribution.

In their seminal paper from 1999, Koutsoupias and Papadimitriou [79] proposed to measure the quality of Nash equilibria that were reached with uncoordinated selfish players with respect to the hypothetical social optimum, that was obtained in a fully coordinated manner, when the methodology used is worst-case approach.

For this purpose they have introduced a measure called the *Price of Anarchy* (POA) (also referred to as the *Coordination Ratio* (CR)), which is defined as the *worst* case ratio between the social value of the worst Nash equilibrium and the value of a social optimum, for a minimization problem (or as the *worst* case ratio between the value of a social optimum to the social value of the worst Nash equilibrium, for a maximization problem). Formally, the Price of Anarchy of a game $G$ is defined by

$$PoA(G) = \sup_{s \in NE(G)} \frac{SC(s)}{OPT(G)}.$$

In this approach we are seeing the Nash equilibrium as an approximated solution to a problem, and the quality of Nash equilibria is reflected by the closeness of the value of the least efficient Nash equilibrium to the value of the optimal solution. If the Price of Anarchy of a game is close to 1, we can conclude that all Nash equilibria states of this game are good approximations of an optimal outcome.

Koutsoupias and Papadimitriou have considered in [79] the Price of Anarchy in a job scheduling game. In the expansive volume of work that was motivated by this paper, the Price of Anarchy has been studied in different settings, that include selfish routing with atomic [12, 30, 94, 58] and non-atomic [95, 33] players (that each control a nonnegligible/negligible fraction of the overall traffic, respectively), variants of job scheduling games [36, 84, 13, 61], congestion games [24, 101, 2, 96, 34], network creation/formation games [4, 50, 50, 32], facility location games [104] and many others.

Schulz an Stier Moses [98] suggested to take a less pessimistic approach, and consider the worst case ratio between the value of the *best* Nash equilibrium and the the value of a coordinated social optimum, for a minimization problem (or as the *worst* case ratio between the value of a social optimum to the social value of the best Nash equilibrium, for a maximization problem). This ratio was called the *Price of Stability* (POS) (also known as optimistic Price of Anarchy) by Anshelevich at el. in [7, 6]. Formally, the Price of Stability of a game $G$ is defined by

$$PoS(G) = \inf_{s \in NE(G)} \frac{SC(s)}{OPT(G)}.$$

The Price of Stability has a natural interpretation in many network problems, as in reality, in many networking applications the players are not completely autonomous. They interact with an underlying protocol that proposes a collective solution to all participants, who can either accept to follow it or to defect from it. Best Nash equilibrium is an obvious stable solution that can be proposed by the protocol designer, from which no selfish player would want to unilaterally defect. The Price of Stability in this context measures the minimum penalty in performance required to ensure a stable equilibrium outcome, quantifying the benefit of such protocols.

The POA and POS have become prevalent measures of the quality of Nash equilibria in the following computer science literature.

Recent research by Andelman et al. [5] has initiated a study of measures that allow us to separate the effect of the lack of coordination among the players from the effect of their selfishness. This is achieved by acknowledging the ability of the players to cooperate, and analyzing the Strong Nash equilibria of the game. The measures considered are the *Strong Price of Anarchy* (SPOA) and the *Strong Price of Stability* (SPOS). These measures are defined similarly to the *Price of Anarchy* and the *Price of Stability*, but only Strong equilibria are taken into account. For demonstration of this line of study, see e.g. [55, 3, 41, 48].

In cases where the SPOA is significantly lower than the POA, we may conclude that the efficiency loss is caused mainly by the lack of coordination and not by the selfishness, hence, coordination can significantly improve the situation. On the other hand, if the SPOA and POA yield similar results, the efficiency loss is caused from selfishness alone, and coordination will not help. In case that the SPOS is notably greater than the POS, this implies that coordination among selfish players may cause a deterioration in the efficiency.

## 1.4 Thesis outline and organization

Having motivated the results of this thesis and introduced the necessary terminology, we sketch below the structure of the chapters, and the technical statements of each result.

**Chapter 2: Strict and Weak Pareto Nash Equilibria in the Job Scheduling game.** In this chapter we consider the well-known job scheduling game, where the jobs are controlled by selfish players that each wishes to minimize the load of the machine on which it is executed, while the social goal is to minimize the makespan, that is, the maximum load of any machine. We study this problem on the three most common machines models, identical machines, uniformly related machines and unrelated machines, with respect to both weak and strict Pareto optimal Nash equilibria. These are kinds of equilibria which are stable not only in the sense that no player can improve its cost by changing its strategy unilaterally, but in addition, there is no alternative choice of strategies for the entire set of players where no player increases its cost, and at least one player reduces its cost (in the case of strict Pareto optimality), or where all players reduce their costs (in the case of weak Pareto optimality).

We give a complete classification of the social quality of such solutions with respect to an optimal solution, that is, we find the Price of Anarchy of such schedules as a function of the number of machines, $m$. We also discuss a notion of preserving deviations, which are deviations from a Nash equilibrium where every player keeps the same cost that it had.

In addition, we give a full classification of the recognition complexity of such schedules.

**Chapter 3: Bin Packing of selfish items.** In this chapter we consider the bin packing game where the items are controlled by selfish players. Each player is charged with a cost according to the fraction of the used bin space its item requires. That is, the cost of the bin is split among the players, proportionally to their sizes. The selfish players prefer their items to be packed in a bin that is as full as possible, whereas the social goal is to minimize the number of the bins used in the packing.

We measure the quality of the (pure) Nash equilibria in this game using the standard measures Price of Anarchy and Price of Stability We also consider the recently introduced measures Strong Price of Anarchy and Strong Price of Stability, that are defined similarly, but consider only Strong Nash equilibria.

We give nearly tight lower and upper bounds of 1.6416 and 1.6428, respectively, on the (asymptotic) Price of Anarchy of the bin packing game, improving upon previous result by Bilò [18]. As for the Strong Nash equilibria of the bin packing game, we show that a packing is a Strong Nash equilibrium iff it is produced by the Subset Sum algorithm for bin packing. This characterization implies that the Strong Price of Anarchy of the bin packing game exactly equals the approximation ratio of the Subset Sum algorithm, for which a tight bound is known [46]. Moreover, the fact that any lower bound instance for the Subset Sum algorithm can be converted by a small modification of the item sizes to a lower bound instance on the Strong Price of Stability, implies that in the bin packing game these measures have the same value. Finally, we address the issue of complexity of computing a Strong Nash packing and show that no polynomial time algorithm exists for finding Strong Nash equilibria, unless P=NP.

**Chapter 4: Parametric Bin Packing of selfish items.** In this chapter we study the parametric variant of the bin packing game which was considered in Chapter 2, where the items that are controlled by selfish players are have sizes in the interval $(0, \alpha]$ for some parameter $\alpha \leq 1$.

We study the (pure) Nash equilibria of the parametric Bin Packing game and show nearly tight upper and lower bounds on the Price of Anarchy for all values of $\alpha \leq 1$.

We also consider the Strong Nash equilibria in this game and provide tight bounds for the Strong Price of Anarchy and Strong Price of Stability for all values of $\alpha \leq 1$.

**Chapter 5: Machine Covering on identical machines.** In this chapter we consider a scheduling problem where each job is controlled by a selfish agent, who is interested in minimizing its own cost, which is defined as the total load on the machine that its job is assigned to. We consider the social objective of maximizing the minimum load (cover) over the machines. We study the Price of Anarchy and the Price of Stability in the induced game. Unlike the regular scheduling problem with makespan minimization objective, for which these measures were extensively studied, they have not been considered in the setting before.

Since these measures are unbounded already for two uniformly related machines (see Chapter 5), we focus on identical machines. We show that the Price of Stability is 1, and we derive tight bounds on the Price of Anarchy for $m \leq 6$ and nearly tight bounds for general $m$. Specifically, we show that the Price of Anarchy is at least 1.691 for large $m$ and at most 1.7. Hence, surprisingly, the Price of Anarchy for this problem is less than the Price of Anarchy for the makespan problem, which is 2. To achieve the upper bound of 1.7, we make an unusual use of weighting functions. In addition, we consider the mixed

Price of Anarchy and show that the mixed Price of Anarchy grows exponentially with $m$ for this problem, whereas it is only $\Theta(\log m/\log\log m)$ for the makespan [78, 36].

Finally, we consider a similar setting with a different objective which is minimizing the maximum ratio between the loads of any pair of machines in the schedule. We show that under this objective for general $m$ the Price of Stability is 1, and the Price of Anarchy is 2.

**Chapter 6: Machine Covering on uniformly related machines.** In this chapter we consider the scheduling game introduced oin Chapter 4 in a setting of uniformly related machines.

We study the Price of Anarchy and the Price of Stability for this setting. As on related machines both these values are unbounded, we consider these measures as a function of the parameter $s_{max}$, which is the maximum speed ratio between any two machines.

We show that on related machines, Price of Stability is unbounded $s_{max} > 2$, and the Price of Anarchy is unbounded for $s_{max} \geq 2$. We study the remaining cases and show that while the Price of Anarchy tends to grow to infinity as $s_{max}$ tends to $2$, the Price of Stability is at most $2$ for any $s_{max} \leq 2$. Finally, we analyze the Price of Anarchy and Price of Stability for the case $m = 2$.

**Chapter 7** concludes this thesis with a short summary of our contributions. We describe possible extensions of the models studied in the framework of this thesis that seem interesting to investigate, and suggest directions for further research.

## 1.5 Publications related to this thesis

Some of the results presented in this thesis are published in parts as joint work in the Proceedings of the *16th Annual European Symposium on Algorithms* (ESA'08) [43], Proceedings of the *10th International Conference on Autonomous Agents and Multiagent Systems* (AAMAS 2011) [44], Proceedings of the *5th International Workshop on Internet and Network Economics* (WINE'09) [46, 47], and Algorithmica [45].

# Chapter 2

# Strict and Weak Pareto Nash equilibria in the Job Scheduling game

## 2.1 Introduction and motivation

When we design protocols intended for use in a modern communication network such as the Internet, we have to take into account its inherent characteristics.

The Internet consists of multiple independent users, or players, which act in their own self-interest and strive to optimize their private objectives, as opposed to optimizing a global social objective. Obviously, such collective behavior often leads to sub-optimal performance of the system, which is highly undesirable.

However, because of its scale and complexity and in presence of raw economic competition between the parties involved, there is no possibility to introduce a single regulatory establishment enforcing binding commitments on the players.

An important issue to explore in this context is the middle-ground between centrally enforced protocols and completely unregulated anarchy.

Selfishly oriented players may have no incentive to cooperate and follow a predefined protocol and may prefer to selfishly deviate from protocols if it is not aligned with their interests. In light of the above, there is an increased need to design efficient protocols that deploy economic measures that motivate self-interested agents to cooperate. Here cooperation may be defined as any enforceable commitment that makes it rational for the self interested players to choose a given strategic profile (which is being initially offered by a central authority).

In the settings in discussion, any meaningful agreement between the players must be self-enforcing. When deciding which particular strategy profile to offer to the users, the first and most basic requirement one has to consider is its stability, in a sense that no player would have an interest to unilaterally defect from this profile, given that the other players stick to it. This is consistent with the notion of Nash equilibrium (NE) [90], which is a widely accepted concept of stability in non-cooperative game theory. The second requirement is that the profile must be efficient. A fundamental concept of efficiency considered in economics is the Pareto efficiency, or Pareto optimality [89]. This efficiency criterion assures that it is not possible that a group of players can change their strategies so that every player is better off (or no worse off) than before.

One may justifully argue that Nash stability and Pareto optimality should be minimal requirements for any equilibrium concept intended to induce self-enforceability in presence of selfishness.

There are even stronger criteria for self-enforceability, requiring fairness in terms of fair competition without coalitions (like cartels and syndicates), and demanding from the profile to be resilient to groups (or coalitions) of players willing to coordinate their decisions, in order to achieve mutual beneficial outcomes. This is compatible with the definition of Strong Nash equilibrium (SNE) [10]. However, this requirement is sometimes too strong that it excludes many reasonable profiles.

We therefore restrict ourselves to profiles that satisfy the requirements of Nash stability and Pareto efficiency. In a sense, Pareto optimal Nash equilibria can be considered as intermediate concepts between Nash and Strong Nash equilibria; One may think of a Pareto optimal equilibrium as being stable under moves by single players or the grand coalition of all players, but not necessarily arbitrary coalitions. We distinguish between two types of Pareto efficiency. In a *weakly* Pareto optimal Nash equilibrium (WPO-NE) there is no alternative strategy profile beneficial for all players. A *strictly* Pareto optimal Nash equilibrium (SPO-NE) is also stable against deviations in which some players do not benefit but are also not worse off and at least one player improves his personal cost. Obviously, any *strictly* Pareto optimal equilibrium is also *weakly* Pareto optimal, but not wise-versa.

In this work we consider strict and weak Pareto optimal Nash equilibria for scheduling games on three common machine models: identical machines, uniformly related machines and unrelated machines in the setting of pure strategies. This class of games is particularly important to our discussion as it models a great variety of problems in modern networks. Example applications include bandwidth sharing in ATM networks [31], market-based protocols for scheduling or task allocation [105], and congestion control protocols [77].

We investigate the quality of these solution concepts in the job scheduling game by comparison to an optimal solution, adopting a worst-case approach. As both papers that originally suggested to compare the NE and SNE to an optimal solution to study their quality ([79] and [5], respectively) demonstrated this approach first in scheduling games, this gives an additional incentive to consider these solution concepts for this particular game class.

## 2.2 The model

We now define the general job scheduling problem. There are $n$ jobs $J = \{1, 2, \ldots, n\}$ which are to be assigned to a set of $m$ machines $M = \{M_1, \ldots, M_m\}$. We study three models of machines, that differ in the relation between the processing times of jobs on different machines. In the most general model of *unrelated* machines, job $1 \leq k \leq n$ has a processing time of $p_{ik}$ on machine $M_i$, i.e., processing times are machine dependent. In the *uniformly related* (or *related*) machine model, each machine $M_i$ for $1 \leq i \leq m$ has a speed $s_i$ and each job $1 \leq k \leq n$ has a positive size $p_k$. The processing time of job $k$ on machine $M_i$ is then $p_{ik} = \frac{p_k}{s_i}$. If $p_{ik} = p_{i'k} = p_k$ for each job $k$ and machines $M_i$ and $M_{i'}$, the machines are called *identical* (in which case it is typically assumed the all speed are equal to $1$).

An assignment or schedule is a function $\mathcal{A} : J \rightarrow M$. The load of machine $M_i$, which is also called the delay of this machine, is $L_i = \sum_{k:\mathcal{A}(k)=M_i} p_{ik}$. The cost, or the *social cost* of a schedule is the maximum delay of any machine, also known as the *makespan*, which we would like to minimize.

The job scheduling game $JS$ is characterized by a tuple $JS = \langle N, (\mathcal{M}_k)_{k\in N}, (c_k)_{k\in N}\rangle$, where $N$ is the set of atomic players. Each selfish player $k \in N$ controls a single job and selects the machine to which it will be assigned. We associate each player with the job it wishes to run, that is, $N = J$. The set of strategies $\mathcal{M}_k$ for each job $k \in N$ is the set $M$ of all machines. i.e. $\mathcal{M}_k = M$. Each job must be assigned to one machine only. Preemption is not allowed. The outcome of the game is an assignment $\mathcal{A} = (\mathcal{A}_k)_{k\in N} \in \times_{k\in N} M_k$ of jobs to the machines, where $\mathcal{A}_k$ for each $1 \leq k \leq n$ is the index of the machine that job $k$ chooses to run on. Let $\mathcal{S}$ denote the set of all possible assignments.

The cost function of job $k \in N$ is denoted by $c_k : \mathcal{S} \rightarrow \mathbb{R}$. The cost $c_k^i$ charged from job $k$ for running on machine $M_i$ in a given assignment $\mathcal{A}$ is defined to be the load observed by machine $i$ in this assignment, that is $c_k(i, \mathcal{A}_{-k}) = L_i(\mathcal{A})$, when $\mathcal{A}_{-k} \in \mathcal{S}_{-k}$; here $\mathcal{S}_{-k} = \times_{j\in N\setminus\{k\}}\mathcal{S}_j$ denotes the actions of all players except for player $k$. The goal of the selfish jobs is to run on a machine with a load which is as small as possible. Similarly, for $K \subseteq N$ we denote by $\mathcal{A}_K \in \mathcal{S}_{-K}$ the set of strategies of players outside of $K$ in a strategy profile $\mathcal{A}$, when $\mathcal{S}_{-K} = \times_{j\in N\setminus K}\mathcal{S}_j$ is the action space of all players except for players in $K$. The social cost of a strategy profile $\mathcal{A}$ is denoted by $SC(\mathcal{A}) = \max_{1\leq k\leq n} c_k(\mathcal{A})$.

We will next provide formal definitions of Nash, Weak/Strict Pareto Nash and Strong Nash equilibria in the job scheduling game, using the notations given above.

**Definition 3.** *(Nash equilibrium) A strategy profile $\mathcal{A}$ is a (pure) Nash equilibrium (*NE*) in the job scheduling game $JS$ if for all $k \in N$ and for any strategy $\bar{\mathcal{A}}_k \in M$, $c_k(\mathcal{A}_k, \mathcal{A}_{-k}) \leq c_k(\bar{\mathcal{A}}_k, \mathcal{A}_{-k})$.*

It was shown that job scheduling games always have (at least one) pure Nash equilibrium [57, 49]. We denote the set of Nash equilibria of an instance $G$ of the job scheduling game by NE($G$).

**Definition 4.** *(Strong Nash equilibrium) A strategy profile $\mathcal{A}$ is a Strong Nash equilibrium (*SNE*) in the job scheduling game $JS$ if for every coalition $\phi \neq K \subseteq N$ and for any set of strategies $\bar{\mathcal{A}}_K \in \times_{j \in K} \mathcal{M}_j$ of players in $K$, there is a player $i \in K$ such that $c_i(\bar{\mathcal{A}}_K, \mathcal{A}_{-K}) \geq c_i(\mathcal{A}_K, \mathcal{A}_{-K})$.*

Existence of Strong Nash equilibrium in job scheduling games was proved in [5]. We denote the set of Strong Nash equilibria of an instance $G$ of the job scheduling game by SNE($G$).

Clearly SNE($G$)$\subseteq$NE($G$), as coalitions of size 1 can not improve by changing their strategy.

**Definition 5.** *(Weak/Strict Pareto optimal profile) A strategy profile $\mathcal{A}$ is weakly Pareto optimal (WPO) if there is no strategy profile $\bar{\mathcal{A}}$ s.t. for all $k \in N$, $c_k(\bar{\mathcal{A}}) < c_k(\mathcal{A})$.*

*A strategy profile $\mathcal{A}$ is strictly Pareto optimal (SPO) if there is no strategy profile $\bar{\mathcal{A}}$ and $k^* \in N$ s.t. for all $k \in N \backslash k^*$, $c_k(\bar{\mathcal{A}}) < c_k(\mathcal{A})$ and $c_{k^*}(\bar{\mathcal{A}}) \leq c_{k^*}(\mathcal{A})$.*

We denote by SPO($G$) and WPO($G$), respectively, the sets of strictly and weakly Pareto optimal profiles of an instance $G$ of the job scheduling game. Clearly, SPO($G$)$\subseteq$ WPO($G$).

A strategy profile $\mathcal{A} \in$ NE($G$)$\cap$ WPO($G$) is called Weak Pareto optimal Nash equilibrium (WPO-NE), and a strategy profile $\mathcal{A} \in$NE($G$) $\cap$ SPO($G$) is called Strict Pareto optimal Nash equilibrium (SPO-NE), and these are the profiles that we put our focus on.

We note that every strong equilibrium is also weakly Pareto optimal, as the requirement in Definition 4 applies to the grand coalition of all players. Hence SNE($G$)$\subseteq$ WPO($G$). The existence of Strong Nash equilibria in job scheduling games assures the existence of weak Pareto optimal Nash equilibria.

On the other hand, in general, neither Nash equilibria nor Strong Nash equilibria are necessarily strictly Pareto optimal. Existence of strict Pareto optimal Nash equilibria in scheduling games (among others) was proved in [67].

An important issue concerns the quality of these solution concepts. As there is a discrepancy between the private goals of the players and the global social goal, we would like to measure the loss in the performance of the system as it is reflected by the closeness of the costs of these concepts to the cost of the optimal solution, when the accepted methodology is worst-case approach.

The quality measures which consider Nash equilibria are the Price of Anarchy and the more optimistic Price of Stability, which are defined as the worst-case ratio between the social cost of the worst/best NE schedule to the social cost of the optimal schedule, which is denoted by OPT. Formally,

**Definition 6.** *(Price of Anarchy and Stability) The Price of Anarchy ($PoA$) of the job scheduling game $JS$ is defined by*

$$PoA(JS) = \sup_{G \in JS} \sup_{\mathcal{A} \in NE(G)} \frac{SC(\mathcal{A})}{\text{OPT}(G)}.$$

*If instead we consider the best Nash equilibrium of every instance, this leads to the definition of the Price of Stability ($PoS$):*

$$PoS(JS) = \sup_{G \in JS} \inf_{\mathcal{A} \in NE(G)} \frac{SC(\mathcal{A})}{\text{OPT}(G)}.$$

This concept is applied analogously to Strong Nash equilibria as well as to weakly/strictly Pareto optimal Nash equilibria yielding the Strong Price of Anarchy $SPoA(JS)$ and the Strong Price of Stability $SPoS(JS)$ as well as the weak and strict Pareto Prices of Anarchy $WPO\text{-}PoA(JS)$, $SPO\text{-}PoA(JS)$ and Stability $WPO\text{-}PoS(JS)$, $SPO\text{-}PoS(JS)$.

By definition, it is clear that $SPoA(JS) \leq WPO\text{-}PoA(JS) \leq PoA(JS)$. As any *strictly* Pareto optimal NE is also a *weakly* Pareto optimal NE, it must be the case that $WPO\text{-}PoA(JS) \geq SPO\text{-}PoA(JS)$. However, we can show that in the job scheduling game there is no immediate relation between the $SPO\text{-}PoA(JS)$ and the $SPoA(JS)$, as there are Strong Nash equilibria that are not *strictly* Pareto optimal, while there are *strictly* Pareto optimal Nash equilibria that are not strong equilibria.

Some natural questions that arise in this context are whether the Pareto Prices of Anarchy are significantly smaller than the standard Price of Anarchy, whether the weak Pareto Price of Anarchy is much larger than the Strong Price of Anarchy, and finally, whether there is any relation between the Strong Price of Anarchy and the strict Pareto Price of Anarchy. In other words, does the requirement that the equilibrium must be Pareto optimal leads to greater efficiency, and is the further demand that the equilibrium must be stable against arbitrary coalitions helpful. We address these questions in our work, and provide conclusive answers.

## 2.3 Related work and our contributions

Pareto efficiency of resource assignments is a well referred issue in economics, especially in welfare economics. Pareto efficiency is a highly desirable trait, however Dubey [39] has shown that Nash equilibria may generally be Pareto inefficient based on the difference between the conditions to be satisfied by Nash equilibria and those to be satisfied by Pareto optima.

Job scheduling is a classical problem in combinatorial optimization. The analysis of job scheduling in the algorithmic game theory context was initiated by Koutsoupias and Papadimitriou in their seminal work [79], which was followed by many others (see e.g. [36, 84, 5, 55]). In our overview of the known results we will limit our discussion only to results concerning the model defined above in the setting of pure strategies. We will begin with the results on quality measures that concern Nash equilibria of the game.

For $m$ identical machines, the $PoA$ is $2 - \frac{2}{m+1}$ which can be deduced from the results of [56] (the upper bound) and [99] (the lower bound). For related machines the $PoA$ is $\Theta(\frac{\log m}{\log \log m})$ [78, 36, 79]. In the model of unrelated machines the $PoA$ is unbounded [13], which holds already for two machines. From the results of [49] it is evident that in all three models the $PoS$ is 1.

The study of quality measures that concern Strong Nash equilibria of this game was initiated by Andelman at el. [5]. For identical machines, they proved that the $SPoA$ equals the $PoA$, which in turn equals $2 - \frac{2}{m+1}$. For related machines, Fiat et al. [55] showed that the $SPoA$ is $\Theta(\frac{\log m}{(\log \log m)^2})$. Surprisingly, the $SPoA$ for this problem is bounded by the number of machines $m$, as shown in [55], and this is tight [5]. Andelman at el. also showed that $SPoS$ is 1.

The previous work on Pareto efficiency of Nash equilibria in algorithmic game theory was mainly concerned with weak Pareto equilibria, probably since a solution which is not weakly Pareto optimal is clearly unstable. A textbook in economics states the following: "The concept of Pareto optimality originated in the economics equilibrium and welfare theories at the beginning of the past century. The main idea of this concept is that society is enjoying a maximum ophelimity when no one can be made better off without making someone else worse off" [83]. In practice, however, the strict Pareto is a stronger and more meaningful efficiency notion, as it captures an important aspect of human social behavior; When something could be done to make at least one person better off without hurting anyone else, most people would agree we should do it. Another issue is that the weak Pareto suggests that some assignment is socially preferable to another by everyone. In reality, such unanimity of preferences among all persons is very rare. To conclude, both concepts are important, and we focus on both of them in this work.

Pareto optimality of Nash equilibria has been studied in the context of congestion games, see Chien and Sinclair [23] and Holzman and Law-Yone [69]. The former gave conditions for uniqueness and for weak and strict Pareto optimality of Nash equilibria, and the latter characterized the weak Pareto Prices of Anarchy and Stability. The existence, and complexity of recognition and computation of weak Pareto Nash equilibria in congestion games was considered recently by Hoefer and Skopalik in [100].

Milchtaich [86] has considered the topology of networks having NE that are always Pareto efficient, however his work concerns the case of non-atomic players, where the processing time of each player is negligible compared to the total processing time.

In [67] Harks at el. show that a class of games that have a *Lexicographical Improvement Property* (which our game indeed has) admits a generalized strong ordinal potential function. They use this to show existence of Strong Nash equilibria with certain efficiency and fairness properties in these games, strict Pareto efficiency included. Which they do by arguing that a player wise cost-lexicographically minimal assignment is also strictly Pareto optimal (and so it is optimal with respect to the social goal function as well).

Weak Pareto Nash equilibria in routing and job scheduling games were considered recently in [11] by Aumann and Dombb. As a measure for quantifying the distance of a best/worst Nash equilibrium from being weakly Pareto efficient, they use the smallest factor by which any player improves its cost when we move to a different strategy profile, which they refer to as "Pareto inefficiency". They do not consider however the quality of Pareto optimal Nash equilibria with respect to the social goal.

Among other results, it is shown in [11] that any Nash equilibrium assignment is necessarily weakly Pareto optimal for both identical and related machines. Moreover, for any machine model, any assignment which achieves the social optimum must be weakly Pareto optimal. One such assignment is one whose sorted vector of machine loads is lexicographically minimal is necessarily weakly Pareto optimal (see also [5, 49]). We consider these issues for SPO-NE assignments. We show that while the property of identical machines remains true, this is not the case for related machines, that is, not every Nash equilibrium assignment is strict Pareto optimal. For unrelated machines, while there always exist an assignment which is a social optimum and a SPO-NE, assignments with lexicographically minimal sorted vector of machine loads are not necessarily strictly Pareto optimal. In this chapter we fully characterize the weak and strict Pareto Prices of Anarchy of the job scheduling game in cases of identical, related and unrelated machines. The characterization of the Prices of Stability follows from previous work as explained above.

Next, we consider the complexity of recognition of weak and strict Pareto optimality of NE. Note that the recognition of NE can be done in polynomial time for any machine model by examining potential deviations of each job. As for strong equilibria, it was shown by Feldman and Tamir [52] that it is NP-hard to recognize an SNE for $m \geq 3$ identical machines and for $m \geq 2$ unrelated machines. For two identical machines, they showed that any NE is a SNE, so recognition can be done in polynomial time (for $m \geq 3$, it was shown in [5] that not every NE is a SNE). For the only remaining case of two related machines, it was shown [42] that recognition is again NP-hard. We show that the situation for Pareto optimal equilibria is slightly different. In fact, recognition of WPO-NE or SPO-NE can be done in polynomial time for identical machines and related machines. For unrelated machines, we show that the recognition of WPO-NE is NP-hard in the strong sense and the recognition of SPO-NE is NP-hard.

We reflect upon the differences between the results for weak and strict Pareto equilibria also compared to strong equilibria, and make conclusions regarding the relations between the quality measures in this game. See Table 2.1 below for a summary of the results.

We also discuss a notion of preserving deviations, which are deviations from a Nash equilibrium where every player keeps the same cost that it had before.

| | # of machines | Strict Pareto | | | Weak Pareto | | |
|---|---|---|---|---|---|---|---|
| | | SPO-PoA | SPO-PoS | Recognition | WPO-PoA | WPO-PoS | Recognition |
| identical | $m$ | $2 - \frac{2}{m+1}$ | **1** [67] | P | $2 - \frac{2}{m+1}$ | **1** [5, 11] | P |
| related | $m$ | $\Theta(\frac{\log m}{\log \log m})$ | **1** [67] | P | $\Theta(\frac{\log m}{\log \log m})$ | **1** [5, 11] | P |
| unrelated | $m = 2$ | 2 | **1** [67] | NP-hard | 2 | **1** [5, 11] | NP-hard |
| | $m \geq 3$ | $m$ | | | $\infty$ | | |

Table 2.1: Summary of Results

## 2.4 Pareto Prices of Anarchy in the Job Scheduling game

### 2.4.1 Identical and Related machines

A result from [11, 86] shows that any NE schedule for identical and related machines is weakly Pareto optimal. This result implies that $WPO\text{-}PoA(JS) = PoA(JS)$. For the case of identical machines, they give an even stronger result: every schedule where every machine receives at least one job is weakly Pareto optimal. Note that if $n < m$, then a schedule is weakly Pareto optimal if and only if at least one machine has a single job (to obtain strict Pareto optimality for this case, or to obtain a NE, each job needs to be assigned to a different machine).

In the strict Pareto case, while the general result for identical machines still holds, and the set of NE schedules is equal to the set of SPO-NE schedules for identical machines (as we prove next), it is not necessarily true for related machines. We exhibit an example of a schedule which is a NE but it is not strictly Pareto optimal.

Consider a job scheduling game with two related machines of speeds 1,2 and two jobs of size 2. There are two types of pure NE schedules: in the first one, both jobs are assigned to the fast machine, and in the other one job runs on each machine. The first one is not a SPO-NE, as switching to a schedule of the second type strictly reduces the cost for one of the jobs, while not harming the other. Moreover, the sorted machine load vector of the first type of schedules is $(2, 0)$, while the load vector of the second type is $(2, 1)$, so the schedule with the lexicographically minimal machine load vector is not a SPO-NE (even though it is a SNE).

This difference in the results for related machines is explained by the fact that conditions for weak Pareto allow Pareto improvements where not all jobs strictly improve while the strict Pareto does not. If a NE schedule has an empty machine, and a job arrived to such a machine as a result of a deviation to a different schedule, where all jobs strictly reduce their costs, then the reduction in the cost of this job contradicts the original schedule being a NE. However, if the job only needs to maintain its previous cost, then there is no contradiction.

We will prove the following theorem, extending the result of [11] which will allow us to claim that for identical machines, $SPO\text{-}PoA(JS) = PoA(JS)$.

**Theorem 7.** *Any schedule for identical machines, where no machine is empty, is strictly Pareto optimal.*

This is a stronger result than the one in [11], since it deals with strict Pareto. The idea of the proof goes along the lines of [11], but we need to modify it so that it applies for the stronger conditions of strict Pareto optimal schedules. First we prove the following property, which we will also use to characterize the $WPO\text{-}PoA$ for related machines.

**Theorem 8.** *Consider a schedule $X$ that is not a SPO-NE, and denote the set of non-empty machines (which receive at least one job) in $X$ by $\mu_X$. Let $Y$ be a different schedule where no job has a larger cost than it has in $X$ and at least one job has a smaller cost. Denote the set of non-empty machines in $Y$ by $\mu_Y$. Then,*

$$\sum_{i \in \mu_X} s_i < \sum_{i \in \mu_Y} s_i,$$

*where $s_i$ is the speed of machine $i$.*

**Proof.** Consider a transition from schedule $X$ to schedule $Y$, and denote by $x_i^j$ the sum of the sizes of jobs that are moved from machine $i \in \mu_X$ to machine $j \in \mu_Y$ (for $j = i$, this gives the sum of sizes of jobs that are assigned to this machine in both schedules). Let $\ell_t$, for $t \in \mu_X$, be the sum of sizes of jobs that run on machine $t$ in $X$, and let $\ell'_t$, for $t \in \mu_Y$, be the sum of sizes of jobs that run on machine $t$ in $Y$. We extended the definition so that if $t \notin \mu_X$, then $\ell_t = 0$, and if $t \notin \mu_Y$, then $\ell'_t = 0$.

Consider the total sum of sizes of jobs assigned to a machine in $X$ or in $Y$, then the following two claims hold:

**Claim 9.** *For every $i \in \mu_X$, $\sum\limits_{j \in \mu_Y} x_i^j = \ell_i$ or alternatively, $\sum\limits_{j \in \mu_Y} \frac{x_i^j}{\ell_i} = 1$.*

**Claim 10.** *For every $j \in \mu_Y$, $\sum\limits_{i \in \mu_X} x_i^j = \ell'_j$, or alternatively, $\sum\limits_{i \in \mu_X} \frac{x_i^j}{\ell'_j} = 1$.*

By the definition of the costs in $Y$ compared to $X$, the following claim holds:

**Claim 11.** *If $x_i^j > 0$, then*

$$\frac{\ell'_j}{s_j} \leq \frac{\ell_i}{s_i},$$

*and there exist $i \in \mu_X$,$i \in \mu_Y$ such that*

$$\frac{\ell'_j}{s_j} < \frac{\ell_i}{s_i}.$$

The following holds:

**Claim 12.** *For every $i \in \mu_X$, $j \in \mu_Y$:*

$$\frac{x_i^j}{\ell_i} \leq \frac{s_j}{s_i} \cdot \frac{x_i^j}{\ell'_j},$$

*and there exist $i, j$ such that*

$$\frac{x_i^j}{\ell_i} < \frac{s_j}{s_i} \cdot \frac{x_i^j}{\ell'_j}.$$

**Proof.** As $i \in \mu_X$, $\ell_i > 0$, as $j \in \mu_Y$, $\ell'_i > 0$. If $x_i^j > 0$, it is derived from Claim 11, if $x_i^j = 0$ it holds trivially. Since there is at least one job for which the cost in $Y$ is strictly smaller than its cost in $X$, then the second property must hold. □

Summing up the inequalities in Claim 12 over all $j \in \mu_Y$, in combination with Claim 9, we get that for any $i \in \mu_X$:

$$1 = \sum_{j \in \mu_Y} \frac{x_i^j}{\ell_i} \leq \sum_{j \in \mu_Y} \frac{s_j}{s_i} \cdot \frac{x_i^j}{\ell'_j}, \tag{2.1}$$

where there is at least one $i \in \mu_X$ for which this inequality is strict. Equivalently,

$$s_i \leq \sum_{j \in \mu_Y} \frac{x_i^j \cdot s_j}{\ell'_j} \ . \tag{2.2}$$

Summing up the inequality in (2.2) over all $i \in \mu_X$ combined with the fact that for some $i$ this inequality is strict, changing the order of summation, and using Claim 10 we get:

$$\sum_{i \in \mu_X} s_i < \sum_{i \in \mu_X} \sum_{j \in \mu_Y} \frac{x_i^j \cdot s_j}{\ell'_j} = \sum_{j \in \mu_Y} \sum_{i \in \mu_X} \frac{x_i^j \cdot s_j}{\ell'_j} = \sum_{j \in \mu_Y} s_j \tag{2.3}$$

which concludes our proof. □

We now return to the proof of Theorem 7.

**Proof.** We show that any schedule $X$ for identical machines where $\mu_X$=$M$ is strictly Pareto optimal. Assume by contradiction that this is not the case, and hence there exists a different schedule $Y$ where at least one job improves, while all the other jobs are not worse off. As the machines in question are identical, $s_1 = s_1 = \ldots = s_m$ holds, thus $\sum_{i \in \mu_X} s_i = m$ and $\sum_{j \in \mu_Y} s_j \leq m$. By Theorem 8 we get that $m = \sum_{i \in \mu_X} s_i < \sum_{j \in \mu_Y} s_j \leq m$, which is a contradiction, and we conclude that such $Y$ cannot exist. ☐

**Corollary 13.** *Every schedule on identical machines which is a* NE *is also a* SPO-NE. *Thus in this case* $SPO$-$PoA$=$WPO$-$PoA$=$PoA$.

**Proof.** Consider a NE schedule. If there is an empty machine, then each machine has at most one job (otherwise, if some machine has two jobs then any of them can reduce its cost by moving to an empty machine), and thus each job has the smallest cost that it can have in any schedule. Otherwise, the property follows from Theorem 7. ☐

We next consider related machines and prove that the three measures are equal in this case as well.

**Theorem 14.** *In the job scheduling game on related machines* $SPO$-$PoA$=$WPO$-$PoA$=$PoA$.

**Proof.** As any SPO-NE is also a WPO-NE, and every WPO-NE is a NE, the following sequence of inequalities holds: $SPO$-$PoA \leq WPO$-$PoA \leq PoA$. We will prove that this is actually a sequence of equalities. It is enough to prove that $PoA \leq SPO$-$PoA$. We will do it by showing that the lower bound example for the $PoA$ given in [36] is also a lower bound for the $SPO$-$PoA$, by proving that it is strictly Pareto optimal.

For completeness, we first present the lower bound of [36]. Consider a job scheduling game on $m$ related machines. The machines are partitioned into $k+1$ groups, each group $j$ ,$0 \leq j \leq k$ has $N_j$ machines. The sizes of the groups are defined in inductive manner: $N_k = \Theta(\sqrt{m})$, and for every $j < k$: $N_j = (j+1) \cdot N_{j+1}$ (and thus $N_0 = k! \cdot N_k$). The total number of machines $m = \sum_{j=0}^{k} N_j = \sum_{j=0}^{k} \frac{k!}{(k-j)!} \cdot N_k$. It follows that $k \sim \frac{\log m}{\log \log m}$. The speed of each machine in group $j$ is $s_j = 2^j$.

A schedule is defined as follows: each machine in group $j$ has $j$ jobs, each with size $2^j$. Each such job contributes 1 to the load of its machine. The load of each machine in group $N_j$ is then $j$, and therefore the makespan which is accepted on the machines in group $N_k$ is $k$. Note that all the machines in group $N_0$ are empty.

We denote this schedule by $X$. It was proven in [36] that $X$ is a pure NE. We claim that it is also strictly Pareto optimal.

**Claim 15.** $X$ *is strictly Pareto optimal.*

**Proof.** Assume by contradiction that $X$ is not a SPO-NE, so there exists another schedule $Y$ where at least one job improves, and all the other jobs are not worse off. Observe that all the machines in group $N_0$ necessarily remain empty in $Y$; each job that runs on a machine in group $N_j$ for $1 \leq j \leq k$ pays a cost of $j$ in $X$, and if it is assigned on a machine from group $N_0$ in $Y$ it has to pay a cost of $2^j$, and $2^j > j$ for $j \geq 1$, which makes it strictly worse off. This means that $\mu_Y \subseteq \mu_X$. On the other hand, according to Theorem 8 which we proved earlier, $\sum_{i \in \mu_X} s_i < \sum_{i \in \mu_Y} s_i$ must hold, and we get a contradiction. Hence, the schedule in this example is strictly Pareto optimal. □

An optimal schedule has a makespan of $2$. To obtain such a schedule, we move all jobs from machines in $N_j$ ($j \cdot N_j$ jobs, each of size $2^j$) to machines in $N_{j-1}$, for $1 \leq j \leq k$. Every machine gets at most one job, and the load on all machines is less or equal to $\frac{2^j}{2^{j-1}} = 2$. The $SPO\text{-}PoA$ is therefore $\Omega(\frac{\log m}{\log \log m})$.

We conclude that $SPO\text{-}PoA$=$WPO\text{-}PoA$=$PoA$. □

It was proved in [55] that schedule $X$ is not a SNE, as a coalition of all $k$ jobs from a machine in group $N_k$ with 3 jobs from each of $k$ different machines from group $N_{k-2}$ can jointly move in a way that reduces the costs of all its members. In addition to determining the SPO-NE, this example illustrates the point that in the job scheduling game not every SPO-NE is necessarily a SNE. We saw that for related machines, the $SPO\text{-}PoA$=$WPO\text{-}PoA$ are the same as the $PoA$, while the $SPoA$ is lower.

### 2.4.2 Unrelated machines

We saw that already for related machines, not every SNE is a SPO-NE and vice versa. However, the results which we find for the $SPO\text{-}PoA$ on unrelated machines are similar to those which are known for the $SPoA$, that is, the $SPO\text{-}PoA$ is equal to $m$ for any number of machines $m$. Interestingly, the $WPO\text{-}PoA$ for the setting $m = 2$ is exactly 2, like the $SPO\text{-}PoA$, but for $m \geq 3$ it is unbounded like the $PoA$.

**Theorem 16.** *There exists a job scheduling game with 2 unrelated machines, such that $WPO\text{-}PoA \geq 2$. For any, $m \geq 3$ there exists a job scheduling game with $m$ unrelated machines, such that $WPO\text{-}PoA$ is unbounded.*

**Proof.** Consider a job scheduling game with two unrelated machines and two jobs, where $p_{11} = p_{21} = p_{12} = 1$ and $p_{22} = 2$. A schedule where job $1$ is assigned to $M_1$ and job $2$ is assigned to $M_2$ with a makespan of 2 is a WPO-NE; No job would benefit from moving to a different machine, and job $1$ will not profit by switching to a different schedule. In an optimal schedule for this game, job $k$, $k \in \{1, 2\}$ is assigned to $M_k$, and the makespan is 1. We get that $WPO\text{-}PoA \geq 2$.

Now, consider a job scheduling game with $m \geq 3$ unrelated machines and $n = m$ jobs, where for each job $k$, $1 \leq k \leq m$: $p_{kk} = \varepsilon$, and $p_{jk} = 1$ for all $j \neq k$, for some arbitrary small positive $\varepsilon$. A schedule where job 1 is assigned to run on $M_1$, job $m$ is assigned to $M_2$ and each job $k$ for $2 \leq k \leq m-1$ is assigned to $M_{k+1}$, is a WPO-NE. It is weakly Pareto optimal since job 1 cannot decrease its cost by changing to any other assignment. The only way that another job could decrease its cost would be by moving to the machine where its cost is $\varepsilon$, but the load on all those machines is 1. Therefore, the schedule is a NE. The makespan of this schedule is 1. An optimal schedule for this game, where each job $1 \leq k \leq m$ is assigned to machine $M_k$, has a makespan of $\varepsilon$. In total, we have $WPO\text{-}PoA \geq \frac{1}{\varepsilon}$, which is unbounded letting $\varepsilon$ tend to zero. □

We next prove a matching upper bound for $m = 2$.

**Theorem 17.** *For any job scheduling game with 2 unrelated machines, $WPO\text{-}PoA \leq 2$.*

**Proof.** Consider a schedule on two unrelated machines which is a WPO-NE. Without loss of generality, assume that the load of $M_1$ is not larger than the load of $M_2$, and denote the loads of the machines are by $L_1$ and $L_2$, respectively. The makespan of this schedule is then $L_2$. We show $L_2 \leq 2\text{OPT}$. We first show $L_1 \leq \text{OPT}$. If $L_2 \geq L_1 > \text{OPT}$ then an optimal schedule has the property that every job has a smaller cost in it than it has in the current schedule (a cost of at most $\text{OPT} < L_1 \leq L_2$), in contradiction to the fact that this schedule is a WPO-NE.

To complete the proof, we upper bound $L_2$. If $L_2 \leq \text{OPT}$, then we are done, otherwise, $L_2 > \text{OPT}$, and there must exist a job $k$ assigned to $M_2$ which is assigned to $M_1$ in an optimal schedule (since the load resulting from jobs assigned to $M_2$ in an optimal schedule is no larger than OPT). Thus, $p_{1k} \leq \text{OPT}$, and in the alternative schedule, where this job moves to $M_1$, the new load of $M_1$ is at most $L_1 + p_{1k} \leq 2\text{OPT}$. However, we know that the given schedule is a NE, which means that $L_1 + p_{1k} \geq L_2$, giving $L_2 \leq 2OPT$. Therefore, $WPO\text{-}PoA \leq 2$. □

From Theorems 16 and 17 we conclude that for $m = 2$, $WPO\text{-}PoA = 2$, and for $m \geq 3$, $WPO\text{-}PoA = \infty$.

We prove next that like the $SPoA$, the $SPO\text{-}PoA$ is $m$. We should note that the previous results for the $SPoA$ cannot be used here. As we saw, the sets of SNE and SPO-NE have no particular relation. The proofs used for the $SPoA$ do not hold for the $SPO\text{-}PoA$ and need to be adapted. The lower bound of $m$ on the $SPoA$ by Andelman et al. [5] is not strictly Pareto optimal (see below), and in the proof of the upper bound by Fiat et al. [55] the claim is proved by considering alternative schedules where the jobs which change their strategies are proper subsets of jobs (so other jobs may increase their costs).

**Theorem 18.** *The $SPO\text{-}PoA$ for $m$ unrelated machines in any job scheduling game is at most $m$.*

**Proof.** Consider a schedule $\mathcal{A}$ on $m$ unrelated machines which is a SPO-NE. Assume that the machines are sorted by non-increasing order of loads, that is, $L_1 \geq L_2 \geq \ldots \geq L_m$. The makespan of $\mathcal{A}$ is therefore $L_1$.

First, note that $L_m \leq OPT$. If $L_m > OPT$ then an optimal schedule has the property that every job has a smaller cost in it, contradicting the strict Pareto optimality of $\mathcal{A}$. Next, we will prove that $L_i - L_{i+1} \leq OPT$ holds for any $1 \leq i \leq m-1$. Assume by contradiction that there exists $i$ so that $L_i - L_{i+1} > OPT$. We let $L_{i+1} = \delta$. By our assumption $L_i > \delta + OPT$ holds.

Now, consider another schedule $\mathcal{A}'$, where each one of the jobs from machines $M_j$ for $1 \leq j \leq i$ in $\mathcal{A}$ is running on the machine on which it runs in an optimal schedule (all the other jobs hold their positions).

We observe that none of these jobs runs on machines $M_{i+1}, \ldots, M_m$ in $\mathcal{A}'$ (or in the optimal schedule under consideration); The processing time of each such job in $\mathcal{A}'$ is at most OPT, and as $L_k \leq \delta$ for $i+1 \leq k \leq m$, its cost in $\mathcal{A}$ if it switches to the machines out of $M_{i+1}, \ldots, M_m$ on which its processing time is at most OPT, then the load of this machine would be at most $\delta + OPT$, while its cost in $\mathcal{A}$ was strictly larger than $\delta + OPT$, contradicting $\mathcal{A}$ being a NE.

We conclude that these jobs are scheduled in $\mathcal{A}'$ on machines $M_1, \ldots, M_i$, where the load of each one of the machines is at most OPT, and that the loads and the allocations on machines $M_{i+1}, \ldots, M_m$ do not change from $\mathcal{A}$ to $\mathcal{A}'$.

This means that in $\mathcal{A}'$ the costs of all jobs from machines $M_1, \ldots, M_i$ in $\mathcal{A}$ are strictly improved, and the costs of all jobs from machines $M_{i+1}, \ldots, M_m$ in $\mathcal{A}$ do not change, which contradicts $\mathcal{A}$ being a SPO-NE. Hence, such $i$ does not exist. Applying this inequality repeatedly, we get that $L_1 \leq L_m + (m-1)$OPT, which in combination with the fact that $L_m \leq OPT$ gives us $SPO\text{-}PoA \leq m$. □

We now provide a matching lower bound.

**Theorem 19.** *There exists an instance of job scheduling game with $m$ unrelated machines for which $SPO\text{-}PoA \geq m$.*

**Proof.** Consider a job scheduling game with $m$ unrelated machines and $n = m$ jobs, where for each job $k$, $2 \leq k \leq m$: $p_{kk} = k - k\varepsilon$, $p_{k(k-1)} = 1$ and $p_{ik} = \infty$ for all $i \neq k-1, k$. For job 1, $p_{11} = 1 - \varepsilon$ (for some small positive $\varepsilon < \frac{1}{m}$), $p_{m1} = 1$ and $p_{i1} = \infty$ for all $i \neq 1, m$, as illustrated in the matrix below (rows describe the processing times of jobs, entries not explicitly stated are $\infty$).

$$P^t = \begin{pmatrix} 1-\varepsilon & & & & & 1 \\ 1 & 2-2\varepsilon & & & & \\ & 1 & 3-3\varepsilon & & & \\ & & 1 & \ddots & & \\ & & & \ddots & (m-1)-(m-1)\varepsilon & \\ & & & & 1 & m-m\varepsilon \end{pmatrix}$$

In an optimal schedule for this game each one of the jobs $2 \leq k \leq m$ runs alone on machine $M_{k-1}$ and job 1 runs on $M_m$, which yields a makespan of 1.

On the other hand, a schedule where each one of the jobs $1 \leq k \leq m$ runs alone on machine $M_k$ has a makespan of $m - m\varepsilon$. We will show that this schedule is a SPO-NE. The schedule is a NE, since for each job, moving to the only additional machine on which its processing time is not infinite increases it cost by at least $\varepsilon$. Consider an alternative schedule where no job increases its cost. Job 1 is currently assigned on a machine with load $1-\varepsilon$, which is the minimal possible cost for it, and this minimum is unique. Thus any alternative schedule must keep job 1 assigned alone to the first machine. We can prove by induction on the indices of jobs that every job has to stay assigned to its current machine alone; once job $k$ must stay on its machine alone, job $k+1$ does not have an alternative machine, and adding another job to the machine that it is assigned to ($M_k$) would increase its cost. Thus such a schedule does not exist. This gives that $SPO\text{-}PoA \geq m$. □

From Theorems 18 and 19 we conclude that for for any $m$, $SPO\text{-}PoA = m$.

This is a proper place to mention that the lower bound example from [5] showing that $SPoA \geq m$ looks similar to our example at a first glance. The difference in processing times is in the definition $p_{kk} = k$, for $1 \leq k \leq m$. However, this example does not apply here, as the schedule of cost $m$ which it gives is not strictly Pareto optimal; if we switch to the optimal schedule, where job 1 runs on $M_m$ and each job $2 \leq k \leq m$ runs on $M_{k-1}$, all jobs $2 \leq k \leq m$ strictly improve their costs and job 1 is not worse off.

This is another example which demonstrates the fact that in the job scheduling game we consider not every SNE is necessarily a SPO-NE. However, we showed that this is the case already for related machines.

## 2.5 Preserving deviations

For an instance of the job scheduling game there may exist many strictly Pareto-optimal schedules, and a number of them can be NE. There can be a number of schedules for which the unsorted vector of the costs of jobs is the same, that is, for every player, its cost is the same, and therefore the makespan is the same in all those schedules. Deciding which of them to achieve can be controversial among players, as these schedules are indistinguishable in their quality and efficiency both from the players' and the social points of view. We would like to investigate then how much two such schedules may vary from one to another.

For this purpose we discuss "preserving deviations", which are deviations from a NE to a different schedule, where every player keeps the same cost that it had. Clearly, if the original schedule is a SPO-NE or a WPO-NE, then the new schedule resulting by such deviation also has this property.

In this section we prove that for such deviation for identical machines, the vector of machines loads does not change, while for related machines it can change significantly.

**Theorem 20.** *For identical machines, a preserving deviation of a* NE *schedule does not change its sorted vector of machines loads.*

**Proof.** Let $\mathcal{A}$ be a NE schedule on $m$ identical machines, and consider a deviation that makes no one worse off which results in a different schedule $\mathcal{A}'$. If there is at least one empty machine in $\mathcal{A}$ then every non-empty machine has a single job (otherwise the schedule is not a NE). Any preserving deviation will result in the same situation (some empty machines, and one job assigned to each non-empty machine), so clearly the sorted machine load vector remains the same.

Let the vector of machines loads of $\mathcal{A}$ be $\ell = (\ell_1, \ldots, \ell_m)$, and let the vector of strictly positive machines loads of $\mathcal{A}'$ be $\ell' = (\ell'_1, \ldots, \ell'_{m'})$, where $m' \leq m$.

Assume that the machines in $\mathcal{A}$ and $\mathcal{A}'$ are sorted in a non-increasing order of loads, that is $\ell_1 \geq \ldots \geq \ell_m$ and $\ell'_1 \geq \ldots \geq \ell'_{m'}$. Assume that the machines are partitioned into $k \leq m$ sets of machines of the same load (load groups) having loads $L_1 \geq L_2 \geq \ldots \geq L_k$, and let the number of machines in each of these load groups be $m_i$ for $1 \leq i \leq k$. We denote the total size of jobs assigned to machines in load group $1 \leq i \leq k$ by $W_i$, that is, $W_i = m_i \cdot L_i$.

Now consider the schedule $\mathcal{A}'$, and recall $m' \leq m$. We denote by $T_i$ for $1 \leq i \leq k$ the suffix of machines in $\mathcal{A}'$ having load of at most $L_i$, and the number of machines in $T_i$ by $t_i$. As the highest cost of any job in $\mathcal{A}'$ can be at most $L_1$, $t_1 = m' \leq m$ holds. We denote the total size of jobs executed by machines in $T_i$ by $W'_i$.

We prove the following claim.

**Claim 21.** *For* $0 \leq i \leq k-1$, $t_{k-i} \geq \sum_{j=0}^{i} m_{k-j}$.

**Proof.** The proof is by induction on $i$. We show that the property holds for $i = 0$, that is, we show that $t_k \geq \sum_{j=0}^{0} m_{k-j} = m_k$ holds. The work executed by machines in load group $L_k$ in $\mathcal{A}$ is $W_k = m_k \cdot L_k$. The work $W'_k$ executed by machines in $T_k$ in $\mathcal{A}'$ is at most $t_k \cdot L_k$. As the machines are identical, in order not to be worse off, all jobs running on machines in load group $L_k$ must be executed on machines of $T_k$ in $\mathcal{A}'$, which yields $W_k \leq W'_k$. This gives $m_k \cdot L_k \leq t_k \cdot L_k$, hence $m_k \leq t_k$.

We assume that $t_{k-p} \geq \sum_{j=0}^{p} m_{k-j}$ holds for $0 \leq p \leq i-1$, and prove that $t_{k-i} \geq \sum_{j=0}^{i} m_{k-j}$. The total work executed by machines in load groups $L_{k-i}, \ldots, L_k$ in $\mathcal{A}$ is $\sum_{j=0}^{i} W_{k-j} = \sum_{j=0}^{i} m_{k-j} \cdot L_{k-j}$. The work $W'_{k-i}$ executed by machines in $T_{k-i}$ in $\mathcal{A}'$ is at most
$t_k L_k + \sum_{j=1}^{i} (t_{k-j} - t_{k-j+1}) \cdot L_{k-j} = \sum_{j=0}^{i-1} t_{k-j} \cdot (L_{k-j} - L_{k-j-1}) + t_{k-i} L_{k-i}$. As the machines are identical, in order not to be worse off all jobs running on machines in load groups $L_{k-i}, \ldots, L_k$ in $\mathcal{A}$ must be executed on machines of $T_{k-i}$ in $\mathcal{A}'$, which yields $\sum_{j=0}^{i} W_{k-j} \leq W'_{k-i}$. We get that

$$\sum_{j=0}^{i} m_{k-j} \cdot L_{k-j} \leq \sum_{j=0}^{i-1} t_{k-j} \cdot (L_{k-j} - L_{k-j-1}) + t_{k-i} L_{k-i}.$$

As $L_{k-j} - L_{k-j-1} < 0$ for any $0 \leq j \leq k-2$, applying the inductive hypothesis for any $0 \leq j \leq i-1$, we get

$$\begin{aligned} \sum_{j=0}^{i} m_{k-j} \cdot L_{k-j} &\leq \sum_{j=0}^{i-1} (\sum_{h=0}^{j} m_{k-h}) \cdot (L_{k-j} - L_{k-j-1}) + t_{k-i} L_{k-i} \\ &= \sum_{j=0}^{i-1} m_{k-j} \cdot L_{k-j} - L_{k-i} \cdot (\sum_{j=0}^{i-1} m_{k-j}) + t_{k-i} \cdot L_{k-i} \, . \end{aligned}$$

Subtracting $\sum_{j=0}^{i-1} m_{k-j} \cdot L_{k-j}$ from both sides of this inequality, and dividing the result by $L_{k-i} > 0$ gives $m_{k-i} \leq -\sum_{j=0}^{i-1} m_{k-j} + t_{k-i}$. Adding $\sum_{j=0}^{i-1} m_{k-j}$ to both sides gives us $\sum_{j=0}^{i} m_{k-j} \leq t_{k-i}$, which concludes the proof. □

Applying this claim for $i = k-1$, we get that $t_1 \geq \sum_{j=0}^{k-1} m_{k-j} = m$, which proves that $t_1 = m$, that is, $\mathcal{A}'$ uses the same number of machines as $\mathcal{A}$, and $m' = m$.

We now prove that the load observed by any machine in $\mathcal{A}'$ is at most the load observed by the respective machine in $\mathcal{A}$.

**Claim 22.** *For any* $1 \leq i \leq m$, $\ell'_i \leq \ell_i$.

**Proof.** This is equivalent to proving that $\ell'_i \leq \ell_i$ holds for the last $m$ machines in $\mathcal{A}'$.

The proof is by induction on the number of last machines in $\mathcal{A}'$.

This property holds for the last $m_k$ machines in $\mathcal{A}'$: as $t_k \geq m_k$ by Claim 21, and by definition all machines in $T_k$ have a load of at most $L_k$, which is exactly the load of the last $m_k$ machines in $\mathcal{A}$.

We assume that this property holds for the last $\sum_{j=0}^{p-1} m_{k-j}$ machines in $\mathcal{A}'$ for some $1 \leq p \leq k-1$ , and prove that it holds for the last $\sum_{j=0}^{p} m_{k-j}$ machines.

The last $\sum_{j=0}^{p-1} m_{k-j}$ machines in $\mathcal{A}$ have loads of at most $L_{k-p}$. By inductive hypothesis, the last $\sum_{j=0}^{p-1} m_{k-j}$ machines in $\mathcal{A}'$ have this property, and so it is left to show that so do the following $m_{k-p}$ machines. By Claim 21, $t_{k-p} \geq \sum_{j=0}^{p} m_{k-j}$, hence machines in $T_{k-p}$ include these $m_{k-p}$ machines. As, by definition, machines in $T_{k-p}$ have a load of at most $L_{k-p}$, these $m_{k-p}$ machines have a load of at most $L_{k-p}$, which concludes the proof. □

We now prove, that for all $1 \leq i \leq m$ the inequality in Claim 22 has to be an equality.

**Claim 23.** *For any* $1 \leq i \leq m$, $\ell'_i = \ell_i$.

**Proof.** By Claim 22, for any $1 \leq i \leq m$, $\ell'_i \leq \ell_i$ holds. Therefore, $\sum_{i=1}^{m} \ell_i \leq \sum_{i=1}^{m} \ell'_i$. As the machines are identical, the total sum of loads in $\mathcal{A}'$ necessarily remains the same as in $\mathcal{A}$. However, $\sum_{i=1}^{m} \ell_i = \sum_{i=1}^{m} \ell'_i$ can hold only if $\ell'_i = \ell_i$ for all $1 \leq i \leq m$. □

The conclusion from Claim 23 is that $t_k = m_k$, $t_i - t_{i+1} = m_i$ and all machines in $T_i \backslash T_{i+1}$ for $1 \leq 1 \leq k-1$ have a load of exactly $L_i$.

Since any schedule on identical machines is a SPO-NE (and therefore a WPO-NE), then such a deviation from schedule $\mathcal{A}$ to $\mathcal{A}'$ is a preserving deviation, since once no job is worse off, then no job can be better off. □

From Theorem 20, we can conclude that in the model of identical machines, preserving deviations leave the schedule almost unchanged: the loads of the machines remain exactly the same, and the only changes in the assignments of the jobs are within the load group of the machine where each job previously ran.

For related machines though, two different schedules where all players have the same costs, may vary greatly both in the assignments of the jobs and in their load vectors. This is true even for schedules which are strictly or weakly Pareto optimal. We demonstrate this by the following example. Recall the SPO-NE schedule $X$ that gives a lower bound on the $SPO$-$PoA$ from the proof of Theorem 14 in Section 2.1. We give a similar example. We take an instance with $k = 4$ and $N_4 = 1$. There are five groups of machines, $N_0 = 24$ machines with speed 1, $N_1 = 24$ machines with speed 2, $N_2 = 12$ machines with speed 4, $N_3 = 4$ machines with speed 8 and $N_4 = 1$ machines with speed 16. The total number of machines is then 65. Now, consider a schedule $X$ for this instance, as in the lower bound. Each machine of group 0 is empty, each machine of group 1 is assigned one job of size 2, each machine of group 2 is assigned two jobs of size 4, each machine of group 3 is assigned three jobs of size 8 and each machine of group 4 is assigned four jobs of size 16. Each job in group 1 pays a cost of 1, each job in group 2 pays a cost of 2, each job in group 3 pays a cost of 3 and each job in group 4 pays a cost of 4. Its vector of machines loads is $x = (\overbrace{4}^{1}, \overbrace{3, \ldots, 3}^{4}, \overbrace{2, \ldots, 2}^{12}, \overbrace{1, \ldots, 1}^{24}, \overbrace{0, \ldots, 0}^{24})$. Now, consider a different schedule $Y$, where each machine of group 0 is empty, each machine in group 1 is assigned one job of size 4, four of machines in group 2 are assigned with one job of size 16 and the remaining 8 machines with two jobs of size 2 each, each machine in group 3 is assigned three jobs of size 8 and the single machine in group 4 is assigned eight jobs of size 2. It is easy to verify that in transition to schedule $Y$ each one of the jobs keeps its original cost, however the vector of machines loads of schedule $Y$, $y = (\overbrace{4,\ldots,4}^{4}, \overbrace{3,\ldots,3}^{4}, \overbrace{2,\ldots,2}^{24}, \overbrace{1,\ldots,1}^{9}, \overbrace{0,\ldots,0}^{24})$ is very different from $x$.

## 2.6 Recognition of weakly and strictly Pareto optimal equilibria

In this section we consider the computational complexity of SPO-NE and WPO-NE for all machine models. Specifically, we investigate the problem of recognition of such schedules. We prove the following theorems.

**Theorem 24.** *There exists a polynomial time algorithms which receives a schedule on related machines (or on identical machines) and check whether the schedule is a* SPO-NE *and whether it is a* WPO-NE.

**Proof.** Consider a schedule $\mathcal{A}$, and recall that one can determine in polynomial time whether a given schedule is a NE. Since any NE on identical machines is a WPO-NE and a SPO-NE, the recognition of such schedules is equivalent to recognition of NE. This is also the case for related machines and WPO-NE.

For the recognition of SPO-NE on related machines, we use the following algorithm. First, check whether the schedule is a NE (if not, then output a negative answer). If the schedule is a NE and it does not contain an empty machine, return a positive answer. Otherwise, for every job $k$, such that $k$ is assigned to a machine which has at least two jobs assigned to it, test if moving it to an empty machine of maximum speed does not increase its cost. If there exists a job for which the cost is not increased, return a negative answer, and otherwise, a positive answer. Note that if there exists an empty machine, but no machine has two jobs assigned to it, then the returned answer is positive.

Now we prove correctness of the last algorithm. If there are no empty machines then any NE is a SPO-NE (by Theorem 8). For the remaining cases of the algorithm, we prove the following claim.

**Claim 25.** *Given $\mathcal{A}$, which is a* NE, *there exists an alternative schedule $\mathcal{A}'$ where no job increases its cost and at least one job reduces its cost if and only if there exists a job $k$ which is assigned to a machine with at least one other job in $\mathcal{A}$, and moving it to an empty machine of maximum speed does not increase its cost.*

**Proof.** We first assume that such a job $k$ exists. Consider the schedule $\tilde{\mathcal{A}}$ in which $k$ is assigned to a machine of maximum speed which is empty in $\mathcal{A}$, and the rest of the assignment is the same as in $\mathcal{A}$. There is at least one job which is assigned to the same machine as $k$ in $\mathcal{A}$, whose cost is strictly reduced (since the load of its machine decreases when $k$ is moved to another machine). The cost of $k$ does not increase, and any job assigned to any machine other than the machine of $k$ in $\mathcal{A}$ and the machine of $k$ in $\tilde{\mathcal{A}}$ keeps its previous cost.

Next, assume that $\mathcal{A}'$ exists, and assume that among such schedules, $\mathcal{A}'$ has a minimum number of jobs which are assigned not to the same machine as in $\mathcal{A}$. Using Theorem 8, we get that $\mathcal{A}$ necessarily has an empty machine $M_{i'}$ which is non-empty in $\mathcal{A}'$. Let $k$ be a job assigned to $M_{i'}$ in $\mathcal{A}'$ and let $M_i$ be the machine to which it is assigned in $\mathcal{A}$.

If machine $M_i$ does not have an additional job in $\mathcal{A}$, and since its cost on $M_{i'}$ (possibly with additional jobs) is no larger, we get $s_{i'} \geq s_i$. However, the schedule is a NE, so $k$ cannot reduce its cost by moving to an empty machine. Therefore, its cost on $M_{i'}$ is the same as its cost on $M_i$, $s_{i'} = s_i$ and $k$ is assigned to $M_{i'}$ alone in $\mathcal{A}'$. The jobs assigned to $M_i$ in $\mathcal{A}'$ are not assigned to $M_{i'}$ or to $M_i$ in $\mathcal{A}$. This is true since $M_{i'}$ is empty in $\mathcal{A}$ and $M_i$ only has the job $k$ in $\mathcal{A}$. We construct a schedule $\hat{\mathcal{A}}$ where the jobs assigned to $M_i$ and $M_{i'}$ in $\mathcal{A}'$ are swapped and the other jobs are assigned to the same machines as in $\mathcal{A}$. The number of jobs assigned to a different machine from their machines in $\mathcal{A}$ is reduced by $1$ (due to $k$ being assigned to the same machines in $\hat{\mathcal{A}}$ and $\mathcal{A}$), which contradicts the choice of $\mathcal{A}'$.

Thus, there exists an additional job $k'$ assigned to $M_i$ in $\mathcal{A}$. Since moving $k$ to some empty machine does not increase its cost, then moving it to an empty machine with maximum speed clearly does not increase its cost. ☐

Given the claim, if every non-empty machine has a single job then the schedule is a SPO-NE. Otherwise, the algorithm tests the existence of a job $k$ as in the claim. ☐

**Theorem 26.** *i. The problem of checking whether a given schedule on unrelated machines is a* WPO-NE *is strongly co-NP-complete. ii. The problem of checking whether a given schedule on unrelated machines is a* SPO-NE *is co-NP-complete.*

**Proof.** Given a schedule and an alternative schedule, checking whether the alternative schedule implies that the given schedule is not a NE or not (weakly or strictly) Pareto optimal can be done in polynomial time, and therefore the problems are in co-NP.

To prove hardness of the recognition of WPO-NE, we reduce from the 3-PARTITION problem, which is strongly NP-hard. In this problem we are given an integer $B$ and $3M$ integers $a_1, a_2, \ldots, a_{3M}$, where $\frac{B}{4} < a_k < \frac{B}{2}$ for $1 \leq k \leq 3M$, $\sum_{k=1}^{3M} a_k = MB$, and we are asked whether there exists a partition of the integers into $M$ sets, where the sum of each subset is exactly $B$. We construct an input with $m = 4M$ machines. There are $4M$ jobs, $3M$ of them are based on the instance of 3-PARTITION and the last $M$ jobs are dummy jobs. For $1 \leq k \leq 3M$, we have $p_{ik} = B + 1$ for $1 \leq i \leq 3M$, and $p_{ik} = a_k$ for $3M + 1 \leq i \leq 4M$. For $3M + 1 \leq k \leq 4M$, we have $p_{ik} = B$ for $1 \leq i \leq 3M$, and $p_{ik} = B + 1$ for $3M + 1 \leq i \leq 4M$. The given schedule is one where job $k$ is assigned to machine $k$. All machines have a load of $B + 1$, so the schedule is a NE. We show that the schedule is weakly Pareto optimal if and only if a 3-partition as required does not exist. Assume first that a 3-partition exists. We define an alternative schedule. In this schedule, each one of the last $M$ machines runs one subset of jobs of the first $3M$ jobs, out of the $M$ subsets of the 3-partition. The sum of the corresponding subsets of numbers in the input of 3-PARTITION is $B$ and therefore, their total processing time on such a machine is $B$. Each dummy job runs on a different machine out of the first $3M$ machines, having a cost of $B$. Thus, all jobs have a smaller cost in the alternative schedule, so the original one is not Pareto optimal.

On the other hand, if there exists an alternative schedule where all jobs reduce their costs, then all the first $3M$ jobs must be assigned to the last $M$ machines (since on the other machines even if such a job is assigned to alone it still has a cost of $B$). For job $k$, no matter which such machine receives it, it has a processing time of $a_k$ on it, so all jobs have a total processing time of $MB$. Since all numbers are integers, the only way that every job reduces its load is that each machine will have a load of exactly $B$, which implies a 3-partition.

To prove hardness of the recognition of SPO-NE, we can use the reduction of [52] showing that the recognition of SNE is hard. For completeness we present an alternative reduction. To prove hardness of the recognition of SPO-NE, we reduce from the PARTITION problem, which is NP-hard. In this problem we are given an integer $B$ and $N$ integers $a_1, a_2, \ldots, a_N$, where, $\sum_{k=1}^{N} a_k = 2B$, and we are asked whether there exists a partition of the integers into two sets, where the sum of each subset is exactly $B$. We construct an input with $m = 2$ machines (it is possible to use the same input for any larger number of machines, giving all jobs infinite processing times on every machine except for the first two machines). We have $N + 2$ jobs. Job $k$, for $1 \leq k \leq N$, $p_{1k} = a_k + \frac{1}{2N}$ while $p_{2k} = a_k$. Job $N + 1$ has $p_{1(N+1)} = B$ and $p_{2(N+1)} = B + \frac{1}{2}$. Job $N + 2$ has $p_{1(N+2)} = \infty$ and $p_{2(N+1)} = B$ so it must be assigned to $M_2$. We are given the schedule where the first $N$ jobs are assigned to $M_1$ and the two last jobs are assigned to $M_2$. The loads of both machines are $2B + \frac{1}{2}$, thus this schedule is a NE. If there exists a partition, consider the alternative schedule where each machine receives one subset of jobs whose total size in the original input is $B$, and job $N + 1$ is assigned to $M_1$. Let $K_1$ be the cardinality of the set of jobs assigned to $M_1$ in the alternative schedule. Then the resulting load of $M_1$ is $2B + \frac{K_1 - 1}{2N}$. Since $M_2$ receives at least two jobs, then $K_1 \leq N$, so the load is strictly below $2B + \frac{1}{2}$. The load of $M_2$ is exactly $2B$. Thus, the original schedule is not strictly (or weakly) Pareto optimal. On the other hand, if the original schedule is not strictly (or weakly) Pareto optimal, then in an alternative schedule, job $N + 1$ must be assigned to $M_1$, and the total processing time of jobs assigned with it must be strictly below $B + 1$. The total processing time of jobs assigned to $M_2$ must be strictly below $B + 1$ as well, and so there are two sets whose sizes (in the original input) are at most $B$, which implies a partition.

Note that this reduction can be used to prove the (weak) co-NP-completeness of the recognition of WPO-NE schedules. Thus both problems are hard for any number of machines. □

## 2.7 Summary and Conclusions

In this chapter we have studied the quality and complexity of the strict and weak Pareto optimal Nash equilibria in job scheduling games, in the settings of identical, related and unrelated machines.

We found that in the models of identical and related machines, strict and weak Pareto optimal Nash equilibria can be as bad as pure Nash equilibria, however in the model of unrelated machines, while for weak Pareto optimal Nash equilibria and $m \geq 3$, this is still the case, strict Pareto optimal Nash equilibria (and even weak Pareto optimal equilibria, for $m = 2$) are as good as Strong Nash equilibria with respect to the Price of Anarchy. This implies that for unrelated machines, cooperation between all players (as opposed to cooperation between subsets of players) still gives solutions of high quality.

As for identical and related machines, recognition of weakly or strictly Pareto optimal equilibria can be done in polynomial time, unlike strong equilibria. Despite the slightly worse quality of such equilibria compared to strong equilibria (due to the results for the Price of Anarchy on related machines), we conclude that weak and strict Pareto optimal equilibria are of interest for identical and related machines.

# Chapter 3

# Bin Packing of selfish items

## 3.1 Introduction and motivation

In this chapter, we consider the well-known Bin Packing problem. It was introduced in the early 1970's [74, 103], and was extensively studied ever since (see e.g. [75], [76], [28] for surveys on this problem). The basic, one-dimensional Bin Packing problem consists of packing a set of objects with sizes in (0,1] into a set of unit-capacity bins while using as few bins as possible. Among other important real-life applications, such as multiprocessor scheduling, optimization of file storage on disks and stock cutting, the Bin Packing problem can be met in a great variety of network problems arising in large-scale communication networks. For example, the packet scheduling problem (the problem of packing a given set of packets into a minimum number of time slots for fairness provisioning), the bandwidth allocation problem (signals have usually a small size and several of them can be transmitted in the same frame[1] so as to minimize bandwidth consumption) and the problem of packing the data for Internet phone calls into ATM packets (filling fixed-size frames to maximize the amount of data that they carry), to mention only a few. Other than the obvious practical significance in the study of the Bin Packing problem from a game-theoretic perspective, there is a pure theoretical interest to this study, as Bin Packing is a problem of fundamental theoretical significance, serving as one of the main test grounds for new algorithmic ideas and models of analysis for several decades. In fact, Bin Packing is one of the first problems to which approximation algorithms were suggested and analyzed with comparison to the optimal algorithm. Therefore, the study of this important problem from a game-theoretic standpoint is clearly well motivated.

[1]frame is the basic unit in media access control layer of the Internet Protocol Suite.

## 3.2 The model

The Bin Packing problem consists of packing a set $N$ of items, each item $i \in N$ having a size $a_i \in (0,1]$, into a set of unit-capacity bins while using as few bins as possible. The induced bin packing game *BP* is defined by a tuple $BP = \langle N, (B_i)_{i \in N}, (c_i)_{i \in N} \rangle$, where $N$ is the set of selfish players, and the bins are of equal capacity. Each player $i \in N$ controls a single item with size $a_i \in (0,1]$ and selects the bin to which this item is packed. We identify the player with the item he wishes to pack. Thus, the set of players corresponds to the set of items. The set of strategies $B_i$ for each item $i \in N$ is the set of all possible bins. Each item can be assigned to one bin only. Splitting items among several bins is not allowed. The outcome of the game is a particular assignment $b = (b_j)_{j \in N} \in \times_{j \in N} B_j$ of the items to bins, that respects the capacity of the bins. Let $X$ denote the set of all possible assignments. All the bins have the same fixed cost which equals their capacity and the cost of a bin is proportionally shared among all the items it contains. The cost function of item $i$ is $c_i$. If we scale the cost and the size of each bin to one, the cost paid by item $i$ for choosing to be packed in bin $\mathcal{B}_j$ such that $j \in B_i$ is defined by $c_i(j, b_{-i}) = \dfrac{a_i}{\sum_{k:b_k=j} a_k}$ , when $b_{-i} \in X_{-i}$; i.e, an item is charged with a cost which is proportional to the portion of the bin it occupies in a given packing. We charge an item for being packed in a bin in which it does not fit with an infinite cost. The selfish items are interested in being packed in a bin so as to minimize their cost. Thus, item $i$ packed into $\mathcal{B}_j$ in a particular assignment $(b_j)_{j \in N}$ will migrate from $\mathcal{B}_j$ each time it will detect another bin $\mathcal{B}_{j'}$ such that $c_i(j', b_{-i}) < c_i(j, b_{-i})$. This inequality holds for each $j'$ such that $\sum_{k:b_k=j'} a_k + a_i > \sum_{k:b_k=j} a_k$, thus an item will perform an improving step each time it will detect a strictly more loaded bin in which it fits. At a Nash equilibrium, no item can unilaterally reduce its cost by moving to a different bin (see Figure 3.1(b) for an example). We call a packing that admits the Nash conditions an NE packing. The social cost function that we want to minimize is the number of used bins (which equals the sum of players' individual costs): $SC(b) = \sum_{i \in N} c_i(b)$.

The selfish bin packing problem defined above can be also interpreted as a routing problem. Consider a network consisting of two nodes, a source and a destination, connected by a potentially infinite number of parallel links having the same bandwidth capacity, and a set of users wishing to send a certain amount of unsplittable flow between the two nodes. To establish a link, one has to pay a fixed cost, which equals the capacity of the link. The cost of each link is shared among the users routing their flow on that link according to the normalized fraction of bandwidth utilized by each. For such a reason, users, who are assumed to be selfish, want to route their traffic on the most loaded link available that can accommodate their load. The social goal is to minimize the total number of links used.

In the cooperative version of the game, we consider all possible (non-empty) groups of items $A \subseteq N$. A group can contain a single item. The cost functions of the players are defined the same as in the non-cooperative case. Each group of items is interested to be packed in a way so as to minimize the costs of all group members. Thus, given a particular assignment, all members of group $A$ will perform a joint improving step if there is a configuration in which, for each member, the new bin (it can be one of the already existing bins, or a newly opened) will admit a strictly greater load than the bin of origin. The costs of the non-members may be enlarged as a result of this improvement step. At a Strong Nash equilibrium, no group of items can reduce the costs for all its members by jointly moving to different bins (see Figure 3.1(c) for an example). We call a packing that ac the

sa

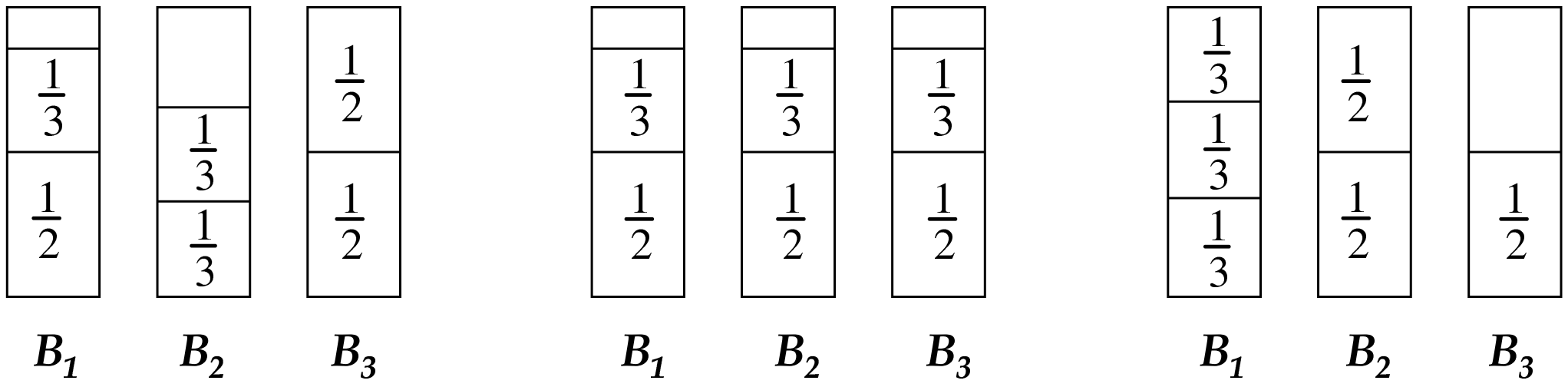

**(a) A non-equilibrium packing** **(b) *NE* packing** **(c) *SNE* packing**

Figure 3.1: (a) A packing that is not an equilibrium, as the item of size $\frac{1}{3}$ on $B_1$ will reduce its cost by migrating to $\mathcal{B}_2$ (b) A packing that is Nash equilibrium but not a Strong Nash equilibrium, since the five items of sizes $\{\frac{1}{2}, \frac{1}{2}, \frac{1}{3}, \frac{1}{3}, \frac{1}{3}\}$ will reduce their cost by deviating (see (c)), (c) A packing which is a Strong Nash equilibrium.

It is well-known that Nash equilibria do not always optimize the social cost function, and our bin packing game is no exception: an equilibrium packing does not necessarily have minimum cost. See Figure 3.2 for example. Note also that not every optimal solution is an equilibrium.

We measure the quality of the equilibria in this game with respect to the social optimum. In the bin packing game, the social optimum is the number of bins used in a coordinated optimal packing. We consider the Prices of Anarchy and Stability, that are prevalent measures of the quality of the equilibria reached with uncoordinated selfish players.

The Price of Anarchy of a game $G \in BP$ is defined to be the ratio between the cost of the worst Nash equilibrium packing and the social optimum. The Price of Stability of a game $G \in BP$ is defined to be the ratio between the cost of the best Nash equilibrium

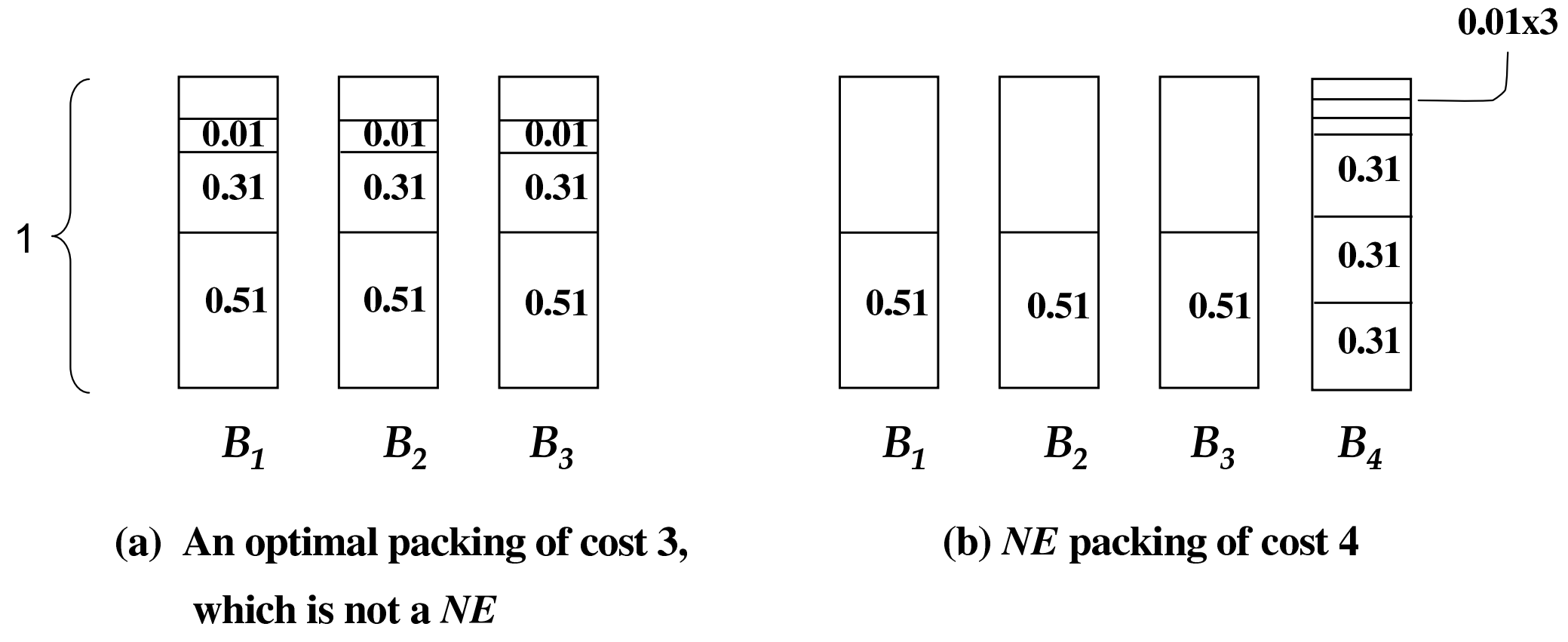

Figure 3.2: An example of a non-optimal *NE* packing

packing and the social optimum. Formally,

$$PoA(G) = \sup_{b \in NE(G)} \frac{SC(b)}{OPT(G)}, \quad PoS(G) = \inf_{b \in NE(G)} \frac{SC(b)}{OPT(G)}.$$

The bin packing problem is usually studied via asymptotic measures, as they are robust against anomalies with a small number of bins in the optimum packing. The asymptotic POA and POS of the bin packing game *BP* are defined by

$$PoA(BP) = \limsup_{OPT(G)\to\infty} \sup_{G \in BP} PoA(G), \quad PoS(BP) = \limsup_{OPT(G)\to\infty} \sup_{G \in BP} PoS(G).$$

We also study stability measures that separate the effect of the lack of coordination between players from the effect of their selfishness. The measures considered are the Strong Prices of Anarchy snd Stability. These measures are defined similarly to the POA and the POS but only Strong equilibria packings are considered. We define the Strong Price of Anarchy of a game $G \in BP$ as the ratio between the cost of the worst Strong Nash equilibrium packing and the social optimum. The Strong Price of Stability is defined as the ratio between the cost of the best Strong Nash equilibrium packing and the social optimum. Formally,

$$SPoA(G) = \sup_{b \in SNE(G)} \frac{SC(b)}{OPT(G)}, \quad SPoS(G) = \inf_{b \in SNE(G)} \frac{SC(b)}{OPT(G)},$$

As before, we define the asymptotic SPOA and SPOS of the bin packing game *BP* as

$$SPoA(BP) = \limsup_{OPT(G)\to\infty} \sup_{G \in BP} SPoA(G), \quad SPoS(BP) = \limsup_{OPT(G)\to\infty} \sup_{G \in BP} SPoS(G).$$

In the following discussion we will often drop the term "asymptotic" when we mention the asymptotic stability measures.

## 3.3 Related work and our contributions

Bilò [18] was the first to study the bin packing problem with selfish items. He proved that the bin packing game admits pure Nash equilibria and provided non-tight bounds on the Price of Anarchy. He used a generalized potential function to prove that the bin packing game converges to a pure Nash equilibrium in a finite sequence of selfish improving steps, starting from any initial configuration of the items; however, the number of steps may be exponential. This result implies that for every instance of bin packing game, among the optimal packings there exists a packing which admits NE; in other words, $\text{POS}(BP) = 1$. It is also implicit from the work of Bilò that the bin packing game admits Strong Nash equilibria, as the optimal packing with the highest potential value is always an SNE.

Other related work includes recent papers by Miyazawa and Vignatti (see [88, 87]), who consider the selfish bin packing problem, and study the convergence time to a Nash equilibrium.

In this chaptr we consider the bin packing game, in a model originally proposed and analyzed by Bilò in [18]. We give improved (and nearly tight) lower and upper bounds of 1.6416 and 1.6428, respectively, on the POA of the bin packing game. This result appeared in [43]. Later, Yu and Zhang [108], independently, used a similar construction to prove a lower bound of the same value on the POA, and claimed an upper bound of 1.6575 on the POA.

We give a characterization of the Strong Nash Equilibria in the bin packing game as the outcomes of the *Subset Sum*[2] algorithm for bin packing. *Subset Sum* is a greedy algorithm that repeatedly solves a one-dimensional knapsack problem for packing each bin in turn. It was originally suggested by Prim and first mentioned by Graham [64], who also gave a lower bound of $\sum_{k=1}^{\infty} \frac{1}{2^k-1} \approx 1.6067$ on its asymptotic worst-case approximation ratio. The first non-trivial upper bound of $\frac{4}{3} + \ln\frac{4}{3} \approx 1.6210$ was proved by Caprara and Pferschy in [19]. The exact asymptotic worst-case approximation ratio of the *Subset Sum* algorithm was recently established to be $1.6067$ [46], which matches Grahams' lower bound.

We also consider the strong measures of stability, and provide tight bounds on the SPOA and the SPOS of this game. We show that the aforementioned *Subset Sum* algorithm in fact produces an assignment that admits strong equilibrium. Therefore, we provide an exponential time deterministic algorithm with guaranteed (asymptotic) worst-case approximation ratio [19] that actually calculates the Strong Nash assignment for each bin. Interestingly, the SPOA equals the SPOS and we prove that this value is equal to the approximation ratio of the *Subset Sum* algorithm. This allows us to characterize the SPOA/SPOS of the game in terms of the approximation ratio of this algorithm.

[2]Also called *fill bin* or *minimum bin slack* in the literature.

Our conclusions can be summarized by the following table:

| | | Lower Bound | Upper Bound |
|---|---|---|---|
| *PoA* | Bilò [18] | 1.6 | 1.6667 |
| | Our results | 1.6416 | 1.6428 |
| *SPoA=SPoS* | Our results | 1.6067 | 1.6067 |
| *PoS* | Bilò [18] | 1 | 1 |

Table 3.1: Summary of Results

Recently, Yu and Zhang [108] have designed a polynomial time algorithm which produces a packing which is a pure NE.

In this work we show that the problem of computing a Strong Nash packing, on the other hand, is NP-hard, hence it is unlikely to come up with a polynomial time algorithm which produces a packing that is an SNE, unless $P = NP$.

## 3.4 Price of Anarchy in the Bin Packing game

In this section we provide a lower bound for the Price of Anarchy of the bin packing game and also prove a very close upper bound.

### 3.4.1 A lower bound: construction

In this section, we present our main technical contribution, which is a lower bound on the *PoA*. We start with presenting a set of items. The set of items consists of multiple levels. Such constructions are sometimes used to design lower bounds on specific bin packing algorithms (see e.g., [80]). Our construction differs from these constructions since the notion of order (in which packed bins are created) does not exist here, and each bin must be stable with respect to all other bins. The resulting lower bound on the POA is different from any bounds known on the asymptotic approximation ratio of well known algorithms for bin packing. Since we prove an almost matching upper bound, we conclude that the POA is probably not related directly to any natural algorithm.

We prove the following theorem.

**Theorem 27.** *The Price of Anarchy of the bin packing game is at least the sum of the following series:* $\sum_{j=1}^{\infty} 2^{-j(j-1)/2}$*, which is equal to approximately* $1.64163$*.*

**Proof.** Let $s > 2$ be an integer. We define a construction with $s$ phases of indices $1 \leq j \leq s$, where the items of phase $j$ have sizes which are close to $\frac{1}{2^j}$, but can be slightly smaller or slightly larger than this value.

Two sequences of positive integers $r_j$ and $d_j$, for $1 \le j \le s$, are used. We choose the number $r_s$ to be an arbitrary sufficiently large value such that $r_s > 2^{s^3}$. We recursively define $r_j = 2^j \cdot r_{j+1} + 1$, for $1 \le j \le s-1$. In addition, we let $d_1 = 0$ and $d_j = r_{j-1} - r_j = (2^{j-1} - 1)r_j + 1$, for $2 \le j \le s$. Let $n = r_1$. Clearly, $r_j \le n$ and $d_j \le n$, for $1 \le j \le s$.

Let $OPT = n$, and note that $n \ge r_s > 2^{s^3}$. We use a sequence of small values, $\delta_j$, $j = 1, \ldots, s$ such that $\delta_j = \frac{1}{(4n)^{3s-2j}}$. Note that this implies $\delta_{j+1} = (4n)^2 \delta_j$ for $1 \le j \le s-1$.

**Observation 28.** *For each* $1 \le j \le s$, $\frac{n}{2^{j(j-1)/2}} - 1 < r_j \le \frac{n}{2^{j(j-1)/2}}$.

**Proof.** We use induction to prove this inequality. This holds for $j = 1$ by definition. We next prove the property for $j+1$ using the property for $j$ (where $j \ge 1$). We have $r_{j+1} = \frac{r_j - 1}{2^j}$. Clearly, using the inductive assumption, $r_{j+1} < \frac{r_j}{2^j} \le \frac{n}{2^{j(j-1)/2+j}} = \frac{n}{2^{j(j+1)/2}}$. On the other hand, $r_{j+1} = \frac{r_j - 1}{2^j} > \frac{\frac{n}{2^{j(j-1)/2}} - 2}{2^j} \ge \frac{n}{2^{j(j+1)/2}} - 1$. $\square$

Phase 1 simply consists of $r_1$ items of size $\sigma_1 = \frac{1}{2} + 2(d_1 + 1)\delta_1$. For $j \ge 2$, phase $j$ consists of the following $2d_j + r_j$ items. There are $r_j$ items of size $\sigma_j = \frac{1}{2^j} + 2(d_j + 1)\delta_j$, and for $1 \le i \le d_j$, there are two items of sizes $\pi_j^i = \frac{1}{2^j} + (2i - 1)\delta_j$ and $\theta_j^i = \frac{1}{2^j} - 2i\delta_j$. Note that $\pi_j^i + \theta_j^i = \frac{1}{2^{j-1}} - \delta_j$.

**Claim 29.** *This set of items can be packed into* $n$ *bins, i.e.,* $OPT \le n$

**Proof.** The optimal packing will contain $d_j$ bins of level $j$, for $2 \le j \le s$, and the remaining bins are of level $s+1$, where a bin of level $j$, contains only items of phases $1, \ldots, j$.

To show that we can allocate these numbers of bins, and to calculate the number of level $s+1$ bins, note that $\sum_{j=2}^{s} d_j = r_1 - r_s = n - r_s$. Thus, the number of level $s+1$ bins is $r_s$.

The packing of a bin of a given level is defined as follows. For $2 \le j \le s$, a level $j$ bin contains one item of each size $\sigma_k$ for $1 \le k \le j-1$, and in addition, one pair of items of sizes $\pi_j^i$ and $\theta_j^i$ for a given value of $i$ such that $1 \le i \le d_j$. A bin of level $s+1$ contains one item of each size $\sigma_k$ for $1 \le k \le j-1$.

We show that every item was assigned into some bin. Consider first items of size $\pi_j^i$ and $\theta_j^i$. Such items exist for $1 \le i \le d_j$, and therefore, every such pair was assigned to a bin together. Next, consider items of size $\sigma_j$ for some $1 \le j \le s$. The number of such items is $r_j$. The number of bins which received such items is $\sum_{k=j+1}^{s} (d_k) + r_s = r_j$.

We further show that the sum of sizes of items in each bin does not exceed 1. Consider a bin of level $j$ for some $2 \le j \le s$. The sum of items in it is

$$\sum_{k=1}^{j-1} \sigma_k + \frac{1}{2^{j-1}} - \delta_j = \sum_{k=1}^{j-1} (\frac{1}{2^k} + 2(d_k + 1)\delta_k) + \frac{1}{2^{j-1}} - \delta_j = 1 + \sum_{k=1}^{j-1} (2(d_k + 1)\delta_k) - \delta_j.$$

As $d_k + 1 \leq n$ (since $r_j$ is a strictly decreasing sequence of integers and $r_1 = n$) and $\delta_i$ is a strictly increasing sequence, we have $2(d_k+1)\delta_k \leq 2n\delta_{j-1}$, and since $j-1 \leq s < n$, $\sum_{k=1}^{j-1} 2(d_k+1)\delta_k < 2n^2\delta_{j-1}$. Using $\delta_j = 16n^2\delta_{j-1}$ we get that the sum is smaller than 1.

Consider a bin of level $s+1$. The sum of items in it is $\sum_{k=1}^{s} \sigma_k = \sum_{k=1}^{s}(\frac{1}{2^k} + 2(d_k+1)\delta_k)$. We have $2(d_k+1)\delta_k \leq 2n\delta_s = \frac{1}{2^{2s-1}n^{s-1}}$. Since $2 < s < n$, we get at most $1 - \frac{1}{2^s} + \frac{n}{2^{2s-1}n^{s-1}} = 1 - \frac{1}{2^s} + \frac{1}{2^{2s-1}n^{s-2}} < 1$. □

We next define an alternative packing, which is a NE. In what follows, we apply a modification to the input by removing a small number of items. Clearly, OPT $\leq n$ would still hold for the modified input.

Our modification to the input is the removal of items $\pi_j^1$ and $\theta_j^{d_j}$ for all $2 \leq j \leq s$. We construct $r_j$ bins for phase $j$ items. A bin of phase $j$ consists of $2^j - 1$ items of phase $j$, as follows. One item of size $\sigma_j = \frac{1}{2^j} + 2(d_i+1)\delta_i$, and $2^{j-1}-1$ pairs of items of phase $j$. A pair of items of phase $j$ is defined to be the items of sizes $\pi_j^{i+1}$ and $\theta_j^i$, for some $1 \leq i \leq d_j - 1$. The sum of sizes of this pair of items is $\frac{1}{2^j} + (2i+1)\delta_j + \frac{1}{2^j} - 2i\delta_j = \frac{1}{2^{j-1}} + \delta_j$. Using $d_j = (2^{j-1}-1)r_j + 1$ we get that all phase $j$ items are packed. The sum of items in every such bin is

$$1 - \frac{1}{2^{j-1}} + (2^{j-1}-1)\delta_j + \frac{1}{2^j} + 2(d_j+1)\delta_j = 1 - \frac{1}{2^j} + \delta_j(2^{j-1} + 1 + 2d_j).$$

**Claim 30.** *The loads of the bins in the packing defined above are monotonically increasing as a function of the phase.*

**Proof.** It is enough to show that $1 - \frac{1}{2^j} + \delta_j(2^{j-1}+1+2d_j) < 1 - \frac{1}{2^{j+1}}$, which is equivalent to proving $\delta_j(2^{j-1}+1+2d_j)2^{j+1} < 1$. Indeed, we have $\delta_j(2^{j-1}+1+2d_j)2^{j+1} < \delta_j(2^{2j} + 2^{j+2}n) < 2\delta_j n^2$, as $n > 2^{s^3}$. Using $\delta_j \leq \delta_s = \frac{1}{2^{2s}n^s} < \frac{1}{16n^3}$ we get $2\delta_j n^2 < 1$. □

**Claim 31.** *The packing as defined above is a valid* NE *packing.*

**Proof.** To show that this a NE, we need to show that an item of phase $j > 1$ cannot migrate to a bin of a level $k \geq j$, since this would result in a load larger than 1, and that it cannot migrate to a bin of phase $k < j$, since this would result in a load smaller than the load of a phase $j$ bin. Due to the monotonicity we proved in Claim 30, we only need to consider a possible migration of a phase $j$ item into a phase $j$ bin, and a phase $j-1$ bin, if such bins exist. Moreover, in the first case it is enough to consider the minimum size item and in the second case, the maximum size item of phase $j$. For phase 1 items, since their sizes are larger than $\frac{1}{2}$, and all other bins are loaded by more than $\frac{1}{2}$, such items clearly cannot migrate.

The smallest phase $j$ item has size $\frac{1}{2^j} - \delta_j(2d_j - 2)$. If it migrates to another bin of this phase, we get a total load of $1 - \frac{1}{2^j} + \delta_j(2^{j-1} + 1 + 2d_j) + \frac{1}{2^j} - \delta_j(2d_j - 2) = 1 + \delta_j(3 + 2^{j-1}) > 1$.

The largest phase $j$ item has size $\frac{1}{2^j} + 2(d_j + 1)\delta_j$. If it migrates to a bin of phase $j-1$, we get a load of $1 - \frac{1}{2^{j-1}} + \delta_{j-1}(2^{j-2} + 1 + 2d_{j-1}) + \frac{1}{2^j} + 2(d_j + 1)\delta_j$. We compare this load with $1 - \frac{1}{2^j} + \delta_j(2^{j-1} + 1 + 2d_j)$, and prove that the first load is smaller. Indeed $\delta_{j-1}(2^{j-2}+1+2d_{j-1}) < \delta_j(2^{j-1}-1)$ since $\delta_j = 16n^2\delta_{j-1}$, $n > 2^{s^3}$ and $(2^{j-2}+1+2d_{j-1}) < 4n < 16n^2(2^{j-1} - 1)$.

The fact $1 - \frac{1}{2^j} + \delta_j(2^{j-1} + 1 + 2d_j) < 1 - \frac{1}{2^{j+1}}$ holds as we proved in Claim 30, also implies that the load of each bin in the NE packing we construct is smaller than 1 for each $2 \leq j \leq s$, therefore this NE packing is valid. $\square$

Finally, we bound the POA as follows. The cost of the resulting NE is $\sum_{j=1}^{s} r_j$. Using Observation 28 we get that $\sum_{j=1}^{s} r_j > \sum_{j=1}^{s} (\frac{n}{2^{j(j-1)/2}} - 1)$ and since $OPT = n >> s$, we get a ratio of at least $\sum_{j=1}^{s} 2^{-j(j-1)/2}$. Letting $s$ tend to infinity as well results in the claimed lower bound. $\square$

**Comment 32.** *In the lower bound construction, in each phase $j$ we define sequences $\pi_j^i$ and $\theta_j^i$ of items rather than just two types of items (items slightly smaller than $\frac{1}{2^j}$ and items slightly larger than $\frac{1}{2^j}$). We use these items to construct pairs of items, where the total sum of each pair is larger than $\frac{1}{2^{j-1}}$ in order to admit a* NE*, while the total size for $\pi_j^i$ and $\theta_j^i$ together has to be smaller than $\frac{1}{2^{j-1}}$ in order to keep the optimal packing valid. Having identical $\pi_j$ and $\theta_j$ items for a phase would violate the* NE *condition, as the largest item of the phase benefits from migrating to a bin of the previous phase.*

See Figure 3.3 for an illustration of the construction of a lower bound for $s = 3$.

### 3.4.2 An upper bound

To bound the POA from above, we prove the following theorem.

**Theorem 33.** *For any instance of the bin packing game $G \in BP$: Any* NE *packing uses at most $1.64286 \cdot OPT(G) + 2$ bins, where $OPT(G)$ is the number of bins used in a coordinated optimal packing.*

**Proof.** Let us consider a configuration $b$ of bins which is yielded by a NE packing. We classify the bins according to their loads into four mutually disjoint groups in the following manner. Group $\mathcal{A}$- contains all bins with an item of size strictly greater than $\frac{1}{2}$ (and possibly other items as well); The rest of the bins contain items of size at most $\frac{1}{2}$. Group $\mathcal{B}$- contains bins with loads at least $\frac{7}{8}$; Group $\mathcal{C}$- contains bins with loads in $[0.8, \frac{7}{8})$; Group $\mathcal{D}$- contains bins with loads in $(0, 0.8)$. We denote the cardinality of these groups by $n_{\mathcal{A}}, n_{\mathcal{B}}, n_{\mathcal{C}}$ and $n_{\mathcal{D}}$, respectively. We list the bins in each group from left to right in non-increasing order w.r.t. their loads.

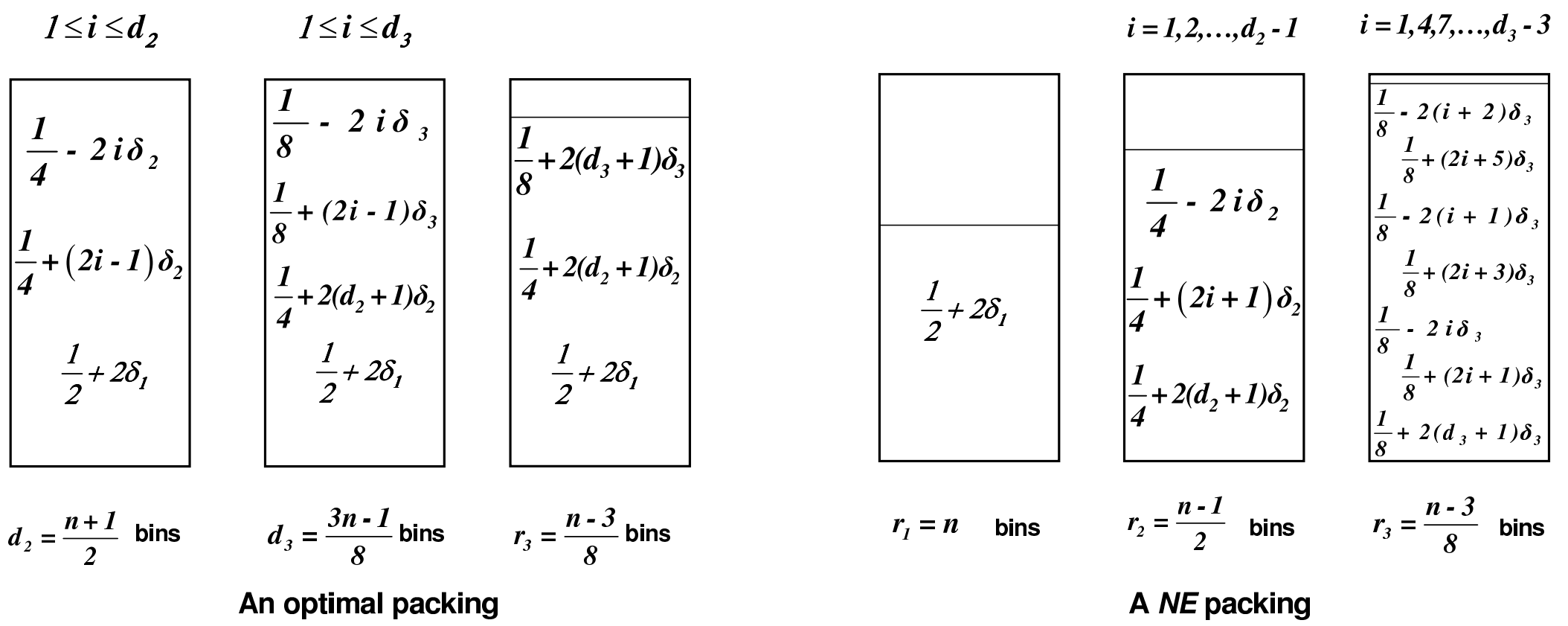

Figure 3.3: A lower bound of 1.625, the construction of Theorem 27 with $s = 3$.

Our purpose is to find an upper bound on the total number of bins in these four groups. This is obtained by formulating an integer linear mathematical program, in which the variables represent the number of bins for each group, and whose optimal solution value yields an upper bound on the POA.

We first show that all the bins in group $\mathcal{D}$ are filled to at least $\frac{2}{3}$, except for at most two bins.

**Claim 34.** *Group $\mathcal{D}$ contains at most one bin with load in $(0, \frac{1}{2}]$ and at most one bin with load in $(\frac{1}{2}, \frac{2}{3})$.*

**Proof.** Assume by contradiction that there exist two bins in group $\mathcal{D}$ that are less than half full. Then, as no bin in $\mathcal{D}$ contains an item of size greater than $\frac{1}{2}$, there would trivially exist an improving step for some item in one of these bins to the other bin, in contradiction to the assumption that the considered configuration yields a NE packing. Thus, we can conclude that all the bins in group $D$ have loads in $(\frac{1}{2}, 0.8)$, except for at most one bin.

Now, let us assume by contradiction that there exist two bins in group $\mathcal{D}$ that have loads in $(\frac{1}{2}, \frac{2}{3})$. Note that the rightmost of these two bins cannot contain any item with size in $(0, \frac{1}{3}]$, as such an item can make an improving step to the other bin, which has at least the same load, as it fits in it. Thus, the rightmost such bin contains only items of sizes in $(\frac{1}{3}, \frac{1}{2}]$. As we have assumed its load is in the range $(\frac{1}{2}, \frac{2}{3})$, we conclude it essentially has to contain exactly two such items, sum of sizes of which exceed $\frac{2}{3}$, and yet again, we derive a contradiction. Thus, only one such bin may exist in group $\mathcal{D}$. ☐

We will split the rest of our analysis into three complementary cases, in accordance to the loads admitted by bins in group $\mathcal{A}$.

**Case (1):** *There are no bins in group $\mathcal{A}$.*
In this case, a formulation of the aforementioned mathematical program is straightforward. By Claim 34 and the definition of the groups, we know that all the bins in groups $\mathcal{B}$, $\mathcal{C}$ and $\mathcal{D}$ (except maybe for two) are filled by more than $\frac{2}{3}$. As there are no bins in group $\mathcal{A}$ present, using $OPT(G) \geq \sum_{i=1}^{n} a_i$ we get:

$$OPT(G) \geq \frac{2}{3}(n_{\mathcal{B}} + n_{\mathcal{C}} + n_{\mathcal{D}} - 2).$$

In total,

$$n_{\mathcal{B}} + n_{\mathcal{C}} + n_{\mathcal{D}} \leq 1.5 \cdot OPT(G) + 2.$$

**Case (2):** *All bins in group $\mathcal{A}$ are loaded by more than 0.573.*
All the bins in group $\mathcal{A}$ have a load of at least $0.573$. By definition, each bin in this group contains exactly one item of size strictly larger than $0.5$. Thus, an optimal packing will have to use at least $n_{\mathcal{A}}$ bins. We also know that by the definition of the groups, bins in groups $\mathcal{B}$ and $\mathcal{C}$ are loaded by at least $\frac{7}{8}$ and $0.8$, respectively. Also, by Claim 34 the bins in group $\mathcal{D}$ (except maybe for two) are filled by more than $\frac{2}{3}$. The suitable mathematical program is thus:

$$\begin{aligned} &OPT(G) \geq n_{\mathcal{A}} \\ &OPT(G) \geq 0.573 n_{\mathcal{A}} + \frac{7}{8} n_{\mathcal{B}} + 0.8 n_{\mathcal{C}} + \frac{2}{3}(n_{\mathcal{D}} - 2). \end{aligned}$$

In total,

$$n_{\mathcal{A}} + n_{\mathcal{B}} + n_{\mathcal{C}} + n_{\mathcal{D}} \leq 1.6405 \cdot OPT(G) + 2.$$

**Case (3):** *There exists at least one bin in group $\mathcal{A}$ that has a load $L$ in the range* $(0.5, 0.573)$.
We first prove a useful general observation, that allows us to characterize the sizes of items that are packed in bins of groups $\mathcal{C}$ and $\mathcal{D}$ in the packing induced by $b$.

**Observation 35.** *Every bin of load less than $y$ in a* NE *packing $b$ does not contain items of size in* $[y - 0.5, 0.427]$, *and a group of bins with load less than $y$, at most one bin (the leftmost one) may contain items of size in* $(0, 1 - y]$.

*Proof.* No group of bins with load less than $y$ can contain items of size in $(0, 1-y]$ (except, maybe, for the leftmost bin in the group), as the leftmost bin in that group is less than $y$ full, and in addition its load is no smaller than the load of any other bin in the group. Hence, if there was an item with size in $(0, 1 - y]$ in any other bin in this group, it could move to the leftmost bin, thus reducing its cost. This contradicts the assumption that $b$ is a NE packing. Now, as there exists a bin in group $\mathcal{A}$ that has a load in $(0.5, 0.573)$, any item $x$, such that $y - 0.5 \leq x \leq 0.427$, that might have been in a bin which is loaded by less than $y$ by assumption, will benefit from migrating to this bin in group $\mathcal{A}$, as it fits in, and the bin will admit a load greater than $y$ as a result of this migration. An existence of such improving step contradicts the assumption that $b$ is a NE packing. Thus, no bin in group that is loaded by less than $y$ contains an item of size in $[y - 0.5, 0.427]$. □

We will further split the analysis into two options:

**Case(3.1)** *There are no bins in group $\mathcal{D}$.*
By the definition of the groups, we know that all the bins in groups $\mathcal{B}$ and $\mathcal{C}$ are loaded by at least $0.8$. We get the following linear program:

$$OPT(G) \geq n_{\mathcal{A}}$$
$$OPT(G) \geq 0.5n_{\mathcal{A}} + 0.8n_{\mathcal{B}} + 0.8n_{\mathcal{C}}.$$

that yields
$$n_{\mathcal{A}} + n_{\mathcal{B}} + n_{\mathcal{C}} \leq 1.625 \cdot OPT(G).$$

As by Claim 34 there can be at most two bins in group $\mathcal{D}$ that are less than $\frac{2}{3}$ full,

$$n_{\mathcal{A}} + n_{\mathcal{B}} + n_{\mathcal{C}} + n_{\mathcal{D}} \leq 1.625 \cdot OPT(G) + 2.$$

**Case(3.2)** *There exists at least one bin in group $\mathcal{D}$.*
In order to formulate the mathematical program, we first prove the following facts.

**Claim 36.** *In the case there exists a bin in group $\mathcal{A}$ that has a load in $(0.5, 0.573)$:*
*a. All the bins in group $\mathcal{D}$ have load in $(\frac{3}{4}, 0.8)$, except for at most one bin (the rightmost bin in $\mathcal{D}$).*
*b. Each bin in group $\mathcal{D}$ contains at least two items of size larger than $\frac{1}{4}$, except for possibly at most two bins (the leftmost and the rightmost bins in $\mathcal{D}$).*

**Proof.** First, we would like to characterize the items that are packed in bins of group $\mathcal{D}$ in the packing induced by $b$.

As no bin in $\mathcal{D}$ contains an item of size greater than $0.5$, using Observation 35 with $y = 0.8$ we conclude that all bins in group $\mathcal{D}$, except for at most one bin (the leftmost bin in $\mathcal{D}$), contain only items with sizes in $(0.2, 0.3) \bigcup (0.427, 0.5]$.

For convenience, in the rest of our discussion we refer to items with sizes in $(0.427, 0.5]$ as items of type $A$, and to items with sizes in $(1 - y, y - 0.5)$ as items of type $B$.

Let us consider what are the possible combinations of the items of types $A$ and $B$ in the bins of group $\mathcal{D}$, all of which, except for possibly two bins as was proven in Claim 34, have loads in $[\frac{2}{3}, 0.8)$:

(i) $A$ – There can be at most one bin that contains a single item of type $A$, as an $A$ type item has size of at most $0.5$, and we know all bins in $\mathcal{D}$ (except maybe for one) are more than $0.5$ full.
(ii) $B$– Not possible, as this item will benefit from moving to a bin from the group $\mathcal{A}$ with load $L \in (0.5, 0.573)$ (which exists).
(iii) $AB$– Not possible, as an item of type $B$ will benefit from moving to a bin from the group $\mathcal{A}$ with load $L \in (0.5, 0.573)$ (which exists), as it fits there, and $L > 0.5$ while the size of type $A$ item is at most $0.5$.
(iv) $AA$– Not possible, as two type $A$ items produce load greater than $0.8$.
(v) $BB$– Not possible, as one of the type $B$ items will benefit from moving to a bin from group $\mathcal{A}$ with load $L \in (0.5, 0.573)$, as it fits there, and its load would be at least $0.7$ in case of migration, which is greater than a load of two $B$ type items (which can get only up to 0.6).
(vi) $BBB$– Possible.
(vii) $BBBB$–Not possible, as any $B$ type item has a size of at least $0.2$, and four such items would incur a load greater than $0.8$. The same applies to $ABB$.

Other combinations are not possible as the items would simply do not fit into a bin of unit capacity. Therefore, the only combination of items that is possible in bins of group $\mathcal{D}$ is a triple of type $B$ items (except for maybe two bins which we consider separately: the bin with a single $A$ type item (that corresponds to the bin with load in $(0, \frac{1}{2}]$ we mentioned earlier in the discussion), and the leftmost bin in $\mathcal{D}$ that may contain items other than of $A$ or $B$ type).

Now we prove that all bins in $\mathcal{D}$, except for at most one bin, are loaded by more than $\frac{3}{4}$. Consider any bin in $\mathcal{D}$ that contains three $B$ type items, with sizes $x, y, z \in (0.2, 0.3)$. We would like to show that $x + y + z > \frac{3}{4}$ must hold. Observe, that each one of $x, y$ and $z$ fits into the bin from group $\mathcal{A}$ with load $L \in (0.5, 0.573)$ (which exists). As we assume the configuration discussed admits a NE packing, none of them would benefit from migrating to this bin. Thus, we can conclude that the following holds:

$$\begin{cases} x + y + z > x + L \\ x + y + z > y + L \\ x + y + z > z + L. \end{cases}$$

This set of inequalities yields $x + y + z > \frac{3}{2}L$, and as $L > \frac{1}{2}$, we get $x + y + z > \frac{3}{4}$.

To prove part (b) of the claim, we will split the $B$ items into two types: items having sizes smaller than 0.25, that is, sizes in $(1-y, 0.25]$ and items having sizes larger than 0.25, that is, sizes in $(0.25, y-0.5)$. We refer to the former as $B_1$ type items and to the latter as $B_2$ type items. There are four different combinations of triples of $B_1$ and $B_2$ type items:

(i) $B_1B_1B_1$– Not possible, as one of the type $B_1$ items will benefit from moving to a bin from group $\mathcal{A}$ with load $L \in (0.5, 0.573)$, as it fits there, and $L > 0.5$, while the size of two $B_1$ is at most $0.5$.

(ii) $B_1B_1B_2$– Same as in case (i), only now $B_2$ type item migrates.

(iii) $B_1B_2B_2$– Possible.

(iv) $B_2B_2B_2$– Possible.

We can see, that both possible combinations include two items of type $B_2$, proving part (b) of Claim 36. Recall, that there can be only two "poorly" loaded bins in $\mathcal{D}$: one with load in $(0, \frac{1}{2}]$- this is the bin with single $A$ type item if such exists, and one with load in $(\frac{1}{2}, \frac{2}{3})$. We can observe that there can not be a bin with load in $(\frac{1}{2}, \frac{2}{3})$ in group $\mathcal{D}$, as any such bin necessarily contains three $B$ type items, but any of the two possible combinations $B_1B_2B_2$ and $B_2B_2B_2$ admits a load strictly greater than $\frac{2}{3}$. We conclude that all bins in $\mathcal{D}$, except for maybe one bin with a load in $(0, \frac{1}{2}]$, are loaded by more than $\frac{3}{4}$, thus finishing the proof of part (a) of Claim 36. □

In this point, we have established that (almost) all bins in group $\mathcal{D}$ have loads in $(0.75, 0.8)$. Again, we consider two complement cases separately:

**Case (3.2.a)** *All bins in group $\mathcal{D}$ have loads in $[0.775, 0.8)$ (except for maybe one bin with load in $(0, \frac{1}{2}]$).*

By Claim 36, we know that each bin in group $\mathcal{D}$ (except for at most two bins) contains at least two items of size greater than $0.25$. An optimal solution will have to use at least $\frac{2(n_{\mathcal{D}}-2)}{3}$ bins to pack these items. Moreover, we can consider each item of size greater than $0.5$ in bin of $\mathcal{A}$ as two items of size greater than $0.25$. As all the bins in groups $\mathcal{B}$ and $\mathcal{C}$ are loaded by at least $0.8$, we obtain the following linear program:

$$\begin{aligned}
&OPT(G) \geq n_{\mathcal{A}}\\
&OPT(G) \geq 0.5n_{\mathcal{A}} + 0.8n_{\mathcal{B}} + 0.8n_{\mathcal{C}} + 0.775(n_{\mathcal{D}}-1)\\
&OPT(G) \geq \frac{2}{3}n_{\mathcal{A}} + \frac{2}{3}(n_{\mathcal{D}}-2).
\end{aligned}$$

In total,

$$n_{\mathcal{A}} + n_{\mathcal{B}} + n_{\mathcal{C}} + n_{\mathcal{D}} \leq 1.6406 \cdot OPT(G) + 1.03125.$$

**Case (3.2.b)** *There exists at least one bin in group $\mathcal{D}$ with load in $(0.75, 0.775)$.*

In order to formulate the mathematical program, we first prove the following fact.

**Claim 37.** *In the case there exists a bin in group $\mathcal{A}$ that has a load in $(0.5, 0.573)$: Each bin in group $\mathcal{C}$, except for possibly one bin (the leftmost bin in $\mathcal{C}$), contains at least two items of size larger than $\frac{1}{4}$.*

**Proof.** First, we would like to characterize the items that are packed in bins of group $\mathcal{C}$ in the packing induced by $b$.

As no bin in $\mathcal{C}$ contains an item of size greater than $0.5$, using Observation 35 with $y = \frac{7}{8}$ we conclude that all bins in group $\mathcal{C}$, except for at most one bin (the leftmost bin in $\mathcal{C}$), contain only items with sizes in $(\frac{1}{8}, \frac{3}{8}) \bigcup (0.427, 0.5]$.

Also, as we assumed that there is at least one bin in group $\mathcal{D}$ having load in $(0.75, 0.775)$, an item with size in $(\frac{1}{8}, 0.225]$ will move to this bin as it fits there, and the load of this bin will be strictly greater than $y$ as a result of the migration. As $b$ is a NE packing, no such items exists in $\mathcal{C}$. We get that all bins in group $\mathcal{C}$, except for at most one bin, contain only items with sizes in $[0.225, \frac{3}{8}) \bigcup (0.427, 0.5]$.

Similarly to above, in the sequel of our discussion we refer to items with sizes in $(0.427, 0.5]$ as items of type $A$, and to items with sizes in $[0.225, y - 0.5)$ as items of type $B$.

Now, let us consider what are the possible combinations of items of types $A$ and $B$ in the bins in group $\mathcal{C}$, all of which have loads in $[0.8, \frac{7}{8})$:

(i) $A$– Not possible, as an $A$ type item has size of at most $0.5$, while all bins in $\mathcal{C}$ admit a load greater than $0.8$.
(ii) $B$– Not possible, as a $B$ type item has size of at most $\frac{3}{8}$, while all bins in $\mathcal{C}$ admit a load greater than $0.8$.
(iii) $AA$– Possible.
(iv) $BB$– Not possible, as two $B$ type items have size of at most $\frac{6}{8}$, while all bins in $\mathcal{C}$ are loaded by more than $0.8$.
(v) $AB$– Not possible, as the $B$ type item will benefit from moving to a bin from the group $\mathcal{A}$ with load $L \in (0.5, 0.573)$ (which exists), as it fits there, and $L > 0.5$ while the size of type $A$ item is at most $0.5$.
(vi) $ABB$– Not possible, as an $A$ type item have size of at least $0.427$ and two $B$ type items have size of at least $0.45$, resulting in a load of $0.877$, which exceeds the maximal load of $\frac{7}{8}$ for a bin in $\mathcal{C}$. The same applies to $BBBB$, as four $B$ type items incur load of at least $0.9$.
(vii) $BBB$– Possible.

Any other combination is not possible, as the items would simply not fit into a bin of unit capacity.

Therefore, a bin in group $\mathcal{C}$ may contain either a pair of $A$ type items or a triple of $B$ type items. For a bin in $\mathcal{C}$ which contains a pair of $A$ type items, Claim 37 trivially holds.

The case where a bin contains a triple of $B$ type items has to be analyzed more thoroughly. We will split the $B$ items into two types: items having sizes smaller than 0.25, that is, sizes in $[0.225, 0.25]$ and items having sizes larger than 0.25, that is, sizes in $(0.25, y - 0.5)$.

Using similar considerations as in Claim 36, we conclude that each bin in $\mathcal{C}$, except for maybe one bin, necessarily contains at least two items of size larger than $\frac{1}{4}$, proving Claim 37.

□

By Claims 36 and 37, we know that each bin in group $\mathcal{C}$ (except for at most one bin) and each bin in group $\mathcal{D}$ (except for at most two bins) contains at least two items of size greater than $0.25$. An optimal solution will have to use at least $\frac{2}{3}(n_{\mathcal{C}} + n_{\mathcal{D}} - 3)$ bins to pack these items. We consider each item of size greater than $0.5$ in bin of $\mathcal{A}$ as two items of size greater than $0.25$. As all the bins in group $\mathcal{B}$ are loaded by at least $\frac{7}{8}$, all bins in group $\mathcal{C}$ by at least $0.8$, and all bins in group $\mathcal{D}$ (except for at most one) by at least $\frac{3}{4}$ (by Claim 36(a)), we get the following program:

$$
\begin{aligned}
&OPT(G) \geq n_{\mathcal{A}} \\
&OPT(G) \geq 0.5 n_{\mathcal{A}} + \frac{7}{8} n_{\mathcal{B}} + 0.8 n_{\mathcal{C}} + 0.75(n_{\mathcal{D}} - 1) \\
&OPT(G) \geq \frac{2}{3}(n_{\mathcal{A}} + n_{\mathcal{C}} + n_{\mathcal{D}} - 3).
\end{aligned}
$$

In total,

$$n_{\mathcal{A}} + n_{\mathcal{B}} + n_{\mathcal{C}} + n_{\mathcal{D}} \leq 1.64286 \cdot OPT(G) + 1.2857.$$

Combining all the possible cases, we get that the number of bins in the packing induced by $b$ which yields a NE is at most $1.64286 \cdot OPT(G) + 2$. □

Note that the choice of the constants that we used to define our groups is not incidental. Constant 0.573 for group $\mathcal{A}$ was chosen s.t. it is large enough for the program for case 1 to yield an upper bound which is smaller than our claimed upper bound $1.64286$, and such that $1 - 0.573$ is large enough so that there can not be a combination of items $ABB$ in both groups $\mathcal{C}$ and $\mathcal{D}$, else we cannot claim that there are always 2 items of size larger than $\frac{1}{4}$ in these groups. Constants $0.8$ and $0.775$ were chosen s.t. they are large enough for the programs for case 3.1 and the for case 3.2.a to yield an upper bound which is smaller than the claimed upper bound of $1.64286$, and s.t. $1 - 0.8$ and $1 - 0.775$ are large enough so that we cannot have four $B$ type items in bins of group $\mathcal{C}$ and $\mathcal{D}$.

From Theorem 33, we get

**Corollary 38.** $PoA(BP) \leq 1.64286.$

**Proof.**

$$\begin{aligned} PoA(BP) &= \limsup_{OPT(G)\to\infty} \sup_{G\in BP} \sup_{b\in NE(G)} \frac{n_{\mathcal{A}} + n_{\mathcal{B}} + n_{\mathcal{C}} + n_{\mathcal{D}}}{OPT(G)} \\ &\leq \limsup_{OPT(G)\to\infty} \frac{1.64286 \cdot OPT(G) + 2}{OPT(G)} = 1.64286. \end{aligned}$$

□

From Theorem 27 and Corollary 38 we conclude that $1.64163 \leq PoA(BP) \leq 1.64286$. We conjecture that the true bound is equal to the lower bound which we provide.

**Conjecture 39.** $PoA(BP) = \sum_{j=1}^{\infty} 2^{-j(j-1)/2} = 1.64163$.

## 3.5 Strong Prices of Anarchy and Stability in the Bin Packing game

This section is dedicated to the analysis of the Strong Price of Anarchy and the Strong Price of Stability of the bin packing game, and to establishing the fact that they admit the same value, which is equal to the (asymptotic) approximation ratio of the *Subset Sum* algorithm.

Consider the *Subset Sum* algorithm for bin packing, that proceeds by filling one bin at a time with a set of items that fills the bin as much as possible.

**Theorem 40.** *For the Bin Packing game, the set of* SNE *and the set of outcomes of Subset Sum algorithm coincide.*

A proof of this result is given in two parts.

**Claim 41.** *The output of the Subset Sum algorithm is always a* SNE.

**Proof.** Let us consider a packing $b$ of a set $N$ of items produced by the *Subset Sum* algorithm, where the output bins are numbered from left to right, in the order in which they are packed by the algorithm. Let $k$ be the number of bins, and let $L_i$ for $i = 1, \dots k$ be the load of bin $\mathcal{B}_i$ in the packing. By the definition of the *Subset Sum* algorithm, we have $L_1(b) \geq L_2(b) \geq \dots \geq L_k(b)$.

Assume by contradiction that this packing is not an SNE. Then, by definition of SNE, there exists a non-empty group $A$ of items $i_1, i_2, \dots, i_m$ in this packing, all of which benefit from migration to different bins, resulting in a packing $b'$. Note that in this cooperative setting, unlike in the non cooperative case, the number of bins may grow as a result of the improving step, since items may reduce their costs by jointly moving into a new bin.

Consider the leftmost bin in packing $b$ that contains an item (or several items) from coalition $A$. Let this bin be $\mathcal{B}_j$, and let this item be $i_r$. Consider the bin $\mathcal{B}_{j^*}$ this item has migrated to. We know that the load observed by $\mathcal{B}_{j^*}$ after the migration is strictly greater than that of $\mathcal{B}_j$. Observe, that $\mathcal{B}_{j^*}$ cannot be one of the bins to the left of $\mathcal{B}_j$ with indices in $\{1, \ldots, j-1\}$, as the loads of these bins were not affected by the migration; Indeed, since they contain no item from $A$, their loads could not be decreased as a result of the step, as no item has left them. Also, their loads could not get increased, because if one of the items $i_h \in A$ would have migrated to one of these bins, say $\mathcal{B}_f$ where $f \in \{1, \ldots, j-1\}$, the contents of that bin together with this item would form a set of items that fits into a unit-capacity bin and observes a greater load than the load of all candidate subsets of items considered by the algorithm for the packing of bin $\mathcal{B}_f$. That contradicts the fact that bins with indices in $\{1, \ldots, f\}$ were filled by the *Subset Sum* algorithm.

We get that $L_i(b) = L_i(b')$ for $i = 1, \ldots j-1$. Thus, $\mathcal{B}_{j^*}$ can be either one of the bins to the right of $\mathcal{B}_j$ with indices in $\{j+1, \ldots, k\}$, or a new bin that was opened as a result of the improving step. As $L_j(b) < L_{j^*}(b')$, and bin $\mathcal{B}_{j^*}$ contains only items from the bins $\mathcal{B}_i$, $i = j, \ldots k$, this contradicts the fact that bin $\mathcal{B}_j$ was packed by the *Subset Sum* algorithm.

Therefore, the load of bin $\mathcal{B}_{j^*}$ after the migration could not be greater than the load of bin $\mathcal{B}_j$ in the original packing, contradicting that all members of $A$ strictly reduce their cost as a result of this migration, as it does not hold for the item $i_r$. □

**Claim 42.** *Any* SNE *is an output of some execution of the Subset Sum algorithm.*

**Proof.** Consider an SNE packing $b$ of a set $N$ of items, where the bins are ordered from left to right in non-increasing order of their loads. Let $k$ be the number of bins in $b$. We will show that there exists a run of the *Subset Sum* algorithm on an input $N$ which produces exactly this packing, by constructing each bin, one at a time, and showing this construction in every step is consistent with the actions of the *Subset Sum* algorithm. The construction is done by induction on the index of the bin in the SNE packing.

Assume that contents of the bins $\mathcal{B}_1, \ldots, \mathcal{B}_{j-1}$ that were packed by the *Subset Sum* algorithm are identical to the contents of the bins with the corresponding indices in the packing $b$, and we now use the *Subset Sum* algorithm to pack the next bin $\mathcal{B}_j$. Denote the load of the $j$-th bin in packing $b$ by $L$. We would like to show that bin $\mathcal{B}_j$ packed by the *Subset Sum* algorithm must have a load $L$, as well. First, observe the load of bin $\mathcal{B}_j$ can not be smaller than $L$, as items that are packed in bins with indices $\{j, \ldots, k\}$ in $b$ are still available to *Subset Sum* algorithm at the point of packing of bin $\mathcal{B}_j$.

Now, we show that the load of bin $\mathcal{B}_j$ can not be larger than $L$, either. Assume, by contradiction, that the load of the items that were packed by the *Subset Sum* algorithm in $\mathcal{B}_j$ is greater than the load observed by the $j$-th bin in $b$. Then, the contents of $\mathcal{B}_j$ forms a coalition of items, all of which are packed in bins with indices $\{j, \ldots, k\}$ in $b$ that observe a load not greater than $L$ (as the bins in $b$ are sorted in non-increasing order of loads), and all of which would benefit from moving to a new bin, in contradiction to the assumption that $b$ is an SNE packing. Thus, the load of bin $\mathcal{B}_j$ has to be exactly $L$.

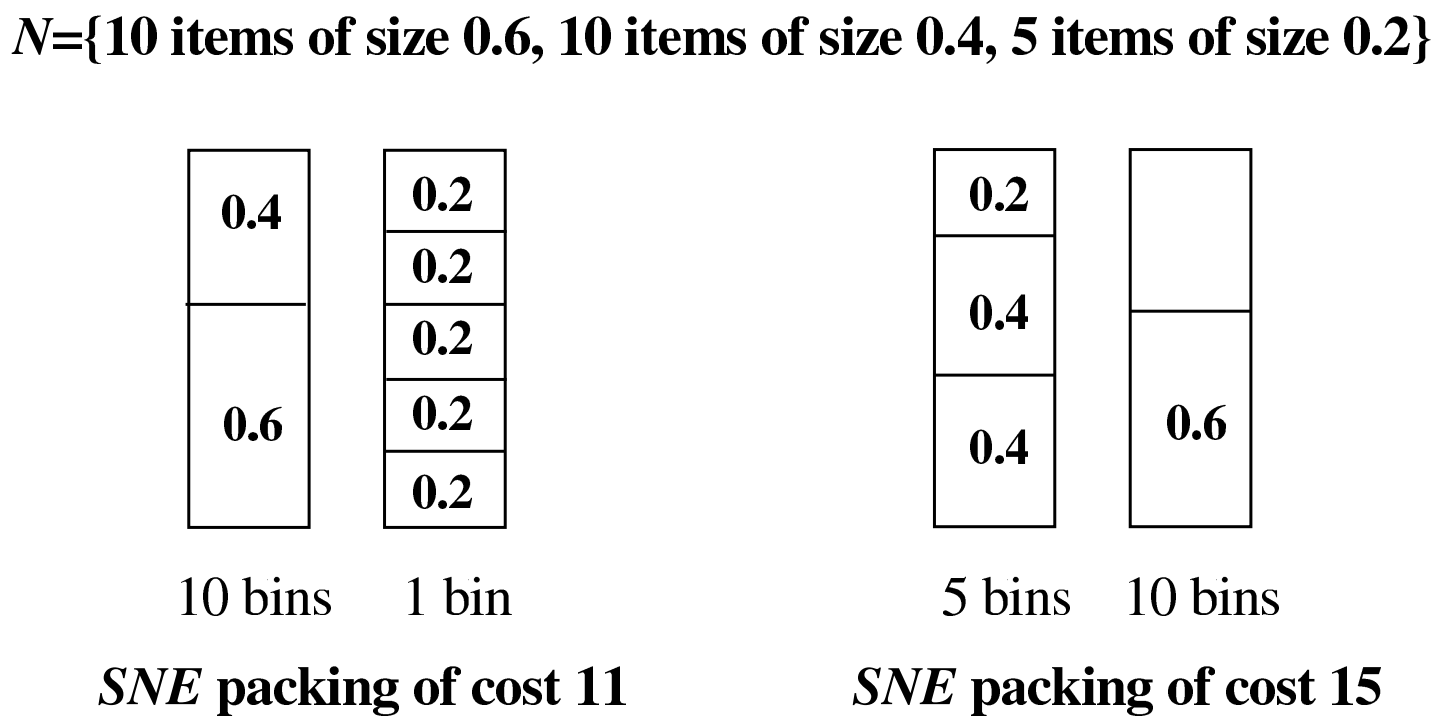

Figure 3.4: An example of two *SNE* packings of the same set of items with different social costs.

As the first $j-1$ bins in both packings are assumed to be identical, the items that are packed in the $j$-th bin in $b$ are still available for the *Subset Sum* algorithm to pack in $\mathcal{B}_j$, and we can consider it taking exactly that action (since the *Subset Sum* algorithm does not have a preference over sets of items of same load). ☐

Note, that not all SNE of a certain instance of the bin packing game necessarily admit the same social cost value, as the output of the *Subset Sum* algorithm for a set $N$ of items may not always be unique. The load and the contents of each consecutive bin is influenced by the choices of the items made by the *Subset Sum* algorithm for filling the previous bins. For example, see Figure 3.4.

Now, we would like to show that for the bin packing game, SPOA equals SPOS, and that this value is equal to the approximation ratio of the *Subset Sum* algorithm. We denote this approximation ratio by $R_{SS}^{\infty}$.

**Theorem 43.** *For the bin packing game introduced above,* $\text{SPOS} = \text{SPOS} = R_{SS}$.

**Proof.** Claim 41 implies that $\text{SPOA} \geq R_{SS}^{\infty}$, and Claim 42 implies that $\text{SPOA} \leq R_{SS}^{\infty}$. In total, this proves $\text{SPOA} = R_{SS}^{\infty}$. It is left to show that $\text{SPOS} = R_{SS}^{\infty}$. In order to show the equivalence between SPOS and $R_{SS}^{\infty}$, we would have to prove the following sequence of inequalities: $\text{SPOS} \leq \text{SPOA} \leq R_{SS}^{\infty} \leq \text{SPOS}$. Recall that for any game the leftmost inequality holds by definition. The second inequality is shown by Claim 42. Thus, $\text{SPOS} \leq R_{SS}^{\infty}$. It therefore suffices to prove the following proposition.

**Proposition 44.** *For the bin packing game,* $R_{SS}^{\infty} \leq \text{SPOS}$.

**Proof.** A way to show this would be to show that every lower bound example for *Subset Sum* can be converted into an example that has a unique SNE (or a unique run of *Subset Sum*, which are equivalent by Claim 42). Then, the best (and the only) SNE of this example will be the result of the execution of *Subset Sum* on this example.

The problem in this approach is that it is not clear which one of the solutions of *Subset Sum* is to be considered, as it is not clear why *Subset Sum* should prefer any one set of items over any other equally respectable set of items. However, it is possible to define a closely related set of items, which does not offer *Subset Sum* any choice over the sets of items it packs at each stage. We achieve this by modifying the sizes of the items in the lower bound example of *Subset Sum*, in a way that determines its preferences over subsets of items that were in the same size in the original example.

So, consider a packing $b$ of a set $N$ containing $n$ items, which is a lower bound example for *Subset Sum*, where the bins are numbered from left to right, in the order in which they were packed by the algorithm. Let $k$ be the number of bins in this packing. Let $\ell = 2^n - 1$ be the number of all possible subsets of items in $N$ (except for the empty set), and let $P$ be the set of the distinct sums $P_i$ of sizes of items in these subsets, arranged in non-increasing order. We add the sum 1 to $P$ even if there does not exist a subset of items in $N$ that sums up to 1. We define $\rho = \min_{1 \le i \le \ell} \dfrac{P_i}{P_{i+1}}$. Denote $\beta = 1 + \varepsilon$, where $\varepsilon$ is a small positive constant, chosen such that $\beta^k < \rho$ holds. For $1 \le i \le k$, we modify the sizes of all the items in bin $\mathcal{B}_i$ in $b$ by a multiplicative factor $\alpha_i = \dfrac{1}{\beta^{i-1}}$, getting a new set of items $N'$. By the definition, $\alpha_1 \ge \alpha_2 \ge \ldots \ge \alpha_k$ holds. Informally, we can say that this factor is defined for each bin in a way that makes the items in the lower indexed bins in $b$ more "desirable" to *Subset Sum* than the items packed in the bins of higher index, by decreasing the sizes of the latter by more than the sizes of the former.

The selection of $\rho$ ensures for all $1 \le i \le \ell$ that $P_i$ would not get below $P_{i+1}$ as a result of this action, which preserves the global order between the sums in $P$. In what follows, we show it also ensures that this modification will not turn a non-feasible subset (i.e., a subset of items total size of which exceeds 1) of items of $N$ into a feasible one. We now show that the aforementioned modification of items in $N$ achieves its purpose, that is, enforces a unique run of *Subset Sum* algorithm on $N'$, and on the other hand does not decrease $R_{SS}^{\infty}$ by proving the following three claims.

We first show that the transformation is defined in a manner such that the sums decrease monotonically, and the order between successive sums is preserved.

**Claim 45.** *Consider two (not necessarily disjoint) sets of items, $U_1$ and $U_2$, such that the total sizes of items in these sets are $Q_1$ and $Q_2$ (respectively) before the modification, and $Q_1'$ and $Q_2'$ (respectively) after the modification. Then $Q_1 > Q_2$ implies $Q_1' > Q_2'$.*

**Proof.** Since the size of items were multiplied by factors no smaller than $\frac{1}{\beta^{k-1}}$, we have $Q'_1 \geq \frac{Q_1}{\beta^{k-1}}$. Since $Q_1 > Q_2$, by the definition of $\rho$, we have $Q_1 \geq \rho Q_2$. Since the modified item sizes are no larger than the original ones, we have $Q_2 \geq Q'_2$. This gives $Q'_1 \geq \frac{Q_1}{\beta^{k-1}} > \frac{Q_1}{\rho} \geq Q_2 \geq Q'_2$. □

From Claim 45 we see that by adding the sum 1 to the set $P$ we ensured that this modification applied on a non-feasible subset of items from $N$ does not make it feasible.

Next, we show that this modification does not decrease the worst-case approximation ratio of the *Subset Sum* algorithm.

**Claim 46.** *The Subset Sum algorithm has an unique run on $N'$, which produces a packing where each bin contains exactly the same set of items packed in the corresponding bin in the packing $b$, after transformation.*

**Proof.** Consider the packing $b$ of the set $N'$ of items produced by *Subset Sum*. The proof is done by induction over the index of the bin in the packing. Assume that each bin $\mathcal{B}_i$ for $1 \leq i \leq j-1$ in this packing contains the modified set of items which were packed in the $i$-th bin in the packing $b$, and consider the next bin $\mathcal{B}_j$ in this packing. We would like to prove that $\mathcal{B}_j$ contains exactly the set of modified items from the $j$-th bin in $b$.

Denote the load of the $j$-th bin in the packing $b$ by $L_j$. Observe, that whilst *Subset Sum* decided to pack the $j$-th bin in $b$ with some subset $U = \{u_1, \ldots, u_r\}$ of items from $N$, there could be more subsets of items with the same sum as $U$ present in $N$, which could have been equally considered by *Subset Sum* as candidates to be packed in this bin, instead of $U$. We claim, that after the transformation was applied on $N$, upon packing the $j$-th bin, the algorithm no longer had a choice over the corresponding subsets, as the modified subset $U' = \{u'_1, \ldots, u'_r\}$ observed a greater load, and thus was preferable to it.

So, consider any subset of items from $N$ other than $U$, say $V = \{v_1, \ldots, v_t\}$, sizes of items in which sum to $L_j$. We show that the transformed set $V' = \{v'_1, \ldots, v'_t\}$ admits a strictly smaller load than that of $U'$. Any subset $V$ that we should consider contains at least one item $v_h$ which was packed in one of the bins with index in $\{j+1, \ldots, k\}$ in $b$ (and no item from bins with indices in $\{1, \ldots, j-1\}$ in $b$). By the definition of the transformation, the sizes of all items $\{u_1, \ldots, u_r\}$ were multiplied by $\dfrac{1}{\beta^{j-1}}$, while all the sizes of the items packed in bins with indices in $\{j+1, \ldots, k\}$ in $b$, and $v_h$ among them, were multiplied by at most $\dfrac{1}{\beta^s}$, such that $s > j-1$. Thus, the sum of the sizes of items in $V'$ is at most $\dfrac{\sum_{i=1}^{t} a_{v_i} - a_{v_h}}{\beta^{j-1}} + \dfrac{a_{v_h}}{\beta^s}$, which is strictly smaller than the sum of the sizes of items in $U'$ which equals $\dfrac{\sum_{i=1}^{r} a_{u_i}}{\beta^{j-1}}$, as $\sum_{i=1}^{r} a_{u_i} = \sum_{i=1}^{t} a_{v_i}$ and $a_{v_h} > 0$, proving the claim. It is left to mention that *Subset Sum* had to choose exactly the set $U'$ having load $\dfrac{L_j}{\beta^{j-1}}$ for packing the bin $\mathcal{B}_j$, that is, that no set of larger total size is available at that time of packing the $j$th bin. Consider the items which are available for packing after bin $j-1$ has been created. For the original sizes of items, the largest total size of a feasible subset of items is $L_j$. Also, by Claim 45, no subset of items with load $L$ could have been created as a result of the transformation, such that $L > \frac{L_j}{\beta^{j-1}}$. ☐

**Claim 47.** *The number of bins in an optimal packing of $N'$ is no larger than the number of bins in an optimal packing of $N$.*

**Proof.** Note that multiplying each item $i_j \in N$, $1 \le j \le n$ by a factor $\frac{1}{\beta^{i-1}}$ for some $1 \le i \le k$, where $\beta > 1$, does not increase the size of this item. Therefore, for each set of items $i_1, \ldots, i_r$ packed in some bin of an optimal packing of $N$ and the corresponding set of items $i'_1, \ldots, i'_r$ received as a result of this multiplication $a_{i_j} \le a_{i'_j}$ holds, and thus $\sum_{j=1}^{r} a_{i'_j} \le \sum_{j=1}^{r} a_{i_j} \le 1$. As this holds for each bin in an optimal packing of $N$, an optimal packing of the set $N'$ uses no more bins than an optimal packing of the set $N$. ☐

Since this holds for any lower bound example, for any $\varepsilon > 0$ there exists an input $I$ for bin packing for which the approximation ratio of the *Subset Sum* algorithm is at least $R^{\infty}_{SS} - \varepsilon$. This would imply $R^{\infty}_{SS} - \varepsilon \le SPoS(G_\varepsilon)$ for a game $G_\varepsilon \in BP$ where $N_\varepsilon = I$ for any $\varepsilon > 0$ and thus $R^{\infty}_{SS} \le SPoS(BP)$ ☐

In total, the above discussion proves SPOA = SPOS = $R^{\infty}_{SS}$. ☐

Theorem 43 implies that the problem of bounding the SPOA and the SPOS of the bin packing game is equivalent to the problem of bounding the approximation worst-case ratio $R^{\infty}_{SS}$ of the well known *Subset Sum* algorithm for bin packing. It was recently proved that $R^{\infty}_{SS} = \sum_{i=1}^{\infty} \frac{1}{2^i - 1} \approx 1.6067$ (see [46]), as it was conjectured in [64]. We conclude that $SPoS(BP) = SPoA(BP) \approx 1.6067$.

## 3.6 The complexity of computing a Strong Nash packing

In this section we prove that given an instance of the bin packing game, the problem of finding a packing that admits Strong Nash is NP-hard. For this purpose we use a polynomial time reduction from the PARTITION problem. This problem is well-known to be NP-hard. The goal in the PARTITION problem is to decide whether a given set $S$ of $n$ integers with total sum $2B$ can be partitioned into two disjoint subsets, where the sum of elements in each such set equals $B$.

Given an instance $S = \{a_1, a_2, \ldots, a_n\}$ of the PARTITION problem we produce an input $N = \{b_1, b_2, \ldots, b_n\}$ to the bin packing game by scaling every element in $S$ by $B$: $b_i = \frac{a_i}{B}$ for $i = 1, \ldots, n$.

**Claim 48.** *$S$ can be partitioned into two distinct subsets such that the sum of elements in each subset is $B$ iff any* SNE *packing of $N$ uses exactly two unit-capacity bins. Thus, there does not exist a polynomial time algorithm which computes an* SNE *packing, unless P=NP.*

**Proof.** Assume that there exists a partition of $S = \{a_1, a_2, \ldots, a_n\}$ into two distinct subsets $S_1$ and $S_2$, such that the sum of elements in each subset is $B$. Let the items in $S_1$ according to this partition be $a_{i_1}, \ldots, a_{i_l}$ and the items in $S_2$ be $a_{i_{l+1}}, \ldots, a_{i_n}$. We have $\sum_{j=1}^{l} b_{i_j} = \sum_{j=1}^{l} a_{i_j}/B = 1$ and $\sum_{j=l+1}^{n} b_{i_j} = \sum_{j=l+1}^{n} a_{i_j}/B = 1$. Consider a SNE packing of the set $N$. As there is (at least one) subset of items in $N$ that have a total sum of $1$, the Subset Sum algorithm (that produces the all possible SNE packings) will pack such a set into one bin, and the remaining subset of items, which also has a total sum of $1$, must be packed in another bin. This implies any packing of $N$ that admits SNE will use $2$ fully loaded bins.

Assume now that there does not exist a partition, in this case, at least three bins must be used for any valid packing of the input items $N$, so clearly no SNE packing can use only two bins. □

## 3.7 Summary and conclusions

In this chapter we have studied the Bin Packing problem, where the items are controlled by selfish agents, and the cost charged from each bin is shared among all the items packed into it, both in non-cooperative and cooperative versions. We have provided improved and almost tight upper and lower bounds on the Price of Anarchy of the induced game.

We also gave a simple deterministic algorithm that computes an SNEassignment for any instance of the Bin Packing game, and proved that the asymptotic worst-case approximation ratio of this algorithm equals the Strong Price of Anarchy and the Strong Price of Stability values of the game, providing tight bound on these measures. We have shown that the problem of computing an SNE assignment is NP-hard, which justifies the fact that this algorithm has exponential running time.

As the SPOA and POA values of the Bin Packing game yield similar results, we conclude that the efficiency loss in the Bin Packing game, as quantified by the POA, is derived from selfishness alone, and coordination will not help. On the other hand, as the POS yields significantly better results than the SPOS, and SPOS equals SPOA which, in turn, has similar value to the POA, it implies that in the Bin Packing game the best equilibrium is less efficient if coordination between the agents is allowed.

# Chapter 4

# Parametric Bin Packing of selfish items

## 4.1 Introduction and Motivation

In this chapter we consider a parametric version of the Bin Packing problem which was discussed in Chapter 3 (see [28] for a comprehensive survey on the Bin Packing problem and its variants). In the classic Bin Packing problem, we are given a set of items $N = \{1, 2, \ldots, n\}$. The $i$th item in $N$ has size $a_i \in (0, 1]$. The objective is to pack the items into unit capacity bins so as to minimize the number of bins used. In the parametric case, the sizes of items are bounded from above by a given value. More precisely, given a parameter $\alpha \leq 1$ we consider inputs in which the item sizes are taken from the interval $(0, \alpha]$. Setting $\alpha$ to 1 gives us the standard Bin Packing problem.

As discussed in Chapter 3, the Bin Packing problem is encountered in a great variety of networking problems, a fact that motivates the study of Bin Packing from a game theoretic perspective. The parametric Bin Packing problem can be met in real applications where the sizes of items are much smaller compared to the respective sizes of the recipient. For example, a common computational model in distributed systems is that the nodes exchange signals via a network. Most often, a signal represents the state of some hardware component and has a signal size ranging from a single bit up to a few bytes. The communication networks used are often based on a broadcast bus where fixed sized frames are transmitted. The amount of data that can be transmitted in each frame is almost always bigger than the size of a signal. Thus, from a resource perspective it would be desirable to transport several signals in each frame so as to reduce the bandwidth consumption.

The Parametric Bin Packing problem also models the problem of efficient routing in networks that consist of parallel links of same bounded bandwidth between two terminal nodes—similar to the ones considered in [18, 43, 79]. As Internet Service Providers often impose a bandwidth consumption policy which restricts the amount of data that can be downloaded/uploaded by each user, placing a restriction on the size of the items allowed to transfer makes the model more realistic.

## 4.2 The model

We study the Parametric Bin Packing problem both in cooperative and non-cooperative versions. In each case the problem is specified by a given parameter $\alpha$. The Parametric Bin Packing game is defined by a tuple $BP(\alpha) = \langle N_\alpha, (B_i)_{i \in N_\alpha}, (c_i)_{i \in N_\alpha} \rangle$. Where $N_\alpha$ is the set of the items, whose size is at most $\alpha$. Each item is associated with a selfish player—we sometimes consider the items themselves to be the players. The set of strategies $B_i$ for each player $i \in N_\alpha$ is the set of all bins. Each item can be assigned to one bin only. The outcome of the game is a particular assignment $b = (b_j)_{j \in N_\alpha} \in \times_{j \in N_\alpha} B_j$ of items to bins. All the bins have unit cost. The cost function $c_i$ of player $i \in N_\alpha$ is defined as follows. A player pays $\infty$ if it requests to be packed in an invalid way, that is, a bin which is occupied by a total size of items which exceeds 1. Otherwise, the set of players whose items are packed into a common bin share its unit cost proportionally to their sizes. That is, if an item $i$ of size $a_i$ is packed into a bin which contains the set of items $B$ then $i$'s payment is $c_i = a_i / \sum_{k \in B} a_k$. Notice that since $\sum_{k \in B} a_k \leq 1$ the cost $c_i$ is always greater than or equal to $a_i$. The social cost function that we want to minimize is the number of used bins (which equals the sum of players' individual costs).

Clearly, a selfish item prefers to be packed into a bin which is as full as possible. In the non-cooperative version, an item will perform an improving step if there is a strictly more loaded bin in which it fits. At a Nash equilibrium, no item can unilaterally reduce its cost by moving to a different bin. We call a packing that admits the Nash conditions an NE packing.

In the cooperative version of the Parametric Bin Packing game, we consider all (non-empty) subgroups of items from $N_\alpha$. The cost functions of the players are defined the same as in the non-cooperative case. Each group of items is interested to be packed in a way so as to minimize the costs for all group members. Thus, given a particular assignment, all members of a group will perform a joint improving step (not necessarily into a same bin) if there is an assignment in which, for each member, the new bin will admit a strictly greater load than the bin of origin. The costs of the non-members may be enlarged as a result of this improving step. At a strong Nash equilibrium, no group of items can reduce the costs of all its members by moving to different bins. We call a packing that admits the Strong Nash conditions an SNE packing.

To measure the extent of deterioration in the quality of Nash packing due to the effect of selfish and uncoordinated behavior of the players (items) in the worst-case we use the Price of Anarchy (POA) and the Price of Stability (POS). These are the standard measures of the quality of the equilibria reached in uncoordinated selfish setting [79, 95]. The POA / POS of an instance $G$ of the Parametric Bin Packing game are defined to be the ratio between the social cost of the worst/best Nash equilibrium and the social optimum, respectively. As packing problems are usually studied via asymptotic measures, we consider asymptotic POA and POS of the Parametric Bin Packing game *BP*$(\alpha)$, that are defined by taking a supremum over the POA and POS of all instances of the Parametric Bin Packing game, for large sets $N_\alpha$.

In addition, we consider stability measures that allow to separate the effect of the lack of coordination between players from the effect of their selfishness [5, 55]. The measures considered are the (asymptotic) Strong Price of Anarchy (SPOA) and the Strong Price of Stability (SPOS). These measures are defined similarly to the (asymptotic) POA and the POS but only strong equilibria are considered.

As we study the SPOA / SPOS measures in terms of the worst-case approximation ratio of a greedy algorithm for bin packing, we define here the *parametric asymptotic worst-case ratio* $R_A^\infty(\alpha)$ of algorithm $A$ by

$$R_A^\infty(\alpha) = \lim_{k\to\infty} \sup_{N \in V_\alpha} \left\langle \frac{A(N)}{OPT(N)} \,\middle|\, OPT(N) = k \right\rangle,$$

where $A(N)$ denotes the number of bins used by algorithm $A$ to pack the set $N$ of items, $OPT(N)$ denotes the number of bins used in the optimal packing of $N$ and $V_\alpha$ is the set of all sets $N$ for which the maximum size of the items is bounded from above by $\alpha$.

## 4.3 Related work and our contributions

The parametric Bin Packing problem is a well studied version of the classic Bin Packing problem (see [75], [28] for surveys on various results attained for this problem).

Bilò [18] was the first to study the classic unrestricted Bin Packing problem from a game theoretic perspective. Among other results, he proved that the Bin Packing game admits a pure Nash equilibrium and provided non-tight bounds on the Price of Anarchy (that apply for the case $\frac{1}{2} < \alpha \leq 1$). His work also shows that the Price of Stability of the Bin Packing game equals to 1. The quality of pure equilibria in this game was further investigated (by us) in [43], where we gave improved and almost tight bounds for the Price of Anarchy.

The quality of strong equilibria in this game in terms of SPOA and SPOS was analyzed in [43].

For this purpose we have considered a natural algorithm for the Bin Packing problem called *Subset Sum* (or SS algorithm for short). In each iteration, this algorithm finds among the unpacked items, a maximum size set of items that fits into a new bin.

Surprisingly, the worst-case approximation ratio of the *Subset Sum* algorithm (denoted by $R^{\infty}_{SS}(1)$) is deeply related to the Strong Price of Anarchy of the Bin Packing game. Indeed, the two concepts are equivalent: Every output of the SS algorithm is a strong Nash equilibrium, and every strong Nash equilibrium is the output of some execution of the SS algorithm. We also proved that in this game SPOA=SPOS (see [43]). We used these facts to show the existence of strong equilibria for the Bin Packing game and to characterize the SPOA and SPOS in terms of this approximation ratio, which was established in [74] (lower bound) and [46] (matching upper bound).

All results from [43] are fully presented in Chapter 3.

Here we consider the parametric Bin Packing problem where item sizes are all in an interval $(0, \alpha]$ for some $\alpha < 1$ from a game theoretic perspective.

In fact, results presented in Chapter 3 for the POA, SPOA and SPOS already apply for the case $\frac{1}{2} < \alpha \leq 1$. In this chapter we complement these results and study the pure Price of Anarchy for the parametric variant and show nearly tight upper bounds and lower bounds on it for any $\alpha \leq \frac{1}{2}$.

The main analytical tool that we use to derive the claimed upper bounds is *weighting functions*—a technique widely used for the analysis of algorithms for various packing problems [74, 80, 103] and other greedy heuristics [71, 72]. The idea of such weights is simple. Each item receives a weight according to its size and its assignment in some fixed NE packing. The weights are assigned in a way that the cost of the packing (the number of the bins used) is close to the total sum of weights. In order to complete the analysis, it is usually necessary to bound the total weight that can be packed into a single bin of an optimal solution.

The tight bound of 1 on the Price of Stability proved in [18] for the general unrestricted Bin Packing game trivially carries over to the parametric case.

We also consider the Strong Prices of Anarchy and Stability of the Bin Packing game for any $\alpha \leq \frac{1}{2}$. As our conclusions regarding the connections between the strong Prices of Anarchy and Stability and the worst-case approximation ratio of SS still apply for the parametric variant of the problem, we again use known results for the approximation ratio of SS where item sizes are all in an interval $(0, \alpha]$ for some $\alpha \leq \frac{1}{2}$ to get tight bounds on the Strong Prices of Anarchy and Stability of the Bin Packing game.

The first non-trivial bounds lower and upper bounds on the worst-case performance of the SS algorithm for $\alpha \leq \frac{1}{2}$ were given by Caprara and Pferschy in [19]. Specifically, they showed that for $t \geq 1$, if $\frac{1}{t+1} < \alpha \leq \frac{1}{t}$ then $R^{\infty}_{SS}(\alpha) \geq \sum_{i=1}^{\infty} \frac{t}{(t+1)^i - 1}$ and

$$R^{\infty}_{SS}(\alpha) \leq \begin{cases} 2 - \frac{4t}{3(t+1)} + \ln \frac{4}{3} & \text{if } t \leq 2 \\ 1 + \ln \frac{t+1}{t} & \text{if } t \geq 3. \end{cases}$$

The exact bounds on the worst-case performance of the SS algorithm for any $\alpha \leq 1$ were established recently by Epstein at. el. in [46] to be $R^{\infty}_{SS}(\alpha) = 1 + \sum_{i=1}^{\infty} \frac{1}{(t+1)2^i - 1}$. Note that the ratio $R^{\infty}_{SS}(\alpha)$ lies strictly between the upper and lower bounds of Caprara and Pferschy for all $\alpha \leq \frac{1}{2}$.

## 4.4 Price of Anarchy in the parametric Bin Packing game

We now provide a lower bound for the Price of Anarchy of the parametric Bin Packing game with bounded size items and, in addition, prove a very close upper bound for each value of $\frac{1}{t+1} < \alpha \leq \frac{1}{t}$ for a positive integer $t \geq 2$, that is, for all $0 < \alpha \leq \frac{1}{2}$. The case $\frac{1}{2} < \alpha \leq 1$ ($t = 1$) was extensively discussed in Chapter 3.

### 4.4.1 A lower bound: construction

In this section we give the construction of a lower bound on *PoA*$(\alpha)$. For each value of $t \geq 2$ we present a set of items which consists of multiple item lists. This construction is related to the one that we gave in Chapter 3 for $\frac{1}{2} < \alpha \leq 1$, though it is not a generalization of the former, which strongly relies on the fact that each item of size larger than $\frac{1}{2}$ can be packed alone in a bin of the NE solution, whereas in the parametric case there are no such items. It is based upon techniques that are often used to design lower bounds on Bin Packing algorithms (see for example [80]), but it differs from these constructions in the notion of order in which packed bins are created (which does not exist here) and the demand that each bin satisfies the Nash stability property. Our lower bound is given by the following theorem.

**Theorem 49.** *For each integer $t \geq 2$ and $\alpha \in (\frac{1}{t+1}, \frac{1}{t}]$, the PoA of the $BP(\alpha)$ game is at least* $\frac{t^2 + \sum_{j=1}^{\infty} (t+1)^{-j} \cdot 2^{-j(j-1)/2}}{t(t-1)+1}$.

**Proof.** Let $s > 2$ be an integer. We define a construction with $s + 1$ phases of indices $0 \leq j \leq s$, where the items of phase $j$ have sizes which are close to $\frac{1}{(t+1) \cdot 2^j}$, but can be slightly smaller or slightly larger than this value. For each $t \geq 2$, $t \in \mathbb{N}$ we use two sequences of positive integers $r^t_j$ and $d^t_j$, for $0 \leq j \leq s$. We choose the number $r^t_s$ to be an arbitrary sufficiently large value such that $r^t_s > 2^{s^3}$. For $0 \leq j \leq s - 1$, we define

recursively

$$r_j^t = 2^j \cdot (t+1) r_{j+1}^t + 1.$$

We define

$$d_j^t = r_{j-1}^t - r_j^t = ((t+1) \cdot 2^{j-1} - 1) r_j^t + 1,$$

for $1 \leq j \leq s$. In addition, we let $n = r_0^t$ and $d_0^t = 0$. Note that $r_1^t = \frac{n-1}{t+1}$, $d_1^t = r_0^t - r_1^t = \frac{nt+1}{t+1}$ (using $r_1^t + d_1^t = r_0^t = n$). Clearly $r_j^t \leq n$ and $d_j^t \leq n$, for $0 \leq j \leq s$.

We construct an input $I$ for which $OPT(I) = t(t-1) \cdot n + n$, and note that $n \geq r_j^s > 2^{s^3}$, $n >> t$. We use a sequence of small values $\delta_j$, $0 \leq j \leq s$, such that $\delta_j = \frac{1}{(4n)^{3s-2j}}$. Note that this implies $\delta_{j+1} = (4n)^2 \delta_j$ for $0 \leq j \leq s-1$.

**Observation 50.** *For each* $1 \leq j \leq s$, $\frac{n}{(t+1)^j \cdot 2^{j(j-1)/2}} - 1 \leq r_j^t \leq \frac{n}{(t+1)^j \cdot 2^{j(j-1)/2}}$.

**Proof.** For $j = 1$ it holds by definition. We next prove the property for $j+1$ using the property for $j$ (where $j \geq 1$). By definition of the sequence $r_j^t$, we have $r_{j+1}^t = \frac{r_j^t - 1}{(t+1) \cdot 2^j}$ for $j \geq 1$. Using the inductive assumption, we get

$$r_{j+1}^t < \frac{r_j^t}{(t+1) \cdot 2^j} \leq \frac{\frac{n}{(t+1)^j \cdot 2^{j(j-1)/2}}}{(t+1) \cdot 2^j} = \frac{n}{(t+1)^{j+1} \cdot 2^{j(j+1)/2}}.$$

On the other hand,

$$r_{j+1}^t = \frac{r_j^t - 1}{(t+1) \cdot 2^j} > \frac{\frac{n}{(t+1)^j \cdot 2^{j(j-1)/2}} - 2}{(t+1) \cdot 2^j} \geq \frac{n}{(t+1)^{j+1} \cdot 2^{j(j+1)/2}} - 1.$$

□

The input set of items for $t \geq 2$ consists of multiple phases. Phase 0 consists of the following sets of items: $nt$ items of size

$$\sigma_{01} = \frac{1}{t+1} + \Delta n t^2 (t-1) + \Delta,$$

$t(t-1)n$ items of size

$$\sigma_{02} = \frac{1}{t+1} - \Delta n t (t-1),$$

and pairs of items of sizes

$$\sigma_{03}^i = \tfrac{1}{t+1} + \Delta n t (t-1) + i\Delta \qquad \text{and} \qquad \sigma_{04}^i = \tfrac{1}{t+1} - i\Delta$$

for $1 \leq i \leq t(t-1)n$, such that $\Delta = \frac{2\delta_0}{nt(t-1)+1}$.

Note that $\sigma_{03}^i + \sigma_{04}^i = \frac{2}{t+1} + \Delta n t (t-1)$. There are also $t(t-1)(t-2)n$ items of size

$$\sigma_{05} = \frac{1}{t+1}.$$

For $1 \le j \le s$, phase $j$ consists of the following $2d_j^t + r_j^t$ items. There are $r_j^t$ items of size

$$\sigma_j = \frac{1}{(t+1)\cdot 2^j} + 2(d_j^t + 1)\delta_j,$$

and for $1 \le i \le d_j^t$, there are two items of sizes

$$\pi_j^i = \frac{1}{(t+1)\cdot 2^j} + (2i-1)\delta_j \qquad \text{and} \qquad \theta_j^i = \frac{1}{(t+1)\cdot 2^j} - 2i\delta_j.$$

Note that $\pi_j^i + \theta_j^i = \frac{1}{(t+1)\cdot 2^{j-1}} - \delta_j$. A bin of level $j$ in the optimal packing contains only items of phases $1, \ldots, j$. A bin of level $s+1$ contains items of all phases. The optimal packing contains $t(t-1)n$ bins of level 0, $d_j^t$ bins of level $j$, for $1 \le j \le s$, and the remaining bins are of level $s+1$. We have $\sum\limits_{j=1}^{s} d_j^t = r_0^t - r_s^t = n - r_s^t$. Thus, if the number of level $s+1$ bins is (at most) $r_s^t$, we have at most $n$ bins of levels $1 \le j \le s+1$, in addition to the $t(t-1)n$ bins of level 0. In total, the packing contains at most $t(t-1)n + n = (t(t-1)+1)n$ bins. The optimal packing of the set of items specified above is defined as follows. A level 0 bin contains $t-2$ items of size $\sigma_{05}$, one item of size $\sigma_{02}$ and, in addition, one pair of items of sizes $\sigma_{03}^i$ and $\sigma_{04}^i$ for a given value of $i$ such that $1 \le i \le t(t-1)n$. For $1 \le j \le s$, a level $j$ bin contains $t$ items of size $\sigma_{01}$ and one item of each size $\sigma_k$ for $1 \le k \le j-1$, and, also, one pair of items of sizes $\pi_j^i$ and $\theta_j^i$ for a given value of $i$ such that $1 \le i \le d_j^t$. A bin of level $s+1$ contains $t$ items of size $\sigma_{01}$ and one item of each size $\sigma_k$ for $1 \le k \le s$.

**Claim 51.** *This set of items $I$ can be packed into $n + t(t-1)n$ bins, i.e., $OPT(I) \le (1 + t(t-1))n$*

**Proof.** First, we show that every item was assigned into some bin. Consider the $nt$ items of size $\sigma_{01}$. Each $t$-tuple of these items is assigned into a bin of level $1 \le j \le s$ together. Consider items of size $\pi_j^i$ and $\theta_j^i$. Such items exist for $1 \le i \le d_j^t$, therefore, every such pair is assigned into a bin (of level $1 \le j \le s$) together. Next, consider items of size $\sigma_j$ for some $1 \le j \le s$. The number of such items is $r_j^t$. The number of bins which received such items is $\sum\limits_{k=j+1}^{s} d_k^t + r_s^t = r_j^t$. As to the items of size $\sigma_{02}$, there are $t(t-1)n$ such items, each item is assigned into one of the $t(t-1)n$ bins of level 0. The items $\sigma_{03}^i$ and $\sigma_{04}^i$ that exist for $1 \le i \le t(t-1)n$. Every such pair is assigned into one of the $t(t-1)n$ level 0 bins together. And, finally consider the $(t-2)\cdot t(t-1)n$ items of size $\sigma_{05}$. Each $(t-2)$ tuple of these items is assigned into one of the $t(t-1)n$ level 0 bins.

We further show that the sum of sizes of items in each bin does not exceed 1. Consider a bin of level 0. The sum of items it contains is:

$$(t-2)\sigma_{05}+\sigma_{02}+\sigma_{03}^i+\sigma_{04}^i = (t-2)\cdot\frac{1}{t+1}+\frac{1}{t+1}-\Delta nt(t-1)+\frac{2}{t+1}+\Delta nt(t-1) = 1.$$

Now, consider a bin of level $j$ for some $1 \leq j \leq s$. The sum of items packed in it is:

$$\begin{aligned}
&t\cdot\sigma_{01}+\sum_{k=1}^{j-1}\sigma_k+\frac{1}{(t+1)\cdot 2^{j-1}}-\delta_j \\
&= t\cdot(\frac{1}{t+1}+\Delta nt^2(t-1)+\Delta)+\sum_{k=1}^{j-1}(\frac{1}{(t+1)\cdot 2^k}+2(d_k^t+1)\delta_k)+\frac{1}{(t+1)\cdot 2^{j-1}}-\delta_j \\
&= \frac{t}{t+1}+t\cdot(\Delta nt^2(t-1)+\Delta)+\frac{1}{(t+1)\cdot 2^{j-1}}-\delta_j+\frac{1}{t+1}\sum_{k=1}^{j-1}\frac{1}{2^k}+2\sum_{k=1}^{j-1}(d_k^t+1)\delta_k \\
&\leq \frac{t}{t+1}+\frac{1}{(t+1)\cdot 2^{j-1}}+\frac{1-(\frac{1}{2})^{j-1}}{t+1}+t^2\cdot 2\delta_0-\delta_j+\sum_{k=1}^{j-1}2(d_k^t+1)\delta_k \\
&= 1+t^2\cdot 2\delta_0+\sum_{k=1}^{j-1}2(d_k^t+1)\delta_k-\delta_j.
\end{aligned}$$

It is left to show that $t^2\cdot 2\delta_0+\sum_{k=1}^{j-1}2(d_k^t+1)\delta_k-\delta_j \leq 0$ holds. As $d_k^t+1 \leq n$ and $\delta_j$ is a strictly increasing sequence, we have $2(d_k^t+1)\delta_k \leq 2n\delta_{j-1}$, and since $j-1 \leq s < n$, $\sum_{k=1}^{j-1}2(d_k^t+1)\delta_k < 4n^2\delta_{j-1}$. Also, as $t < n$, $t^2\cdot 2\delta_0 < 2n^2\delta_{j-1}$. Using $\delta_j = 16n^2\delta_{j-1}$ we get that the sum $t^2\cdot 2\delta_0+\sum_{k=1}^{j-1}2(d_k^t+1)\delta_k$ is smaller than $\delta_j$.

It is left to consider a bin of level $s+1$. The sum of items in it is:

$$\begin{aligned}
t\cdot\sigma_{01}+\sum_{k=1}^{s}\sigma_k &= t\cdot(\frac{1}{t+1}+\Delta nt^2(t-1)+\Delta)+\sum_{k=1}^{s}(\frac{1}{(t+1)\cdot 2^k}+2(d_k^t+1)\delta_k) \\
&= \frac{t}{t+1}+t\cdot(\Delta nt^2(t-1)+\Delta)+\frac{1-(\frac{1}{2})^s}{(t+1)}+\sum_{k=1}^{s}2(d_k+1)\delta_k \\
&= 1-\frac{(\frac{1}{2})^s}{(t+1)}+t\cdot 2\delta_0+\sum_{k=1}^{s}2(d_k^t+1)\delta_k.
\end{aligned}$$

We have $2(d_k^t+1)\delta_k \leq 2n\delta_s = \frac{1}{2^{2s-1}n^{s-1}}$. Since $2 < s < n$, $t < n$ and $t\cdot 2\delta_0 < 2n^2\delta_s$, we

get that the quantity above is at most

$$1-\frac{(\frac{1}{2})^s}{(t+1)}+\frac{n}{2^{2s-1}n^{s-1}}+2n^2\delta_s = 1-\frac{1}{2^s(t+1)}+\frac{1}{2^{2s-1}n^{s-2}}+2n^2\delta_s$$
$$= 1-\frac{1}{2^s(t+1)}+\frac{1}{2^{2s-1}n^{s-2}}+\frac{2n^2}{(4n)^s}$$
$$= 1-\frac{1}{2^s(t+1)}+\frac{1}{2^{2s-1}n^{s-2}}+\frac{1}{2^{2s-1}n^{s-2}} = 1-\frac{1}{2^s(t+1)}+\frac{1}{2^{2(s-1)}n^{s-2}} < 1.$$

□

Before introducing the NE packing for this set of items, we slightly modify the input by removing a small number of items. Clearly, $OPT(I') \leq (1+t(t-1))n$ would still hold for the modified input $I'$. The modification applied to the input is a removal of items $\pi_j^1$ and $\theta_j^{d_j^t}$ for all $1 \leq j \leq s$, the two items $\sigma_{03}^1$ and $\sigma_{04}^{t(t-1)n}$ and $(t-2)$ of the $\sigma_{05}$ items from the input. We now define an alternative packing, which is a NE. There are three types of bins in this packing. The bins of the first type are bins with items of phase $j$, $1 \leq j \leq s+1$. We construct $r_j^t$ such bins. A bin of phase $j$ contains $(t+1)\cdot 2^j - 1$ items, as follows. One item of size $\sigma_j = \frac{1}{(t+1)\cdot 2^j} + 2(d_j^t+1)\delta_j$, and $(t+1)\cdot 2^{j-1} - 1$ pairs of items of phase $j$. A pair of items of phase $j$ is defined to be the items of sizes $\pi_j^{i+1}$ and $\theta_j^i$, for some $1 \leq i \leq d_j^t - 1$. The sum of sizes of this pair of items is $\frac{1}{(t+1)\cdot 2^j}+(2i+1)\delta_j+\frac{1}{(t+1)\cdot 2^j}-2i\delta_j = \frac{2}{(t+1)\cdot 2^j}+\delta_j = \frac{1}{(t+1)\cdot 2^{j-1}}+\delta_j$.

Using $d_j^t = ((t+1)\cdot 2^{j-1}-1)r_j^t+1$ we get that all phase $j$ items, for $1 \leq j \leq s$ are packed. The sum of items in every such bin is $1-\frac{1}{(t+1)\cdot 2^{j-1}}+((t+1)\cdot 2^{j-1}-1)\delta_j + \frac{1}{(t+1)\cdot 2^j}+2(d_j^t+1)\delta_j = 1-\frac{1}{(t+1)\cdot 2^j}+\delta_j((t+1)\cdot 2^{j-1}+1+2d_j^t)$.

The $nt$ bins of the second type in the NE packing contain $(t-1)$ items of size $\sigma_{02} = \frac{1}{t+1}-\Delta nt(t-1)$ and one item of size $\sigma_{01} = \frac{1}{t+1}+\Delta nt^2(t-1)+\Delta$, from the 0 phase bins. The load of each such bin is

$$(t-1)\left(\frac{1}{t+1}-\Delta nt(t-1)\right)+\frac{1}{t+1}+\Delta nt^2(t-1)+\Delta$$
$$=\frac{t}{t+1}-\Delta nt(t-1)^2+\Delta nt^2(t-1)+\Delta = \frac{t}{t+1}+\Delta nt(t-1)(t-(t-1))+\Delta$$
$$=\frac{t}{t+1}+\Delta nt(t-1)+\Delta = \frac{t}{t+1}+\Delta(nt(t-1)+1) = \frac{t}{t+1}+2\delta_0,$$

by definition of $\Delta$. As there are in total $t(t-1)n$ identical items of size $\sigma_{02}$ and $nt$ identical $\sigma_{01}$ items in the input set, we get that all these items are packed in these $nt$ second type bins in the NE packing constructed above.

The $t(t-1)n-1$ bins of third type in the NE packing each contain $(t-2)$ items of size $\sigma_{05} = \frac{1}{t+1}$, and, in addition, one pair of items of sizes $\sigma_{03}^{i+1}$ and $\sigma_{04}^i$, for some $1 \leq i \leq t(t-1)n$ from the phase 0 bins. The sum of sizes of this pair of items is:

$\sigma_{03}^{i+1} + \sigma_{04}^{i} = \frac{1}{t+1} + \Delta nt(t-1) + (i+1)\Delta + \frac{1}{t+1} - i\Delta = \frac{2}{t+1} + \Delta(nt(t-1)+1) = \frac{2}{t+1} + 2\delta_0$. Thus, the total load of such bin is $(t-2)\cdot\frac{1}{t+1} + \frac{2}{t+1} + 2\delta_0 = \frac{t}{t+1} + 2\delta_0$, which equals the load of the bins of the second type in the NE packing. As there are in total $((t-2)\cdot t(t-1)n-(t-2)) = (t-2)(t(t-1)n-1)$ items of size $\sigma_{05}$ and $t(t-1)n-1$ pairs of $\sigma_{03}^{i}$ and $\sigma_{04}^{i}$ items, we conclude that all the items of size $\sigma_{05}$ and $\sigma_{03}^{i}$, $\sigma_{04}^{i}$ are packed in these $t(t-1)n-1$ NE bins of the third type, as defined above.

We now should verify that the sum of sizes of the items packed in the three types of bins in the defined NE packing does not exceed 1. This holds for the second and the third type bins, as:

$$\frac{t}{t+1} + 2\delta_0 < \frac{t}{t+1} + \frac{1}{(4n)^{3s}} < \frac{t}{t+1} + \frac{1}{t+1} = 1.$$

For the bins of the first type, this property directly follows from the inequality proven in the next claim.

**Claim 52.** *The loads of the bins in the packing defined above are monotonically increasing as a function of the phase.*

**Proof.** It is sufficient to show $1 - \frac{1}{(t+1)\cdot 2^j} + \delta_j((t+1)\cdot 2^{j-1} + 1 + 2d_j^t) < 1 - \frac{1}{(t+1)\cdot 2^{j+1}}$ for $1 \le j \le s$, $t \ge 2$ which is equivalent to proving $\delta_j((t+1)\cdot 2^{j-1}+1+2d_j^t)2^{j+1} < \frac{1}{t+1}$. Using $d_j^t < n$, we have: $\delta_j((t+1)\cdot 2^{j-1}+1+2d_j^t)2^{j+1} < \delta_j((t+1)\cdot 2^{2j}+2^{j+2}n) < (t+1)\cdot 2\delta_j n^2$, as $n > 2^{s^3}$. Recall that $s \ge 3$. Using $\delta_j \le \delta_s = \frac{1}{2^{2s}n^s} \le \frac{1}{2n^2(t+1)}$ we get $2\delta_j n^2 < \frac{1}{t+1}$.

For $j = 0$, $\frac{t}{t+1} + 2\delta_0 < 1 - \frac{1}{(t+1)\cdot 2^j} + \delta_j((t+1)\cdot 2^{j-1} + 1 + 2d_j^t)$ holds for all $j \ge 1$, as $2\delta_0 \le \delta_j((t+1)\cdot 2^{j-1} + 1 + 2d_j^t)$, since $t \ge 2$ and $\delta_j$ is a strictly increasing sequence. □

**Claim 53.** *The packing defined above is a valid* NE *packing.*

**Proof.** To show that this is a NE packing, we need to show that an item of phase $j > 0$ cannot migrate to a bin of a level $k \ge j$, since this would result in a load larger than 1, and that it cannot migrate to a bin of phase $k < j$, since this would result in a load smaller than the load of a phase $j$ bin. Due to the monotonicity we proved in Claim 52, we only need to consider a possible migration of a phase $j$ item into a phase $j$ bin, and a phase $j-1$ bin, if such bins exist. Moreover, in the first case it is enough to consider the minimum size item and in the second case, the maximum size item of phase $j$.

For phase 0 items, since the smallest phase 0 item has size $\frac{1}{t+1} - \Delta nt(t-1)$, if it migrates to another bin of this phase, we get a total load of $\frac{t}{t+1} + \Delta(nt(t-1)+1) + \frac{1}{t+1} - \Delta nt(t-1) = 1 + \Delta > 1$, as $\Delta > 0$.

For items of phase $j \ge 1$: The smallest phase $j$ item has size $\frac{1}{(t+1)\cdot 2^j} - \delta_j(2(d_j^t - 1)) =$

$\frac{1}{(t+1)\cdot 2^j} - \delta_j(2d_j^t - 2)$. If it migrates to another bin of this phase, we get a total load of

$$\begin{aligned} 1 - \frac{1}{(t+1)\cdot 2^j} + \delta_j((t+1)\cdot 2^{j-1} + 1 + 2d_j^t) + \frac{1}{(t+1)\cdot 2^j} - \delta_j(2d_j^t - 2) \\ = 1 + \delta_j((t+1)\cdot 2^{j-1} + 1 + 2d_j^t) - 2d_j^t\delta_j + 2\delta_j \\ = 1 + \delta_j(3 + (t+1)\cdot 2^{j-1}) > 1. \end{aligned}$$

The check for the largest item in the phase should be done separately for cases $j = 1$ and $j \geq 2$, because we want to show that the largest item of phase $j = 1$ (in first type bin) cannot migrate into a phase 0 bin (a second or third type bin), while for the largest item of phase $j \geq 2$ we need to show that it cannot move into other bin of first type. For phase $j = 1$: The largest phase item has size $\frac{1}{2(t+1)} + 2(d_1^t + 1)\delta_1$. If it migrates to a bin of phase 0, we get a load of $\frac{t}{t+1} + 2\delta_0 + \frac{1}{2(t+1)} + 2(d_1^t + 1)\delta_1 = \frac{2t+1}{2(t+1)} + 2\delta_0 + 2(d_1^t + 1)\delta_1$. This load is strictly smaller than a load of level 1 which is $1 - \frac{1}{(t+1)\cdot 2} + \delta_1((t+1) + 1 + 2d_1^t) = \frac{2t+1}{2(t+1)} + \delta_1((t+1) + 1 + 2d_1^t)$, as $t \geq 2$ and $\delta_1 > \delta_0$.

For phase $j \geq 2$: The largest phase $j$ item has size $\frac{1}{(t+1)\cdot 2^j} + 2(d_j^t + 1)\delta_j$. If it migrates to a bin of phase $j - 1$, we get a load of

$$\begin{aligned} 1 - \frac{1}{(t+1)\cdot 2^{j-1}} + \delta_{j-1}((t+1)\cdot 2^{j-2} + 1 + 2d_{j-1}^t) + \frac{1}{(t+1)\cdot 2^j} + 2(d_j^t + 1)\delta_j \\ = 1 - \frac{1}{(t+1)\cdot 2^j} + \delta_{j-1}((t+1)\cdot 2^{j-2} + 1 + 2d_{j-1}^t) + 2(d_j^t + 1)\delta_j. \end{aligned}$$

We compare this load with $1 - \frac{1}{(t+1)\cdot 2^j} + \delta_j((t+1)\cdot 2^{j-1} + 1 + 2d_j^t)$, and prove that the first load is smaller. Indeed $\delta_{j-1}((t+1)\cdot 2^{j-2} + 1 + 2d_{j-1}^t) < \delta_j((t+1)\cdot 2^{j-1} - 1)$ since $\delta_j = 16n^2\delta_{j-1}$, $n > 2^{s^3}$ and $((t+1)\cdot 2^{j-2} + 1 + 2d_{j-1}^t) < 4n(t+1) < 16n^2((t+1)\cdot 2^{j-1} - 1)$.
$\square$

Finally, we bound the *PoA* as follows. The cost of the resulting NE packing is $nt + t(t-1)n - 1 + \sum_{j=1}^{s} r_j^t = t^2 n - 1 + \sum_{j=1}^{s} r_j^t$. Using Observation 50 we get that $\sum_{j=1}^{s} r_j^t \geq \sum_{j=1}^{s} (\frac{n}{(t+1)^j \cdot 2^{j(j-1)/2}} - 1)$ and since $OPT(I) = t(t-1)\cdot n + n$ and $n >> s$, we get a ratio of at least $\frac{t^2 + \sum_{j=1}^{s} (t+1)^{-j}\cdot 2^{-j(j-1)/2}}{t(t-1)+1}$. Letting $s$ tend to infinity as well results in the claimed lower bound.

$\square$

The corresponding lower bound values for different values of $\alpha$ are given in Table 4.1.

### 4.4.2 An upper bound

We now provide a close upper bound on *PoA*($\alpha$) for a positive integer $t \geq 2$. The technique used in Chapter 3 can be considered as a refinement of the one we use here. However, here we are required to use additional combinatorial properties of the NE packing. To bound the POA from above, we prove the following theorem.

**Theorem 54.** *For each integer $t \geq 2$, for any instance of the parametric bin packing game $G \in BP(\frac{1}{t})$: Any* NE *packing uses at most $\left(\frac{2t^3+t^2+2}{(2t+1)(t^2-t+1)}\right) \cdot OPT(G) + 5$ bins, where $OPT(G)$ is the number of bins used in a coordinated optimal packing.*

**Proof.** Let us consider a packing $b$ of the items in $N_G$ which admits NE conditions. We classify the bins according to their loads into four groups-$\mathcal{A}$, $\mathcal{B}$,$\mathcal{C}$ and $\mathcal{D}$. The cases $t = 2$ and $t \geq 3$ are treated separately. For $t = 2$: group $\mathcal{A}$- contains bins with loads of more than $\frac{5}{6}$; Group $\mathcal{B}$- contains bins with loads in $(\frac{3}{4}, \frac{5}{6}]$; Group $\mathcal{C}$- contains bins with loads in $(\frac{17}{24}, \frac{3}{4}]$; Group $\mathcal{D}$- contains bins with loads not greater than $\frac{17}{24}$. For $t \geq 3$: group $\mathcal{A}$- contains bins with loads of more than $\frac{2t+1}{2(t+1)}$; Group $\mathcal{B}$- contains bins with loads in $(\frac{t+1}{t+2}, \frac{2t+1}{2(t+1)}]$; Group $\mathcal{C}$- contains bins with loads in $(\frac{t^2-t+1}{t^2}, \frac{t+1}{t+2}]$; Group $\mathcal{D}$- contains bins with loads not greater than $\frac{t^2-t+1}{t^2}$. This partition is well defined, as $\frac{t}{t+1} < \frac{t^2-t+1}{t^2}$, $\frac{t^2-t+1}{t^2} < \frac{t+1}{t+2}$ and $\frac{t+1}{t+2} < \frac{2t+1}{2(t+1)}$ for any $t \geq 3$. We denote the cardinality of these groups by $n_{\mathcal{A}}, n_{\mathcal{B}}, n_{\mathcal{C}}$ and $n_{\mathcal{D}}$, respectively. Hence, $NE = n_{\mathcal{A}} + n_{\mathcal{B}} + n_{\mathcal{C}} + n_{\mathcal{D}}$. We list the bins in each group from left to right in non-increasing order w.r.t. their loads. Our purpose is to find an upper bound on the total number of bins in these four groups.

In the case $n_{\mathcal{D}} < 3$, using the fact that $OPT \geq \sum_{i=1}^{n} a_i$ we get:

- For $t = 2$, this means that all bins in packing $b$ (except for at most 2) have load of at least $\frac{17}{24}$, thus $OPT \geq \frac{17}{24}NE$, and $PoA \leq \frac{24}{17} < \frac{22}{15}$.

- For $t \geq 3$, this means that all bins in packing $b$ (except for at most 2) have load of at least $\frac{t^2-t+1}{t^2}$, thus $OPT \geq \frac{t^2-t+1}{t^2}NE$, and $PoA \leq \frac{t^2}{t^2-t+1} < \frac{2t^3+t^2+2}{(2t+1)(t^2-t+1)}$.

In the rest of the analysis we assume that $n_{\mathcal{D}} \geq 3$. We start with a simple lower bound on the load of the bins (except possibly at most two bins) in a NE.

**Claim 55.** *For a positive integer $t \geq 2$, all the bins in an* NE *packing $b$ (except for maybe a constant number of bins) are at least $\frac{t}{t+1}$ full.*

**Proof.** Consider the well-known First Fit algorithm (FF for short) for bin packing. FF packs each item in turn into the lowest indexed bin to where it fits. It opens a new bin only in the case where the item does not fit into any existing bin. It was shown in [74] that any bin (accept for maybe two) in the packing produced by FF is more than $\frac{t}{t+1}$ full for any $t \geq 2$. For each $N_G$ instance it is possible to define, by reordering the items, an instance for which running the FF algorithm will produce exactly the packing $b$. This can be achieved by going through all bins in $b$, bin by bin, bottom to top, and listing the items in this order. So, as any NE packing $b$ can be produced by a run of FF, it has all the properties of a FF packing, including the one mentioned above. $\square$

Moreover, the fact that any NE packing can be produced by a run of FF implies that the worst-case asymptotic ratio of FF, which is known to be $\frac{t+1}{t}$ for $t \geq 2$ [74], upper-bounds the POA. But, as we show further, the upper-bound we provide on the POA is tighter than this trivial bound for any $t \geq 2$.

From Claim 55 it is evident that all the bins (except for maybe two) in group $\mathcal{D}$ have loads in $(\frac{2}{3}, \frac{17}{24}]$ for $t = 2$, or in $(\frac{t}{t+1}, \frac{t^2-t+1}{t^2}]$ for $t \geq 3$.

**Claim 56.** *For a positive integer $t \geq 2$, in an NE packing $b$, all bins that are filled by less than $\frac{2t+1}{2(t+1)}$ (i.e. bins in groups $\mathcal{B}$, $\mathcal{C}$ and $\mathcal{D}$), except for maybe a constant number of bins, contain exactly $t$ items with sizes in $(\frac{t-1}{t^2}, \frac{1}{t}]$.*

**Proof.** First, consider the bins in group $\mathcal{D}$. For $t \geq 3$, as all bins in $\mathcal{D}$ are filled by no more than $\frac{t^2-t+1}{t^2}$, no bin in this group (except maybe the leftmost bin) contains an item of size in $(0, \frac{t-1}{t^2}]$, as such an item will reduce its cost by moving to the leftmost bin in $\mathcal{D}$ (which is the bin with the largest load in $\mathcal{D}$), contradicting the fact that $b$ is an NE. Hence, all the items in bins (except for maybe one) in group $\mathcal{D}$ have items of sizes in $(\frac{t-1}{t^2}, \frac{1}{t}]$. For $t = 2$, as all bins in $\mathcal{D}$ are filled by no more than $\frac{17}{24}$, no bin in this group (except maybe the leftmost bin) contains an item of size in $(0, \frac{7}{24}]$, as such an item will reduce its cost by moving to the leftmost bin in $\mathcal{D}$, which contradicts the fact that $b$ is an NE. Hence, all the items in bins (except for maybe one) in group $\mathcal{D}$ have items of sizes in $(\frac{7}{24}, \frac{1}{2}]$.

Now, consider the bins in group $\mathcal{C}$. For $t \geq 3$, as all bins in $\mathcal{C}$ are filled by no more than $\frac{t+1}{t+2}$, no bin in this group (except maybe the leftmost bin) contains an item of size in $(0, \frac{1}{t+2}]$, as such an item will reduce its cost by moving to the leftmost bin in $\mathcal{C}$ (which is the bin with the largest load in $\mathcal{C}$), contradicting the fact that $b$ is an NE. Also, no bin in $\mathcal{C}$ contains an item of size $x \in (\frac{1}{t+2}, \frac{t-1}{t^2}]$, as such an item will benefit from moving to a bin in group $\mathcal{D}$ that is loaded by at most $\frac{t^2-t+1}{t^2}$, as it fits there and $x + \frac{t}{t+1} > \frac{t+1}{t+2}$ for any $x > \frac{1}{(t+2)}$. Hence, all the items in bins in group $\mathcal{C}$ have sizes in $(\frac{t-1}{t^2}, \frac{1}{t}]$. For $t = 2$, as all bins in $\mathcal{C}$ are filled by no more than $\frac{3}{4}$, no bin in this group (except maybe the leftmost bin) contains an item of size in $(0, \frac{1}{4}]$, as such an item will reduce its cost by moving to the leftmost bin in $\mathcal{C}$, which contradicts the fact that $b$ is an NE. Also, no bin in $\mathcal{C}$ contains an item of size $x \in (\frac{1}{4}, \frac{7}{24}]$, as such an item will benefit from moving to a bin in group $\mathcal{D}$, as $x + \frac{2}{3} > \frac{3}{4}$ for any $x > \frac{1}{4}$. Hence, all the items in bins (except for maybe one) in group $\mathcal{C}$ have sizes in $(\frac{7}{24}, \frac{1}{2}]$.

Finally, consider the bins in group $\mathcal{B}$. For $t \geq 3$, as all bins in $\mathcal{B}$ are filled by no more than $\frac{2t+1}{2(t+1)}$, no bin in this group (except maybe the leftmost bin) contains an item of size in $(0, \frac{1}{2(t+1)}]$, as such an item will reduce its cost by moving to the leftmost bin in $\mathcal{B}$ (which is the bin with the largest load in $\mathcal{B}$), contradicting the fact that $b$ is an NE. Also, no bin in $\mathcal{B}$ contains an item of size $x \in (\frac{1}{2(t+1)}, \frac{t-1}{t^2}]$, as such an item will benefit from moving to a bin in group $\mathcal{D}$ that is loaded by at most $\frac{t^2-t+1}{t^2}$, as it fits there and $x + \frac{t}{t+1} > \frac{2t+1}{2(t+1)}$ for any $x > \frac{1}{2(t+1)}$. Hence, all the items in bins (except for maybe one) in group $\mathcal{B}$ have sizes in $(\frac{t-1}{t^2}, \frac{1}{t}]$. For $t = 2$, as all bins in $\mathcal{B}$ are filled by no more than $\frac{5}{6}$, no bin in this group (except maybe the leftmost bin) contains an item of size in $(0, \frac{1}{6}]$, as such an item will reduce its cost by moving to the leftmost bin in $\mathcal{B}$, which contradicts the fact that $b$ is an NE. Also, no bin in $\mathcal{B}$ (except maybe the leftmost bin) contains an item of size $x \in (\frac{1}{6}, \frac{7}{24}]$, as such an item will benefit from moving to a bin in group $\mathcal{D}$, as $x + \frac{2}{3} > \frac{5}{6}$ for any $x > \frac{1}{6}$. Hence, all the items in bins (except for maybe one) in group $\mathcal{B}$ have sizes in $(\frac{7}{24}, \frac{1}{2}]$.

We conclude, that any bin in groups $\mathcal{B}$, $\mathcal{C}$ and $\mathcal{D}$, except for maybe a constant number of bins, contain only items of sizes in $(\frac{t-1}{t^2}, \frac{1}{t}]$ for $t \geq 3$, and items of sizes in $(\frac{7}{24}, \frac{1}{2}]$ for $t = 2$.

Now, we show that each one of these bins contains exactly $t$ such items. Note, that by definition of the groups all bins in $\mathcal{B}$, $\mathcal{C}$ and $\mathcal{D}$ (except maybe two) have loads in $(\frac{t}{t+1}, \frac{2t+1}{2(t+1)}]$ for $t \geq 3$, or in $(\frac{2}{3}, \frac{5}{6}]$ for $t = 2$.
If a bin contains at most $t-1$ such items, then it has a load of at most $(t-1) \cdot \frac{1}{t} = \frac{t-1}{t}$ for $t \geq 3$ of at most $\frac{7}{24}$ for $t = 2$, which is less than the assumed load in these bins, so they must have more than $(t-1)$ such items. If a bin contains at least $t+1$ such items, then it has a load of at least $(t+1) \cdot \frac{t-1}{t^2} = 1 - \frac{1}{t^2}$, which is greater than $\frac{2t+1}{2(t+1)}$ for $t \geq 3$, or at least $\frac{7}{8}$ which is greater than $\frac{5}{6}$ for $t = 2$, so they must have less than $(t+1)$ such items.

We conclude that each bin in groups $\mathcal{B}$, $\mathcal{C}$ and $\mathcal{D}$, except for maybe 5 special bins (the leftmost bins in groups $\mathcal{B}$, $\mathcal{C}$ and $\mathcal{D}$ and the two rightmost bins in $\mathcal{D}$) contain exactly $t$ items with sizes in $(\frac{t-1}{t^2}, \frac{1}{t}]$ for $t \geq 3$, or exactly 2 items of sizes in $(\frac{7}{24}, \frac{1}{2}]$ for $t = 2$. $\square$

Note that the expressions that we use to define our groups are intentionally hand-tailored in a way that the bins exhibit the combinatorial properties that we later use to get the upper bound; The bound $\frac{2t+1}{2(t+1)}$ is the average between $\frac{t}{t+1}$ and 1, so every item from a bin that is loaded by at most $\frac{2t+1}{2(t+1)}$ that has size greater than $1 - \frac{2t+1}{2(t+1)} = \frac{1}{2(t+1)}$ would benefit from moving to a bin of group $\mathcal{D}$, if it fits there. As $\frac{t-1}{t^2} > \frac{1}{t+2} > \frac{1}{2(t+1)}$, this allows us to show that bins in groups $\mathcal{B}$, $\mathcal{C}$ and $\mathcal{D}$ do not have items of sizes less or equal to $\frac{t-1}{t^2}$, and hence, have to at most $t$ such items. On the other hand $\frac{t^2-t+1}{t^2}$, $\frac{t+1}{t+2}$ and $\frac{2t+1}{2(t+1)}$ are large enough such that there are at least $t$ items with sizes of at most $t$, which is large enough to yield the asserted upper bound. The bound $\frac{t^2-t+1}{t^2}$ was chosen s.t. it is also large enough for packing $b$ to yield an upper bound which is smaller than our claimed upper bound for the case where $n_{\mathcal{D}} < 3$.

We considered the case $t = 2$ separately, as $\frac{t^2-t+1}{t^2} < \frac{t+1}{t+2}$ holds for $t > 2$.

Henceforth, we call the bins in groups $\mathcal{B}$, $\mathcal{C}$ and $\mathcal{D}$ that contain exactly $t$ items with sizes in $(\frac{t-1}{t^2}, \frac{1}{t}]$ for $t \geq 3$, or exactly 2 items of sizes in $(\frac{7}{24}, \frac{1}{2}]$ for $t = 2$ *regular* bins, and refer to each one of those items as $t$-item.

To derive the upper bound on the total number of bins in the NE packing $b$, we use the *weighting functions* technique. We define for each value of $t \geq 2$ a weighting function $w_t$ on the items, in the following manner. The weight $w_t(x)$ of an item of size $x$ which is packed in a bin of group $\mathcal{A}$ in a packing $b$ is:

$$w_t(x) = \frac{2(t+1)}{2t+1}x.$$

The weight $w_t(x)$ of an item of size $x$ which is packed in a regular bin of load $L < \frac{2t+1}{2(t+1)}$ in a packing $b$ is:

$$w_t(x) = \frac{2(t+1)}{2t+1}x + \frac{(1 - \frac{2(t+1)}{2t+1}L)}{k},$$

where $k$ is the number of items in the bin of $x$. The purpose of the additive term $\frac{(1-\frac{2(t+1)}{2t+1}L)}{k}$ is to complete the weight of any bin in the packing to 1. Clearly, any bin in group $\mathcal{A}$ (which is full by more than $\frac{2t+1}{2(t+1)}$) will have a total weight of at least 1. Any of the less filled bins from groups $\mathcal{B}$, $\mathcal{C}$ and $\mathcal{D}$ will have a weight of 1 as $\frac{2(t+1)}{2t+1} \cdot L + \frac{(1-\frac{2(t+1)}{2t+1}L)}{t} \cdot t = 1$, and each of the $t$ items packed in each one of these bins (except maybe 5 bins) will get an addition of at most $\frac{1-\frac{2(t+1)}{2t+1}\cdot\frac{t}{t+1}}{t} = \frac{1}{t(2t+1)}$.

As we study asymptotic measures, we will disregard these 5 special bins in the rest of the analysis, and consider all the bins with load in $(\frac{t}{t+1}, \frac{2t+1}{2(t+1)}]$ to have exactly $t$ items with sizes in $(\frac{t-1}{t^2}, \frac{1}{t}]$, and the weight of these 5 bins will be taken into account in the additive constant.

Now, we bound from above the weight observed by a bin in the optimal packing of these items. First, note that in a bin of the optimal packing for $t \geq 2$ there can be at most $t+1$ $t$-items from the regular bins of groups $\mathcal{B}$, $\mathcal{C}$ and $\mathcal{D}$. For $t = 2$ the size of these items is greater than $\frac{7}{24}$, and the size of four of these items exceeds 1. For $t \geq 3$ the size of these items is greater than $\frac{t-1}{t^2}$, and the size of $t+2$ of these items, which is at least $(t+2) \cdot \frac{t-1}{t^2} = 1 + \frac{t-2}{t^2}$, exceeds 1.

The weight of a bin in an optimal packing that has a load $S$ and contains $t+1$ $t$-items that come from bins of groups $\mathcal{B}$, $\mathcal{C}$ and $\mathcal{D}$ in $b$, is at most

$$\frac{2(t+1)}{2t+1} \cdot S + (t+1) \cdot \frac{1}{t(2t+1)} \leq \frac{2(t+1)}{2t+1} + \frac{t+1}{t(2t+1)} = \frac{2t^2+3t+1}{t(2t+1)} = \frac{t+1}{t}.$$

The weight of a bin in an optimal packing that has a load $S$ and contains at most $t$ $t$-items that came from bins of groups $\mathcal{B}$, $\mathcal{C}$ and $\mathcal{D}$ in $b$, is at most

$$\frac{2(t+1)}{2t+1} \cdot S + t \cdot \frac{1}{t(2t+1)} \leq \frac{2(t+1)}{2t+1} + \frac{t}{t(2t+1)} = \frac{2t^2+3t}{t(2t+1)}.$$

We claim that in any optimal packing, the fraction of the number of bins that contain $t+1$ $t$-items from bins of groups $\mathcal{B}$, $\mathcal{C}$ and $\mathcal{D}$ out of total number of bins is at most $\frac{t(t-1)}{t^2-t+1}$. To establish this, we consider all the bins in the optimal packing that contain exactly $t+1$ $t$-items from groups $\mathcal{B}$, $\mathcal{C}$ and $\mathcal{D}$ (and maybe additional items as well), let the number of such bins be $N_t$.

If $N_t = 0$, we are done as then the total weight of all the items in $N_G$ is at most $W(N_G) \leq \left(\frac{2t+3}{2t+1}\right) \cdot OPT(G)$. As $n_{\mathcal{A}} + n_{\mathcal{B}} + n_{\mathcal{C}} + n_{\mathcal{D}} - 5 \leq W(N_G)$, we get that $NE \leq \left(\frac{2t+3}{2t+1}\right) \cdot OPT(G) + 5 < \left(\frac{2t^3+t^2+2}{(2t+1)(t^2-t+1)}\right) \cdot OPT(G) + 5$. Else, we prove the following claim.

**Claim 57.** *Among the $N_t \cdot (t+1)$ $t$-items that are packed in $(t+1)$-tuples in the bins of the optimal packing, only at most $(N_t - 1) \cdot t$ are packed together in $t$-tuples, in bins that belong to groups $\mathcal{B}$, $\mathcal{C}$ and $\mathcal{D}$ in the* NE *packing.*

**Proof.** Assume by contradiction that $(N_t + k) \cdot t$ of these items for $k \geq 0$ are packed together in $t$-tuples in bins of groups $\mathcal{B}$, $\mathcal{C}$ and $\mathcal{D}$ in the NE packing. Consider the first $N_t$ such bins in the NE packing. Call them $B_1, B_2, \ldots, B_{N_t}$. In a slight abuse of notation, we use $B_i$ to indicate both the $i$-th bin and its load. Denote the sizes of the remaining $N_t$ $t$-items by $t_1, t_2, \ldots, t_{N_t}$. These items are also packed in bins of groups $\mathcal{B}$, $\mathcal{C}$ and $\mathcal{D}$ in $b$, and share their bin with $t-1$ $t$-items (when at least one of these items is not packed in any of the aforementioned $N_t$ bins in the optimal packing). Obviously, as all these $N_t \cdot (t+1)$ $t$-items fit into $N_t$ unit-capacity bins, $t_1 + \ldots + t_{N_t} + B_1 + \ldots + B_{N_t} \leq N_t$ holds. To derive a contradiction, we use the following observation:

**Observation 58.** *A $t$-tuple of items with sizes in $(\frac{t-1}{t^2}, \frac{1}{t}]$ always has a greater total size than any $(t-1)$-tuple of such items.*

**Proof.** The total size of any $(t-1)$ items with sizes in $(\frac{t-1}{t^2}, \frac{1}{t}]$ is at most $\frac{t-1}{t}$, while the total size of any $t$ items with sizes in $(\frac{t-1}{t^2}, \frac{1}{t}]$ is strictly greater than $\frac{t(t-1)}{t^2} = \frac{t-1}{t}$. □

Thus, any item $t_i$, $1 \leq i \leq N_t$ would be better off sharing a bin with other $t$ items of size in $(\frac{t-1}{t^2}, \frac{1}{t}]$ instead of just $t-1$ such items as it does in the NE packing $b$. For an item which shares a bin with $t-1$ $t$-items we conclude that the only reason it does not move to another bin with $t$ such items in $b$ is that it does not fit there.

So, we know that no item $t_1$, $1 \leq i \leq N_t$ fits in any of the bins $B_1, B_2, \ldots, B_{N_t}$ in $b$. We get that for any $1 \leq i \leq N_t$, for any $1 \leq j \leq N_t$, the inequality $t_i + B_j > 1$ holds. Summing these inequalities over all $1 \leq i \leq N_t$ and $1 \leq j \leq N_t$ we get $t_1 + \ldots + t_{N_t} + B_1 + \ldots + B_{N_t} > N_t$, which is a contradiction. □

Hence, at most $(N_t - 1) \cdot t$ $t$-items out of $N_t \cdot (t+1)$ are packed together in $t$-tuples in bins from groups $\mathcal{B}$, $\mathcal{C}$ and $\mathcal{D}$ in the NE packing $b$. The remaining $t$-items (at least $N_t + t$ in number) are also packed in bins of groups $\mathcal{B}$, $\mathcal{C}$ and $\mathcal{D}$ in $b$, but they share their bin with at most $(t-2)$ other $t$-items from the $N_t$ bins from the optimal packing, and at least one $t$-item that is not packed in one of these $N_t$ bins. In total, there are at least $\frac{N_t+t}{t-1}$ $t$-items that are not packed in one of the $N_t$ bins in discussion, and they are packed with at most $t-1$ other such items in the optimal packing.

Thus, in the optimal packing for any $N_t$ bins with $t+1$ items of size in $(\frac{t-1}{t^2}, \frac{1}{t}]$ there are at least $\frac{N_t+t}{t(t-1)}$ bins that have at most $t$ such items. Letting $N_t$ be very large in comparison to $t$ gives us the claimed proportions. We conclude that in average, the weight of any bin of the optimal packing is at most

$$\frac{t(t-1) \cdot \frac{t+1}{t} + \frac{2t+3}{2t+1}}{t(t-1)+1} = \frac{2t^3+t^2+2}{(2t+1)(t^2-t+1)}.$$

Hence, the total weight of all the items in $N_G$ is at most $W(N_G) \leq \left(\frac{2t^3+t^2+2}{(2t+1)(t^2-t+1)}\right) \cdot OPT(G)$. As $n_{\mathcal{A}} + n_{\mathcal{B}} + n_{\mathcal{C}} + n_{\mathcal{D}} - 5 \leq W(N_G)$, we get that $NE \leq \left(\frac{2t^3+t^2+2}{(2t+1)(t^2-t+1)}\right) \cdot OPT(G) + 5$ □

A more careful consideration of the contents of special bins allows to reduce the additive constant to 2.

**Theorem 59.** *For each integer $t \geq 2$ and $\alpha \in (\frac{1}{t+1}, \frac{1}{t}]$, the PoA of the parametric Bin Packing game $BP(\alpha)$ is at most $\frac{2t^3+t^2+2}{(2t+1)(t^2-t+1)}$.*

**Proof.** The asserted upper bound on the POA follows directly from Theorem 54. □

We conjecture that the true value of the $PoA(\alpha)$ equals our lower bound from Theorem 49, for each $\alpha \leq \frac{1}{2}$.

**Comment 60.** *Note that in our construction of the lower bound for $t \geq 2$ there are also two types of bins in the optimal packing: bins that contain exactly $t+1$ items of sizes in $(\frac{t-1}{t^2}, \frac{1}{t}]$, and bins that contain at most $t$ such items. These items are packed in $t$-tuples in the* NE *packing of the instance. The proportion between the numbers of these two types of bins in the optimal solution is $t(t-1)$ to 1, similarly to our conclusion in the analysis of the upper bound.*

The corresponding upper bound values for different values of $\alpha$ are given in Table 4.1.

We illustrate our almost matching lower and upper bound values for $PoA(\alpha)$ in Figure 4.1 below.

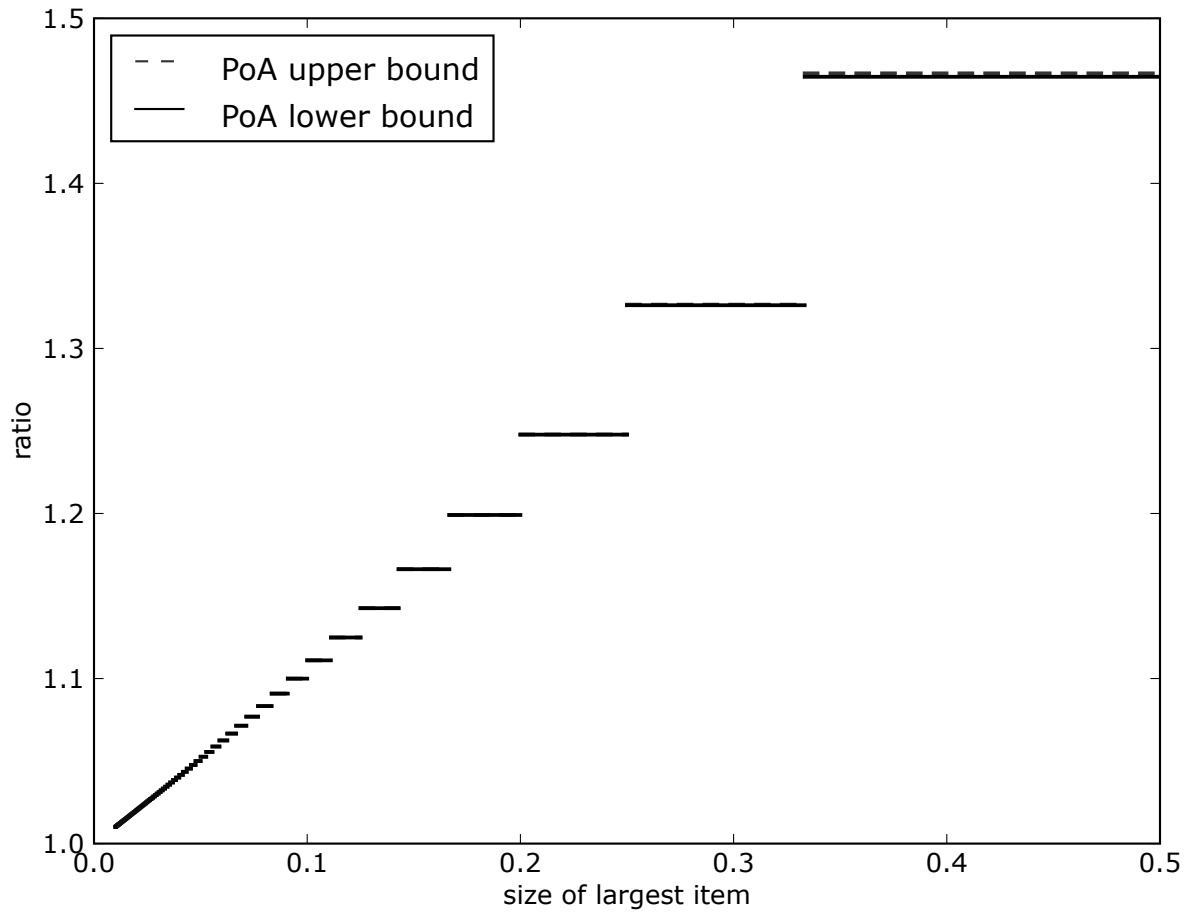

Figure 4.1: Almost matching upper and lower bounds for the $PoA(\alpha)$ of the Parametric Bin Packing game.

## 4.5 Strong Prices of Anarchy and Stability in the Parametric Bin Packing game

In this section we show tight bounds for $SPoA(\alpha)$ and $SPoS(\alpha)$ for each $\alpha \leq \frac{1}{2}$.

It was proved in Chapter 3 that the strong equilibria in the Bin Packing game coincide with the packings produced by the Subset Sum algorithm for Bin Packing. The equivalence of the SPoA, SPoS and the worst-case performance ratio of the Subset Sum algorithm which was also proved in Chapter 3 still applies for the Parametric Bin Packing game; indeed, it holds for all possible lists of items (players), and in particular to lists where all items have size at most $\alpha$. This allows us to characterize the $SPoA(\alpha)/SPoS(\alpha)$ in terms of $R_A^\infty(\alpha)$, which was completely settled in [46]. We formulate this result in the following theorem.

**Theorem 61.** *For each integer $t \geq 2$ and $\alpha \in (\frac{1}{t+1}, \frac{1}{t}]$, the Subset Sum algorithm has an approximation ratio of $R_{SS}^\infty(\alpha) = 1+\sum_{i=1}^\infty \frac{1}{(t+1)2^i-1}$. Furthermore, the $SPoA(\alpha)/SPoS(\alpha)$ of the $BP(\alpha)$ game has the same value.*

## 4.6 Summary and conclusions

In this chapter we have studied the Parametric Bin Packing problem, where the items that have sizes in $(0, \alpha]$, for $\alpha \leq \frac{1}{2}$, are controlled by selfish agents, and the cost charged from each bin is shared among all the items packed into it, both in non-cooperative and cooperative versions. We have provided almost tight upper and lower bounds on the Price of Anarchy of the induced game, and tight bounds on the Strong Prices of Anarchy and Stability for each $\alpha \leq \frac{1}{2}$, the former done by using the equivalence of the strong stability measures in this game to the worst-case performance ratio of the Subset Sum algorithm.

In order to illustrate the results, we report in Table 4.1 the values for the worst-case ratio of the Subset Sum algorithm and the range of possible values for the $PoA$ for various values of $\alpha$. We also include the worst-case approximation ratios of First Fit (FF) and First Fit Decreasing (FFD) algorithms for bin packing for these values of $\alpha$.

| | $R_{FFD}(\alpha)$ [74] | $R_{SS}(\alpha)$[46] | $PoA(\alpha)$ | $R_{FF}(\alpha)$ [74] |
|---|---|---|---|---|
| $t = 1$ | 1.222222 | 1.606695 | [1.641632, 1.642857] [43] | 1.700000 |
| $t = 2$ | 1.183333 | 1.376643 | [1.464571, 1.466667] | 1.500000 |
| $t = 3$ | 1.166667 | 1.273361 | [1.326180, 1.326530] | 1.333333 |
| $t = 4$ | 1.150000 | 1.214594 | [1.247771, 1.247863] | 1.250000 |
| $t = 5$ | 1.138095 | 1.176643 | [1.199102, 1.199134] | 1.200000 |
| $t = 6$ | 1.119048 | 1.150106 | [1.166239, 1.166253] | 1.166667 |
| $t = 7$ | 1.109127 | 1.130504 | [1.142629, 1.142635] | 1.142857 |
| $t = 8$ | 1.097222 | 1.115433 | [1.124867, 1.124871] | 1.125000 |
| $t = 9$ | 1.089899 | 1.103483 | [1.111029, 1.111031] | 1.111111 |
| $t = 10$ | 1.081818 | 1.093776 | [1.099946, 1.099947] | 1.100000 |

Table 4.1: Comparison of the worst-case ratio of *FFD*, *SS*, *FF* and *PoA* as a function of $\alpha$ when $\alpha \leq \frac{1}{t}$, for $t = 1, \ldots, 10$.

# Chapter 5

# Machine Covering with selfish jobs on identical machines

## 5.1 Introduction and motivation

In this chapter we consider a scheduling problem on identical machines where the goal is maximization of the minimum load.

This goal function is motivated by issues of Quality of Service and fair resource allocation. It is useful for describing systems where the complete system relies on keeping all the machines productive for as long as possible, as the entire system fails in case even one of the machines ceases to be active. From the networking aspect, this problem has applications to basic problems in network optimization such as fair bandwidth allocation. Consider pairs of terminal nodes that wish to communicate; we would like to allocate bandwidth to the connections in a way that no link unnecessarily suffers from starvation, and all links get a fair amount of resources. Another motivation is efficient routing of traffic. Consider a network that consists of parallel links between pairs of terminal nodes. Requests for shifting flow are assigned to the links. We are interested in having the loads of the links balanced, in the sense that each link should be assigned a reasonable amount of flow, compared to the other links.

Yet another incentive to consider this goal function is congestion control by fair queuing. Consider a router that can serve $m$ shifting requests at a time. The data pieces of various sizes, need to be shifted, are arranged in $m$ queues (each queue may have a different data rate), each pays a price which equals the delay that it causes in the waiting line. Our goal function ensures that no piece gets a “preferred treatment” and that they all get at least some amount of delay.

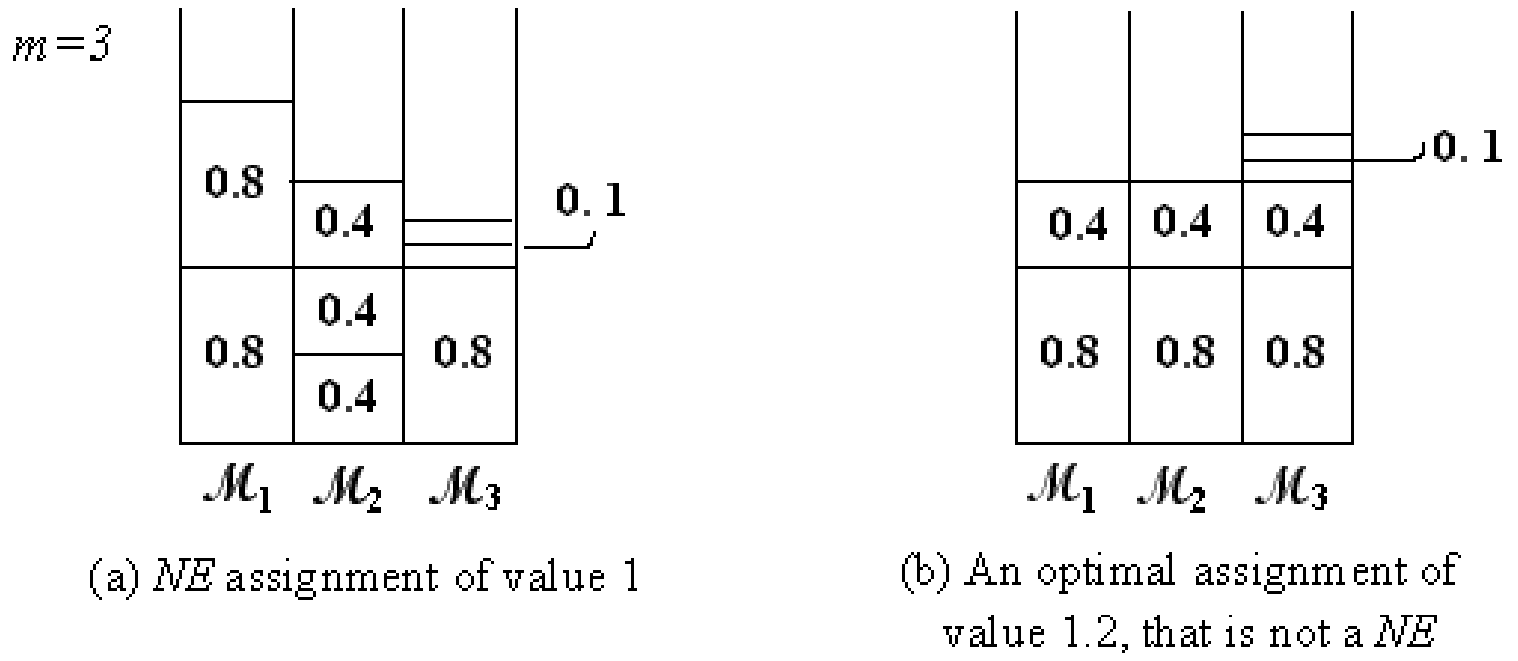

Figure 5.1: An example of two schedules with different social values. This example demonstrates the non-triviality of the problem. There are three jobs of size 0.8, three jobs of size 0.4 and two jobs of size 0.1. The three machines are identical. The assignment on the right hand side is not a Nash equilibrium, since a job of size 0.1 would reduce its delay from 1.4 to 1.3 by migrating to another machine. The social value of this assignment is 1.2. The assignment on the left hand side is a Nash equilibrium, but its social value is only 1.

The problem of maximizing the minimum load, seeing jobs as selfish agents, can be modeled as a routing problem. In this setting, machines are associated with parallel links between a source and a destination. The links have bounded capacities, and a set of users request to send a certain amount of unsplitable flow between the two nodes. Requests are to be assigned to links and consume bandwidth which depends on their sizes. The cost charged from a user for using a link equals to the total amount of the utilized bandwidth of that link. Thus, the selfish users prefer to route their traffic on a link with small load. This scenario is similar to the model proposed by Koutsoupias and Papadimitriou in [79], but our model has a different social goal function. To demonstrate the non-triviality of the problem, see Figure 5.1.

The novelty of our study compared to other work in the area is that the social goal is very different from the private goals of the players. As we study a maximization problem, the Price of Anarchy (POA) in our scheduling model is the worst case ratio between the social value (i.e., minimum delay of any machine, or cover) of an optimal schedule, denoted by OPT, and the value of any Nash equilibrium. If both these values are $0$ then we define the POA to be $1$. The Price of Stability (POS) is the worst case ratio between the social value of an optimal solution, and the value of the *best* Nash equilibrium. Similarly, if both these values are $0$ then we define the POS to be $1$.

In addition, we study the *mixed* POA (MPOA), where we consider mixed Nash equilibria that result from mixed strategies, where the player's choices are not deterministic and are regulated by probability distributions on a set of pure strategies (in our case, over the set $M$ of the machines). A mixed Nash equilibrium is characterized by the property that there is no incentive for any job to deviate from its probability distribution (a deviation is any modification of its probability vector over machines), while probability distributions of other players remain unchanged. The existence of such an equilibrium over mixed strategies for non-cooperative games was shown by Nash in his famous work [90]. The values MPOA and MPOS are defined similarly to the pure ones, but mixed Nash equilibria are being considered instead of pure ones. Clearly, any pure NE is also a mixed NE.

## 5.2 Related work and our results

The non-selfish version of the problem has been well studied (known by different names such as "machine covering" and "Santa Claus problem") in the computer science literature (see e.g. [37, 15, 40]). Various game-theoretic aspects of max-min fairness in resource allocation games were considered before this paper (e.g. in [17, 63, 8, 22, 51, 9, 97]), but unlike the makespan minimization problem POA and POS of which were extensively studied (see e.g. [79, 36, 84]), these measures were not previously considered for the uncoordinated machine covering problem in the setting of selfish jobs. A different model, where machines are selfish rather than jobs with the same social goal was studied recently in [48, 38, 26].

For identical machines, we show that the POS is equal to $1$. As our main result, we study the pure POA and show close bounds on the overall value of the POA (POA $=$ $\sup_m$ POA$(m)$, where POA$(m)$ is the POA on $m$ machines), i.e., that it is at least 1.691 and at most 1.7. This in contrast with the makespan minimization problem, where it is known that the POA for $m$ identical machines is $\frac{2m}{m+1}$, giving an overall bound of 2 [56, 99]. This is rather unusual, as the cover maximization problem is typically harder than the makespan minimization problem, thus it could be expected that the POA for the covering problem would be higher.

For the analysis of our upper bound we use the weighting function technique, which is uncommon in scheduling problems. Moreover, we use not only the weight function but also its inverse function in our analysis. Surprisingly, these lower and upper bounds are approximation ratios of well known algorithms for Bin-Packing (Harmonic [80] and First-Fit [73], respectively). We furthermore prove that the POA is monotonically non-decreasing as a function of $m$. For small numbers of machines we provide the exact values of POA: we find that POA$(2)$ $=$ POA$(3) = 3/2$ and POA$(4) = 13/8 = 1.625$. We show that POA$(m) \geq \frac{5}{3}$ for $m > 5$. As for the MPOA, we show that its value is very large as a function of $m$, and MPOA$(2) = 2$.

In this chapter we focus on identical machines. For uniformly related machines, we show in Chapter 6 that even the POS is unbounded already for two machines with a speed ratio *larger than* 2, and the POA is unbounded for a speed ratio of *at least* 2. The same property holds for $m$ machines (where the speed ratio is defined to be the maximum speed ratio between any pair of machines). This is very different from the results in the situation of the makespan minimization social goal, where the POA is finite [36, 54]. In Chapter 6 we also study the POA and POS for the complementary cases.

In Section 5.7 we present a different model where the social goal is to minimize the ratio between the load of the most loaded machine and the least loaded machine, and we show tight bounds on the POS and POA for any number of identical machines. Similarly to the results of Chapter 6, one can argue that for any number of uniformly related machines these measures are unbounded.

## 5.3 The model

In this section, we define the more general model of scheduling on related machines. A set of $n$ jobs $J = \{1, 2, \ldots, n\}$ is to be assigned to a set of $m$ machines $M = \{M_1, \ldots, M_m\}$, where machine $M_i$ has a speed $s_i$. If $s_i = 1$ for $i = 1, \ldots, m$, the machines are called identical, and this is the case on which we focus here. This is an important and widely studied special case of uniformly related machines. The size of job $1 \leq k \leq n$ is denoted by $p_k$. An assignment or schedule is a function $\mathcal{A} : J \to M$. The load of machine $M_i$, which is also called the delay of this machine, is $L_i = \sum_{k:\mathcal{A}(k)=M_i} \frac{p_k}{s_i}$. The value, or the *social value* of a schedule is the minimum delay of any machine in this schedule, also known as the *cover*. We denote it by $\text{COVER}(\mathcal{A})$. This problem is a dual to the makespan scheduling problem.

The non-cooperative machine covering game $MC$ is characterized by a tuple $MC = \langle N, (\mathcal{M}_k)_{k\in N}, (c_k)_{k\in N}\rangle$, where $N$ is the set of atomic players. Each player $k \in N$ controls a single job of size $p_k > 0$ and selects the machine to which it will be assigned. We associate each player with the job it wishes to run, that is, $N = J$. The set of strategies $\mathcal{M}_k$ for each job $k \in N$ is the set $M$ of all machines. i.e. $\mathcal{M}_k = M$. Each job must be assigned to one machine only. Preemption is not allowed. The outcome of the game is an assignment $\mathcal{A} = (\mathcal{A}_k)_{k\in N} \in \times_{k\in N} M_k$ of jobs to the machines, where $\mathcal{A}_k$ for each $1 \leq k \leq n$ is the index of the machine that job $k$ chooses to run on. Let $\mathcal{S}$ denote the set of all possible assignments. The cost function of job $k \in N$ is denoted by $c_k : \mathcal{S} \to \mathbb{R}$. The cost $c_k^i$ charged from job $k$ for running on machine $M_i$ in a given assignment $\mathcal{A}$ is defined to be the load observed by machine $i$ in this assignment, that is $c_k(i, \mathcal{A}_{-k}) = L_i(\mathcal{A})$, when $\mathcal{A}_{-k} \in \mathcal{S}_{-k}$; here $\mathcal{S}_{-k} = \times_{j\in N\setminus\{k\}} \mathcal{S}_j$ denotes the actions of all players except for player $k$.

The goal of the selfish jobs is to run on a machine with a load which is as small as possible. At an assignment that is a (pure) Nash equilibrium or NE assignment for short, there exists no machine $M_{i'}$ for which $L_{i'}(\mathcal{A}) + \frac{p_k}{s_{i'}} < L_i(\mathcal{A})$ for some job $k$ which is assigned to machine $M_i$ (see Figure 5.1(a) for an example). For this selfish goal of players, a pure Nash equilibrium (with deterministic agent choices) always exists [57, 49]. We can also consider mixed strategies, where players use probability distributions. Let $t_k^i$ denote the probability that job $k \in N$ chooses to run on machine $M_i$. A strategy profile is a vector $p = (t_k^i)_{k \in N, i \in M}$ that specifies the probabilities for all jobs and all machines. Every strategy profile $p$ induces a random schedule. The *expected load* $\mathbb{E}(L_i)$ of machine $M_i$ in setting of mixed strategies is $\mathbb{E}(L_i) = \frac{1}{s_i} \sum_{k \in N} p_k t_k^i$. The *expected cost* of job $k$ if assigned on machine $M_i$ (or its *expected delay* when it is allocated to machine $M_i$) is $\mathbb{E}(c_k^i) = \frac{p_k}{s_i} + \sum_{j \neq k} p_j t_j^i / s_i = \mathbb{E}(L_i) + (1 - t_k^i)\frac{p_k}{s_i}$. The probabilities $(t_k^i)_{k \in N, i \in M}$ give rise to a (*mixed*) Nash equilibrium if and only if any job $k$ will assign non-zero probabilities only to machines $M_i$ that minimize $c_k^i$, that is, $t_k^i > 0$ implies $c_k^i \leq c_k^j$ for any $j \in M$. The social value of a strategy profile $p$ is the *expected minimum load* over all machines, i.e. $\mathbb{E}(\min_{i \in M} L_i)$.

## 5.4 The Price of Anarchy for identical machines

Figure 5.1 clearly demonstrates that not every NE schedule is optimal. We next measure the extent of deterioration in the quality of NE schedules due to the effect of selfish and uncoordinated behavior of the players (jobs), in the worst case. As mentioned before, the measure metrics that we use are the POA and the POS.

Before we present the proof of our bounds on the POA for general $m$, to get the notion of the problem, we prove tight bounds on the the POA for simple cases where $m = 2$, $m = 3$ and $m = 4$. Finally we discuss the MPOA in Section 5.6.

### 5.4.1 Small numbers of machines

Consider a pure NE assignment of jobs to machines, denoted by $\mathcal{A}$, for an instance of the machine covering game. We assume that the social value of $\mathcal{A}$, that is, the load of the least loaded machine in $\mathcal{A}$, is 1. Otherwise, we can simply scale all sizes of jobs in the instances which we consider so that $\text{COVER}(\mathcal{A}) = 1$.

We denote a machine which is loaded by $1$ in $\mathcal{A}$ by $P$. All other machines are called *tall* machines. We would like to estimate the load of $P$ in the optimal assignment. Let $C = \text{COVER}(\text{OPT})$. Obviously, $C \geq 1$, and the total sum of jobs sizes, denoted by $W$, satisfies $W \geq mC$. First, we introduce some assumptions on $\mathcal{A}$. Note that the modifications needed to be applied such that this instance will satisfy these assumptions do not increase $\text{COVER}(\mathcal{A})$, do not violate the conditions for NE and do not decrease $\text{COVER}(\text{OPT})$.

1. Machine $P$ contains only tiny jobs, that is, jobs of arbitrarily small size.
   Since no machine has a smaller load, replacing the jobs on this machine by tiny jobs keeps the schedule as an NE. The value COVER(OPT) may only increase.

2. For a tall machine in $\mathcal{A}$ which has two jobs, both jobs have a size of 1.
   If one of them is larger, then the second job would want to move to $P$, so this case cannot occur. If some such job is smaller, its size can be increased up to 1 without affecting the NE.

3. For every tall machine in $\mathcal{A}$ which contains a single job, this job has a size of exactly $C$.
   A larger size of a job can be decreased without decreasing COVER(OPT), while a smaller size can be increased without disturbing the equilibrium.

**Observation 62.** *On identical machines,* POA*(m) is monotonically non-decreasing as a function of the number of machines $m$.*

**Proof.** An NE with $m-1$ machines becomes one for $m$ machines by adding a machine with a single job of size $C$ both to $\mathcal{A}$ and OPT. □

We now prove tight upper and lower bounds on the POA for $m = 2$, $m = 3$ and $m = 4$ machines.

Our analysis of the upper bound on the POA for small values of $m$ strongly relies on the following two observations, which apply for any $m$.

**Observation 63.** *For an* NE *assignment $\mathcal{A}$, the total size of jobs assigned to a machine which has $t \geq 2$ jobs assigned to it is at most $\frac{t}{t-1}$, which is a decreasing function of $t$.*

**Proof.** Every $t-1$ of these jobs have a size of at most 1, otherwise the remaining job would benefit from moving to $P$. Summing this up over all subsets of $(t-1)$ jobs produces the claimed inequality. □

**Observation 64.** *For all $m$, $C < 2$.*

**Proof.** Let $x$ be the number of tall machines in $\mathcal{A}$ with at least two jobs and $y$ the number of tall machines with a single job. Due to Assumption 1, $x+y = m-1$. From Observation 63 and Assumptions 2 and 3, we get $mC \leq W \leq 2x + Cy + 1 = 2x + C(m-1-x) + 1$, which gives $C \leq \frac{2x+1}{x+1} < 2$ since $x \geq 0$. □

**Theorem 65.** *Consider identical machines. i. For two machines,* POA$(2) = \frac{3}{2}$*. ii. For three machines,* POA$(3) = \frac{3}{2}$*. iii. For four machines,* POA$(4) = \frac{13}{8}$*.*

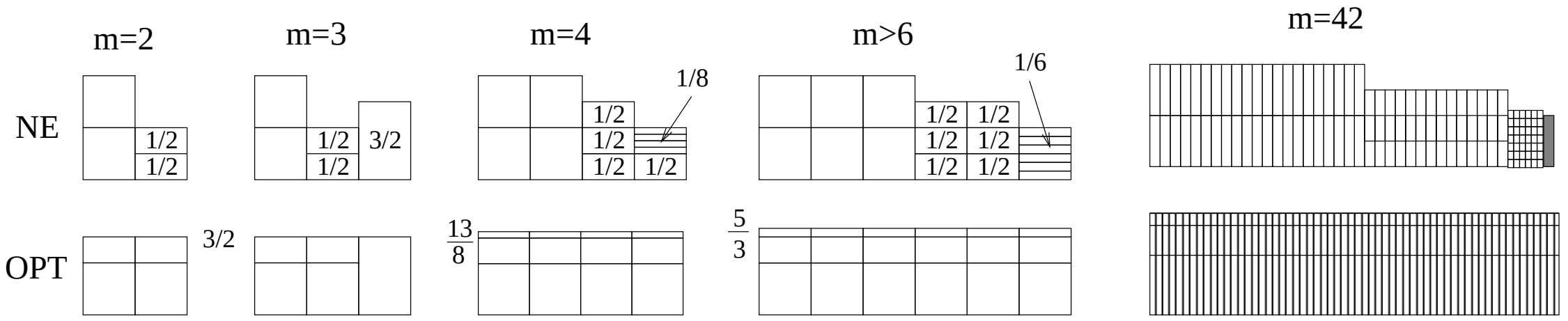

Figure 5.2: Lower bounds for the Price of Anarchy. Machines are on the horizontal axis, jobs are on the vertical axis. The squares in the first few figures represent jobs of size 1. In the figures for $m = 42$, the job sizes are $1, 1/2, 1/6$ and $1/42$.

**Proof.** The lower bounds are shown in Figure 5.2. It is straightforward to verify that the assignments in the top row are Nash equilibria.

i. We show that the POA for two machines is at most $\frac{3}{2}$. Given an assignment, there is exactly one tall machine. If this machine has just one job assigned to it, then $P$ must have a total size of at least $C$. Hence this is an optimal assignment. Otherwise, the tall machine has $t \geq 2$ jobs, and the total size of jobs assigned to $P$ is 1. Using the fact that in an optimal solution each one of the machines is covered by at least $C$ and from Observation 63 for $t = 2$ we get that $2C \leq W \leq 3$ which implies $C \leq \frac{3}{2}$, and thus POA$\leq \frac{3}{2}$.

ii. We show that the POA for three machines is at most $\frac{3}{2}$. We can assume that no tall machine has a single job, as by removing the machine with the job from the NE and the machine with it from OPT (and re-assigning the remaining jobs for OPT), results in a value COVER(OPT) which cannot be smaller, thus reducing us to the case with a smaller number of machines, which we had already considered. There are two cases need to be considered. If at least one tall machine has at least three jobs, using the fact that in an optimal solution each one of the machines is covered by at least $C$, and applying Observation 63 for $t \geq 3$ we get $3C \leq W \leq 2 + \frac{3}{2} + 1 = \frac{9}{2}$ which implies $C \leq \frac{3}{2}$. Otherwise, there are two tall machines that have two jobs, where each job has a size of 1. In total, there are at least four jobs of size 1, so by the pigeonhole principle, the optimal assignment has a machine with two such jobs. Therefore, $2 + 2C \leq W \leq 5$, thus $C \leq \frac{3}{2}$. In total, POA$\leq \frac{3}{2}$.

iii. The POA for four machines is at most $\frac{13}{8}$. The case where a tall machine in $\mathcal{A}$ has a single job and the case where more than one machine has minimum load can be reduced to the case of a smaller number of machines.

There are two additional cases need to be considered. If at least one tall machine has at least three jobs, using the fact that in an optimal solution each one of the machines is covered by at least $C$ and applying Observation 63 for $t \geq 3$ we get $4C \leq W \leq 2 + 2 + \frac{3}{2} + 1 = \frac{13}{2}$ which implies $C \leq \frac{13}{8}$.

Otherwise, there are three tall machines that have two jobs, where each jobs has a size of 1. In total, there are at least six jobs of size 1, so the optimal assignment has at least two machines with two jobs of size 1, or at least one machine with three such jobs. Therefore, $\max\{3+3C, 4+2C\} = 4+2C \le W \le 7$, thus $C \le \frac{3}{2}$, and POA$\le \frac{3}{2}$. In total, POA$\le \frac{13}{8}$. □

We note that similar constructions and observations allow us to find the exact POA for additional small numbers of machines. Thus, e.g., POA$(5) = \frac{13}{8}$ and POA$(6) = \frac{5}{3}$. By Observation 62, this implies POA$(m) \ge 5/3$ for $m \ge 6$, and hence POA $\ge 5/3$.

### 5.4.2 The POA for $m$ identical machines

In this section, we prove that the POA for $m$ machines is at most $1.7$. Consider a pure NE assignment $\mathcal{A}$. Scale the sizes of the jobs such that the social value of this assignment is $1$. Let $C = \text{COVER}(\text{OPT}) \ge 1$. Denote by $P$ a machine of load 1. We state some assumptions on $\mathcal{A}$, and modifications needed to be applied to the instance so it would satisfy these assumptions. To begin with, we still use Assumptions 1 and 2 from before. We no longer use Assumption 3, but instead use the following.

3. This assignment is minimal with respect to the number of machines (among assignments for which $\text{COVER}(\text{OPT}) \ge C$). In particular, no machine in $\mathcal{A}$ has a single job.

   Else, if some machine has a single job, remove this machine and the job from $\mathcal{A}$, and the machine with it from the optimal assignment OPT. Assign any remaining jobs that ran on this machine in OPT arbitrarily among the remaining machines. This gives a new assignment with $\text{COVER}(\text{OPT}) \ge C$, and less machines.

4. Given jobs of sizes $p_1 \le p_2 \le \ldots \le p_t$ assigned to a machine $Q$ in $\mathcal{A}$, then $p_2 + \ldots + p_t = 1$. In fact, $p_2 + \ldots + p_t > 1$ is impossible since this would mean that the job of size $p_1$ has an incentive to move. If the sum is less, enlarge the size $p_t$ to $1 - p_{t-1} - \ldots - p_2$. This does not affect the NE conditions, and keeps the property $\text{COVER}(\text{OPT}) \ge C$.

5. Consider a machine $Q \ne P$ in $\mathcal{A}$ which has $t \ge 3$ jobs assigned to it. Let $a, b$ denote the sizes of the smallest and largest jobs on it, respectively. Then, $b < 2a$.

   Otherwise, if we have an assignment where $b \ge 2a$, replace $b$ with two jobs of size $\frac{b}{2}$. This modification preserves the NE, as the new jobs do not have an incentive to move; Let $T$ denote the total size of jobs on machine $Q$. As $a$ does not want to move to $P$, $T \le 1 + a$ holds. As in this case $a \le \frac{b}{2}$, we have $T \le 1 + \frac{b}{2}$, whereas $1 + \frac{b}{2}$ would have been the load of $P$ if the job $\frac{b}{2}$ moved there.

**Claim 66.** *No job has a size larger than 1.*

**Proof.** This follows from the fact that there is no machine in $\mathcal{A}$ with a single job. □

**Claim 67.** *There is no job of size in $[\frac{2}{3}, 1)$ assigned to a machine $Q$ ($Q \neq P$) in $\mathcal{A}$.*

**Proof.** If there is such job, then it has at least two jobs assigned together with it, each of size greater than $\frac{1}{3}$ (due to assumption 5), which contradicts assumption 4. $\square$

We define a weight function $w(x)$ on sizes of jobs.

$$w(x) = \begin{cases} \frac{1}{2} & \text{, for} \quad x = 1 \\ \frac{x}{2-x} & \text{, for} \quad x \in (\frac{1}{2}, \frac{2}{3}) \\ \frac{x}{x+1} & \text{, for} \quad x \in (0, \frac{1}{2}] \end{cases}$$

For illustration of the function see Figure 5.3.

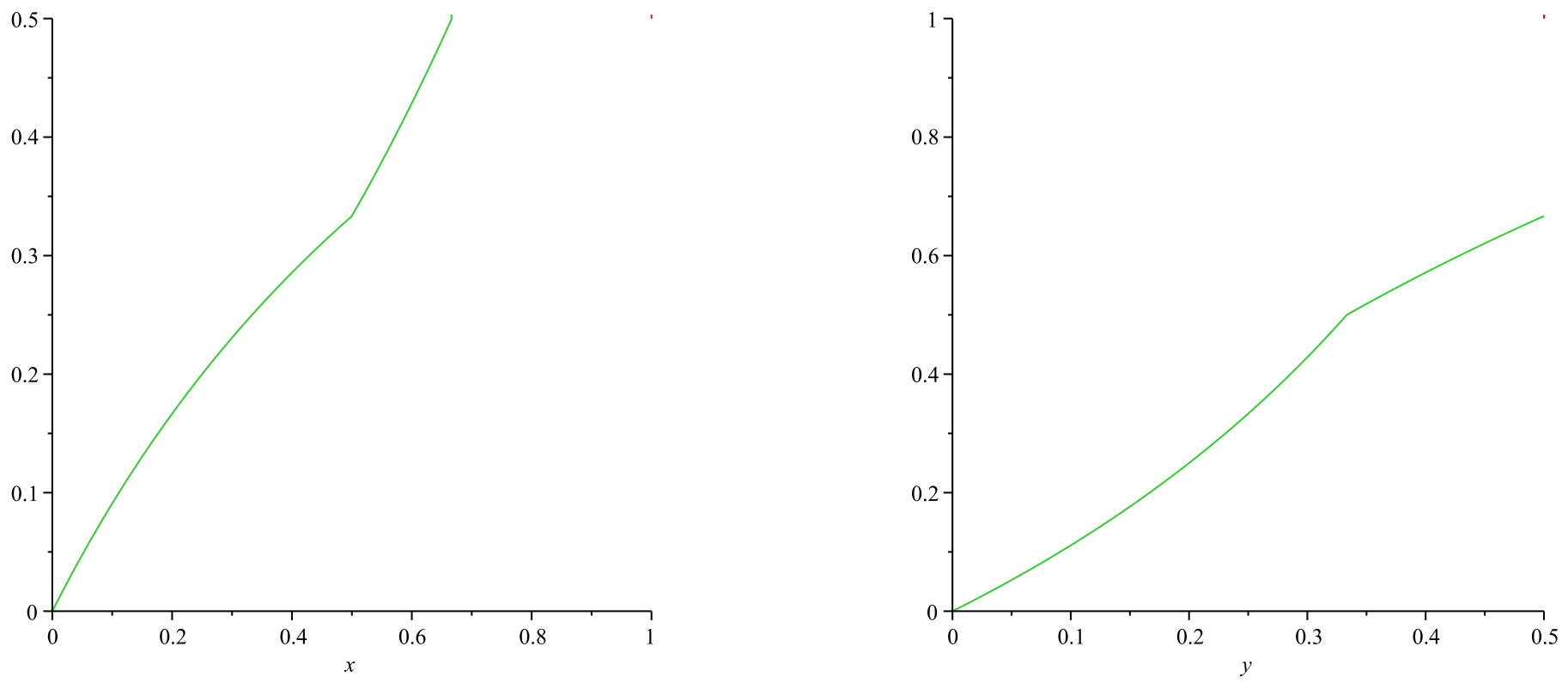

Figure 5.3: The functions $w(x)$ (left hand side) and $f(y)$ (right hand side).

The motivation for the weight function is to define the weight of a job $j$ to be at least the fraction of its size out of the total size of jobs assigned to the same machine in $\mathcal{A}$. In fact, for a job $j$ of size $x$, the total size of jobs assigned in an NE to the same machine as $j$ is no larger than $1 + x$. Moreover if $x > \frac{2}{3}$, then by our assumptions, it is possible to prove that the total size of these jobs is at most $2 - x$.

Its inverse function $f(y)$ is

$$f(y) = \begin{cases} 1 & \text{, for} \quad y = \frac{1}{2} \\ \frac{2y}{y+1} & \text{, for} \quad y \in (\frac{1}{3}, \frac{1}{2}) \\ \frac{y}{1-y} & \text{, for} \quad y \in (0, \frac{1}{3}] \end{cases}$$

Note that $f(y)$ is continuous at $\frac{1}{3}$ but not at $\frac{1}{2}$. Both functions are monotonically increasing.

We now state several claims, which follow from the properties of this weight function defined above.

**Claim 68.** *The total weight (by $w$) of jobs on $P$ in $\mathcal{A}$ is less than* $1$.

**Proof.** Follows from the fact that $P$ has a load 1 and for any $x$, $w(x) < x$. $\square$

**Claim 69.** *The total weight (by $w$) of jobs assigned to a machine $Q \neq P$ in $\mathcal{A}$ is at most* $1$.

**Proof.** By Assumption 3, each machine $Q$ in $\mathcal{A}$ has at least two jobs. The claim clearly holds for a machine with two jobs, as by Assumption 2 both these jobs have size 1, and by the definition of $w$ have a total weight of $\frac{1}{2} + \frac{1}{2} = 1$. For a machine which has at least three jobs assigned to it, there are two possible cases. If there are no jobs with size in $(\frac{1}{2}, \frac{2}{3})$ (i.e., all jobs are of size in $(0, \frac{1}{2}]$), then let $T$ be the total size of jobs on $Q$. For each job of size $x_i$, as the job does not want to move to machine $P$, $x_i + 1 \geq T$ holds. Combining this with the definition of $w$ for jobs of size in $(0, \frac{1}{2}]$, we get that $w(x_i) = \frac{x_i}{x_i+1} \leq \frac{x_i}{T}$. Summing this up over all the jobs on $Q$ proves the claim.

Else, for a job $x_i \in (\frac{1}{2}, \frac{2}{3})$ (there can be only one such job assigned to $Q$, by Assumption 5), we have $T \leq 2 - x_i$. Otherwise, let $a$ be the size of the smallest job on $Q$. Then, by Assumption 4, $T = 1 + a$. As $T > 2 - x_i$, we get $x_i + a > 1$. Therefore, as there is an additional job of size of at least $a$ assigned to $Q$, we get that the total size of all jobs except for the smallest job is more than 1, contradicting Assumption 4.

Combining this with the definition of $w$ for jobs of size in $(\frac{1}{2}, \frac{2}{3})$, we get that $w(x_i) = \frac{x_i}{2-x_i} \leq \frac{x_i}{T}$ for this job too. □

**Claim 70.** *There is a machine in the optimal assignment with a total weight strictly smaller than* $1$.

**Proof.** The total weight of all jobs is less than $m$ by Claims 68 and 69. □

**Claim 71.** *The total size of any set of jobs with total weight below* $1$ *is at most* $1.7$.

**Proof.** To prove the claim, we use the property that there is at most one item of weight $1$. If it exists, then there is at most one item of weight larger than $\frac{1}{3}$. If such an item of weight $1$ does not exist, then the ratio between size and weight is at most $1.5$. We find the supremum possible weight in each one of three cases. Consider a set of jobs $I$ of a total weight strictly below $1$. Note that for any $x$, $w(x) \leq \frac{1}{2}$. If there is no job of weight $\frac{1}{2}$ in $I$, then since $\frac{2}{y+1} \leq \frac{3}{2}$ for $y \geq \frac{1}{3}$ and $\frac{1}{1-y} \leq \frac{3}{2}$ for $y \leq \frac{1}{3}$, the size of any job is at most $\frac{3}{2}$ times its weight, and thus the total size of the jobs in $I$ does not exceed $\frac{3}{2}$.

Otherwise, there is exactly one job of weight $\frac{1}{2}$, and its size is $1$. We therefore need to show that the total size of any set of jobs $I'$ which has total weight below $\frac{1}{2}$ is at most $0.7$. There is at most one job of weight in $(\frac{1}{3}, \frac{1}{2})$ in $I'$. If there is one such job we show that without loss of generality, there is at most one other job in $I'$, and its weight is in $(0, \frac{1}{3}]$). Else, we show that there are at most two jobs, and their weights are in $(0, \frac{1}{3}]$).

Note that $f_1(y) = \frac{y}{1-y}$ is a convex function, thus, for any pair of jobs of weights $\alpha, \beta \in (0, \frac{1}{3}]$, $f_1(0) + f_1(\alpha + \beta) \geq f_1(\alpha) + f_1(\beta)$ holds. As we can define $f_1(0) = 0$, it turns into $f_1(\alpha + \beta) \geq f_1(\alpha) + f_1(\beta)$.

Thus, any two jobs of total weight of at most $\frac{1}{3}$ can be combined into a single job while as a result, their total size cannot decrease. This is due to the convexity of $f_1$ and the fact that it is monotonically increasing. The replacement may only increase the weight, and respectively, the size. If there exists a job of weight larger than $\frac{1}{3}$, then the total weight of jobs of weight at most $\frac{1}{3}$ is at most $\frac{1}{6}$, so they can all be combined into a single job. Moreover, among any three jobs of a total weight of at most $\frac{1}{2}$, there exists a pair of jobs of total weight no larger than $\frac{1}{3}$, which can be combined as described above, so if there is no job of weight larger than $\frac{1}{3}$, still jobs can be combined until at most two jobs remain. Thus, there are only two cases to consider.

**Case 1** There is one job of weight in $(0, \frac{1}{3}]$ and one job of weight in $(\frac{1}{3}, \frac{1}{2})$.

Since the inverse function $f$ is monotonically increasing as a function of $y$ and their weight does not exceed $\frac{1}{2}$, we can assume that their total weight is $\frac{1}{2} - \gamma$ for a negligible value of $\gamma > 0$ (by increasing the weight the job of the smaller weight, which may only increase the total size). Letting $d < \frac{1}{6}$ denote the weight of the smaller job (since if $d \geq \frac{1}{6}$ then $\frac{1}{2} - \gamma - d < \frac{1}{3}$), we get a size of at most

$$\frac{d}{1-d} + \frac{2(\frac{1}{2} - \gamma - d)}{\frac{1}{2} - \gamma - d + 1} < \frac{d}{1-d} + \frac{2-4d}{3-2d}$$

(by letting $\gamma \to 0$). This function is increasing (as a function of $d$) so its greatest value is for $d \to \frac{1}{6}$ and it is 0.7. As the inverse function $f$ is monotonically increasing, this case also encompass the case where there is only one job of weight in $(\frac{1}{3}, \frac{1}{2})$, and no jobs of weight in $(\frac{1}{3}, \frac{1}{2})$.

**Case 2** There are at most two jobs, where each job has a weight in $(0, \frac{1}{3}]$. If there is at most one job, then its size is at most $\frac{1}{2}$. We therefore focus on the case of exactly two such jobs. Recall that the total size of the two jobs is larger than $\frac{1}{3}$ (since they cannot be combined). The total weight of these jobs is less than $\frac{1}{2}$, so we can assume that their total weight is $1/2 - \gamma$ for a negligible value of $\gamma > 0$ (increasing the respective size). Let the weight of the large of these jobs be $d > \frac{1}{6}$. We get (by letting $\gamma \to 0$) a total size of at most

$$\frac{d}{1-d} + \frac{\frac{1}{2} - \gamma - d}{\frac{1}{2} + d + \gamma} < \frac{d}{1-d} + \frac{1-2d}{2d+1},$$

where $\frac{1}{6} < d \leq \frac{1}{3}$. This function is monotonically decreasing in $(\frac{1}{6}, \frac{1}{4}]$ and increasing in $(\frac{1}{4}, \frac{1}{3}]$, and its values at the endpoints $\frac{1}{6}$ and $\frac{1}{3}$ are both 0.7. $\square$

**Theorem 72.** *For covering identical machines, the* POA *is at most* $1.7$.

**Proof.** This follows from Claims 70 and 71. □

We conjecture that the true bound is equal to the upper bound which we provided above.

**Theorem 73.** *For covering identical machines, the* POA *is at least* $1.691$.

**Proof.** We first define a sequence $t_i$ of positive integers, which is often used in the literature for analysis and proving of lower bounds for online bin packing algorithms. Let $t_1 = 1$ and $t_{i+1} = t_i(t_i + 1)$ for $i \geq 1$. The sequence starts with $\{1, 2, 6, 42, 1806, \ldots\}$ and grows rapidly.

Let $m = t_k$ for an integer $k$. Consider the following assignment $\mathcal{A}$, that has $\frac{m}{t_i+1}$ machines with $t_i + 1$ jobs of size $\frac{1}{t_i}$, (for $1 \leq i < k$) and one machine (i.e., $\frac{m}{t_k}$ machines) with $t_k$ jobs of size $\frac{1}{t_k}$. For an example, see the rightmost diagrams in Figure 5.2. We assume that the machines are sorted in a non-increasing order w.r.t. their load. We define the load class $i$, $1 \leq i \leq k$ as the subset of $\frac{m}{t_i+1}$ machines with the same load $L_i = \frac{t_i+1}{t_i} = 1 + \frac{1}{t_i} > 1$ in this assignment. As $t_i$ is an increasing sequence of integers, it follows that $L_i$ is monotonically non-increasing as a function of $i$. Since $L_k = 1$, the social value of this assignment is 1. We now verify that it is a Nash equilibrium. As $L_i > 1$ for any $1 \leq i < k$, no job will benefit from leaving the machine of class $L_k$. It is enough to show that any job assigned to a machine of a class $L_i$ ($1 \leq i < k - 1$) would not benefit from moving to the machine of class $k$. Since machine $i$ has $t_i + 1$ jobs of size $\frac{1}{t_i}$, the migration of such a job to the machine of load 1 would again result in a load of $1 + \frac{1}{t_i}$, thus the job would not benefit from the migration.

In the socially optimal assignment, each machine has a set of jobs of distinct sizes, 1, $\frac{1}{2}$, $\frac{1}{6}$, $\ldots$, $\frac{1}{t_{k-1}}$, $\frac{1}{t_k}$. The social value of this assignment is $\sum_{i=1}^{k} \frac{1}{t_i}$. Thus, the POA equals $\sum_{i=1}^{k} \frac{1}{t_i}$. For $k \to \infty$, this value tends to $h_\infty = \sum_{i=1}^{\infty} \frac{1}{t_i} = 1.69103\ldots$, the well-known worst-case ratio of the Harmonic algorithm for bin packing. As the POA is monotonically non-decreasing as a function of the number of the machines, we conclude that this is a lower bound for any number of machines larger than $t_k$. In particular, the overall POA for identical machines is at least $1.69103$. □

## 5.5 The Price of Stability for *m* identical machines

**Theorem 74.** *On identical machines,* POS $= 1$ *for any* $m$.

**Proof.** We show that for every instance of the machine covering game, among the optimal assignments there exists an optimal assignment which is also an NE. Our proof technique is based upon the technique which was used in [49, 57] to prove that in job scheduling games where the selfish goal of the players is run on the least loaded machine (like in our machine covering game), any sequence of improvement steps converges to an NE.

We first define a complete order relation on the assignments, and then show that an optimal assignment which is the “highest” among all optimal assignments with respect to this order is always an NE.

**Definition 75.** *A vector* $(l_1, l_2, \ldots, l_m)$ *is larger than* $(l'_1, l'_2, \ldots, l'_m)$ *with respect to the inverted lexicographic order, if for some* $i$, $l_i > l'_i$ *and* $l_k = l'_k$ *for all* $k > i$*. An assignment* $s$ *is called larger than* $s'$ *according to the inverted lexicographic order if the vector of machine loads* $L(s) = (L_1(s), L_2(s), \ldots, L_m(s))$*, sorted in non-increasing order, is larger in the inverted lexicographic order than the vector* $L(s') = (L_1(s'), L_2(s'), \ldots, L_m(s'))$*, sorted in non-increasing order. We denote this relation by* $s \succ_{L^{-1}} s'$*.*

The inverted lexicographic order $\succ_{L^{-1}}$ defines a total order on the assignments.

**Lemma 76.** *For any instance of the machine covering game, a maximal optimal schedule w.r.t. the inverted lexicographic order is an* NE*.*

**Proof.** Let $\mathcal{A}^*$ be an optimal assignment such that no other optimal assignment $\mathcal{A}$ satisfies $\mathcal{A} \succ_{L^{-1}} \mathcal{A}^*$. We show that $\mathcal{A}^*$ is an NE assignment. Order the machines in $\mathcal{A}^*$ according to a non-increasing order of loads. Assume by contradiction that $\mathcal{A}^*$ is not NE, that is, there is at least one job that would benefit from moving to another (lower) machine. Consider the rightmost machine with such a job, call it $M_i$ and let the job be denoted by $p_r$. Job $p_r$ is selfish, and will move to the least-loaded machine (or one of the least-loaded machines if there are several such machines) in $\mathcal{A}^*$. Denote this machine by $M_j$, $j > i$, and the resulting assignment by $\mathcal{A}'$.

**Case 1** If there is no machine in $\mathcal{A}^*$ other than $M_j$ having load that equals to $L_j(\mathcal{A}^*)$.

First, observe that $L_i(\mathcal{A}^*) - p_r > L_j(\mathcal{A}^*)$, otherwise job $p_r$ would not have migrated. Note that this migration has an effect on the loads of only two machines $M_i$ and $M_j$ in $\mathcal{A}^*$.

If machine $M_j$ with load $L_j(\mathcal{A}') = L_j(\mathcal{A}^*) + p_r$ is the lowest machine in $\mathcal{A}'$ as well, as the loads of the other machines remain unchanged, except for $L_i(\mathcal{A}')$, which is larger than $L_j(\mathcal{A}^*)$, thus $\min_q L_q(\mathcal{A}') > \min_q L_q(\mathcal{A}^*)$ contradicting the optimality of the assignment $\mathcal{A}^*$. If $M_j$ is not the lowest machine in $\mathcal{A}'$, there is a new lowest machine $M_t$ in $\mathcal{A}'$ such that $L_t(\mathcal{A}') > L_j(\mathcal{A}^*)$ (as there was no additional machine with load $L_j(\mathcal{A}^*)$ in $\mathcal{A}^*$), again contradicting the optimality of $\mathcal{A}^*$.

**Case 2** If there are $k \geq 2$ machines in $\mathcal{A}^*$ having load that equals to $L_j(\mathcal{A}^*)$.

After the migration the social value of $\mathcal{A}'$ remains the same as the social value of $\mathcal{A}^*$ was, as there is (at least one) other machine besides $M_j$ with load $L_j(\mathcal{A}^*)$ in $\mathcal{A}^*$, and its load is not influenced by the migration. As there are $k$ machines with the same lowest load $L_j(\mathcal{A}^*)$ in $\mathcal{A}^*$, in the new assignment $\mathcal{A}'$ the loads of the $k-1$ least loaded machines do not change, and either $M_j$ with load $L_j(\mathcal{A}') = L_j(\mathcal{A}^*) + p_r > L_j(\mathcal{A}^*)$ becomes the $k$-th least loaded machine, or other machine $M_t$, $t < k$ such that $L_t(\mathcal{A}^*) > L_j(\mathcal{A}^*)$ becomes the $k$-th least loaded machine. Thus, by definition 75 $\mathcal{A}' \succ_{L^{-1}} \mathcal{A}^*$, contradicting the maximality of $\mathcal{A}^*$ with respect to the inverted lexicographic order.

Thus $\mathcal{A}^*$ is an NE assignment. □

Since for any set of $n$ jobs there are finitely many possible assignments, among the assignments that are optimal with respect to our social goal there exists at least one which is maximal w.r.t. the total order $\succ_{L^{-1}}$, and according to Lemma 76 this assignment is an NE. As no NE assignment can have a strictly greater social value than the optimal one, we conclude that POS $= 1$. □

## 5.6 The mixed Price of Anarchy

In the setting of mixed strategies we consider the case of identical machines, similarly to [79]. In that work, it was shown that the mixed POA for two machines is $\frac{3}{2}$. In this section we prove that the mixed POA for two machines is equal to $2$.

We start by showing that for $m$ identical machines, the mixed POA can be exponentially large as a function of $m$, unlike the makespan minimization problem, where the mixed POA is $\Theta(\frac{\log m}{\log \log m})$ [79, 36].

**Theorem 77.** *The mixed* POA *for* $m$ *identical machines is at least* $\frac{m^m}{m!}$.

**Proof.** Consider the following instance $G \in MC$ of the Machine Covering game. $N = \{1, 2, \ldots\}$ such that $p_1 = \ldots = p_n = 1$, and let $M_j = \{M_1, \ldots, M_m\}$, for $j = 1, \ldots, n$. Each of the jobs $p_i$, $i = 1, \ldots, n$ chooses each machine with probability $t_i^j = 1/m$. Each jobs sees the same expected load for each machine, and thus has no incentive to change its probability distribution vector. We get a schedule having a non-zero cover, where each job chooses to run on a different machine, with a probability of $\frac{m!}{m^m}$. So, for the mixed Nash equilibrium the expected minimum load is $\frac{m!}{m^m}$. But the coordinated optimal solution achieved by deterministically allocating each job to its own machine has a social value $\text{COVER}(\text{OPT}(G)) = 1$, and so it follows that the $\text{MPOA}(G) = \frac{m^m}{m!}$. We conclude that the MPOA $\geq \frac{m^m}{m!}$. □

**Theorem 78.** *The mixed* POA *for two identical machines is exactly* $2$.

**Proof.** The lower bound follows from Theorem 77.

Consider a specific job $i$ and let $q_i$ be the probability that the job (which has a size of $p_i$) is assigned to the most loaded machine of the two. It was shown in [79] that the expected maximum load over the two machines, call it $\mathbb{E}(\text{MAKESPAN})$, is at most $\sum_k q_k p_k = (\frac{3}{2} - q_i)\sum_k p_k + (2q_i - \frac{3}{2})p_i$. As there are two machines, this implies that the expected minimum load over the two machines, $\mathbb{E}(\text{COVER}) = \sum_k (1 - q_k)p_k = \sum_k p_k - \sum_k q_k p_k = \sum_k p_k - \mathbb{E}(\text{MAKESPAN})$, is at least $\sum_k p_k - (\frac{3}{2} - q_i)\sum_k p_k + (2q_i - \frac{3}{2})p_i = (q_i - \frac{1}{2})\sum_k p_k - (2q_i - \frac{3}{2})p_i)$.

Note that $\text{COVER}(\text{OPT}) \leq \frac{\sum_k p_k}{2}$, which is the case if we can distribute the weight of each job in an equal manner over the two machines. Also, if there exists a job $p_i$ such that $p_i > \sum_{k \neq i} p_k$, then $\text{COVER}(\text{OPT}) \leq \sum_{k \neq i} p_k$. In total, $\text{COVER}(\text{OPT}) \leq \min\{\sum_{k \neq i} p_k, \frac{\sum_k p_k}{2}\}$.

One of the following may occur:

1. There exists a job $p_i$ such that $q_i > \frac{3}{4}$. Therefore, $\mathbb{E}(\text{COVER}) \geq (q_i - \frac{1}{2})\sum_{k \neq i} p_k + (1 - q_i)p_i = (1 - q_i)\sum_k p_k + \sum_{k \neq i} p_k(2q_i - \frac{3}{2}) \geq (2(1 - q_i) + 2q_i - \frac{3}{2})\text{COVER}(\text{OPT}) \geq \frac{1}{2}\text{COVER}(\text{OPT})$.
2. For any job $p_i$, $q_i \leq \frac{3}{4}$. Therefore, $\mathbb{E}(\text{COVER}) = \sum_k (1 - q_k)p_k \geq \frac{1}{4}\sum_k p_k \geq \frac{1}{4} \cdot 2\text{COVER}(\text{OPT}) = \frac{1}{2}\text{COVER}(\text{OPT})$.

In both cases, we get that $\mathbb{E}(\text{COVER}) \geq \frac{1}{2}\text{COVER}(\text{OPT})$, which proves our claim. □

## 5.7 The maximum envy ratio objective

In this chapter we investigated up until now the tradeoff between fairness and optimality of a NE schedule, when the fairness was considered with respect to maximizing the minimum load in the schedule. However, there are different accepted notions of fairness of a job allocation. Other suitable fairness criterion that we can consider in the same setting is the so-called maximum envy ratio.

The envy-ratio of job $i$ for job $j$ is the cost of $i$ over the cost of $j$. As before, the set of players is the set of $n$ jobs to be scheduled on $m$ identical machines. Every job has a positive processing time and the load of every machine is the sum of the processing times of the jobs assigned to it. The cost of a job equals the load of the machine it is assigned on. The objective is to schedule the jobs so as to minimize the ratio of the maximum load over the minimum load of the machines in the assignment.

Going back to the application of data routing on identical parallel links, this objective aims to reduce the "envy" of the job (or jobs) that suffers the greatest delay towards the job (or jobs) that suffers the smallest delay in the network (assuming that each machine has at least one job assigned to it).

For the study of this problem, we use the following notations.

$$L_{max}(\mathcal{A}) = \max_{1\le i\le m} L_i \quad \text{and} \quad L_{min}(\mathcal{A}) = \min_{1\le i\le m} L_i \ .$$

The envy ratio of schedule $\mathcal{A}$ is defined by

$$e(\mathcal{A}) = \frac{L_{max}(\mathcal{A})}{L_{min}(\mathcal{A})} \ .$$

Accordingly, we denote the envy ratio of an optimal schedule OPT (with respect to the envy ratio) by

$$e(\text{OPT}) = \frac{L_{max}(\text{OPT})}{L_{min}(\text{OPT})} \ .$$

This objective was previously considered in the context of scheduling by [29]. It is shown that Graham's greedy algorithm [65] has an approximation ratio of $1.4$ for the envy-ratio problem.

This fairness criterion was considered in a game theoretic setting by Lipton at. el. in [81]. They consider the envy-ratio objective for the problem of envy-free allocation of indivisible goods. In the special case where all the players have the same utility function for each good, one can think of the players as identical machines and the set of goods as a set of jobs, and then this problem is equivalent to minimizing the ratio of the maximum completion time over the minimum completion time.

The measures POA and POS were not previously considered for this objective in the setting of selfish jobs. We show tight bounds of $2$ on the POA for any $m \ge 2$ and that the value of the POS is $1$ for any $m \ge 2$.

### 5.7.1 Price of Anarchy for $m$ identical machines

We start with an upper bound.

**Theorem 79.** *For the problem of envy ratio minimization on identical machines, for any fixed $m \ge 2$, the POA is at most $2$.*

**Proof.** Consider a NE assignment $\mathcal{A}$ on $m$ machines. Denote the largest load observed by a machine in $\mathcal{A}$ by $L_{max}$, and the smallest load by $L_{min}$. If all machines in $\mathcal{A}$ are assigned at most one job then the schedule is optimal. Consider a machine $M_i$ in $\mathcal{A}$ which has at least two jobs assigned to it, having a maximum load among such machines. Consider the size of a smallest job running on $M_i$, denoted by $p$. As this job is assigned with at least one more job, $p \le \frac{1}{2}L_i$. As this schedule is a NE, $L_{min} + p > L_i$ holds, since the job does not benefit from moving to a machine with load of $L_{min}$. Hence $L_{min} > L_i - p \ge \frac{1}{2}L_i$.

If $L_i = L_{max}$, then it immediately follows that $e(\mathcal{A}) \leq 2$, and as the envy ratio of the optimal schedule $e(\text{OPT}) \geq 1$, we get that POA $\leq 2$. Otherwise, let $1 \leq k \leq m-1$ denote the number of machines in $\mathcal{A}$ having a load strictly above $L_i$. By definition of $i$, each such machine has a single job. We denote the set of those jobs by $J_{long}$, and the other jobs by $J_{short}$. $J_{long}$ must contain at least one job of size at least $L_{max}$ and so $L_{max}(\text{OPT}) \geq L_{max}$, since any schedule must assign this job. In OPT there are at least $m-k$ machines which have no jobs of $J_{long}$, and the total size of jobs in $J_{short}$ is at most $(m-k)L_i$. Therefore, $L_{min}(\text{OPT}) \leq L_i$. We get $e(\text{OPT}) \geq \frac{L_{max}(\text{OPT})}{L_{min}(\text{OPT})} \geq \frac{L_{max}}{L_i}$ and POA $\leq \frac{e(\mathcal{A})}{e(\text{OPT})} \leq \frac{L_{max}/L_{min}}{L_{max}/L_i} = \frac{L_i}{L_{min}} \leq 2$. □

We next provide a matching lower bound, which is derived using the construction of [99].

**Theorem 80.** *For the problem of envy ratio minimization on identical machines, for any fixed $m \geq 2$, the* POA *is at least* $2$.

**Proof.** Consider the following game instance with $m$ machines, the set of the jobs contains $m(m-1)$ jobs of size $\frac{1}{m}$ and two jobs of size $1$. One can verify that the schedule described in figure 5.4, where each machine $M_1, \ldots, M_{m-1}$ runs $m$ of the jobs of size $\frac{1}{m}$ and has a load of $1$, and machine $M_m$ runs the two jobs of size $1$ is a NE; moving one of the "heavy" jobs of size $1$ from $M_m$ to any other machine will not decrease their cost. The load of $M_m$ is $2$ while all other loads are $1$. Thus, the envy ratio of this schedule is exactly $2$ for any value of $m$. The optimal envy ratio is obtained by moving one of the two heavy jobs from $M_m$ to $M_1$ and equally distributing the $m$ jobs of $M_1$ among all machines. This results in a schedule with an envy ratio of $1$. The POA of this instance of the game is $2$. □

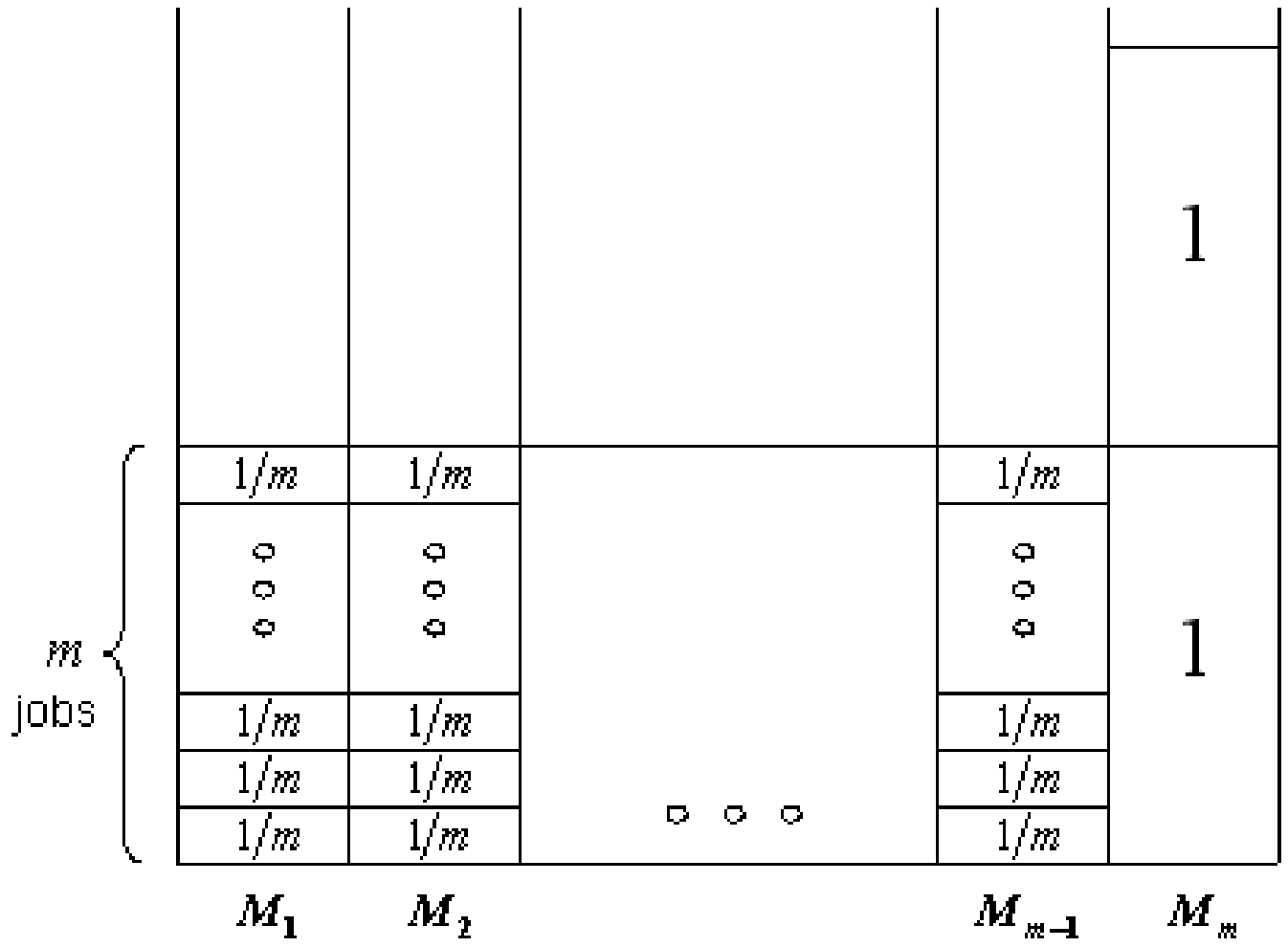

Figure 5.4: A $NE$ schedule with envy ratio $2$.

### 5.7.2 Price of Stability for $m$ identical machines

**Theorem 81.** *On identical machines,* POS $= 1$ *for any* $m$*.*

**Proof.** We show that for every instance of the game, among the optimal assignments there exists one which is also an NE. Our arguments resemble the ones we used to show that POS $= 1$ in the machine covering game.

We will define a complete order relation on the assignments, and then show that an optimal assignment which is the "lowest" among all optimal assignments with respect to this order is always an NE.

**Definition 82.** *A vector* $(l_1, l_2, \ldots, l_m)$ *is smaller than* $(l'_1, l'_2, \ldots, l'_m)$ *with respect to the lexicographic order, if for some* $i$*,* $l_i < l'_i$ *and* $l_k = l'_k$ *for all* $k < i$*. An assignment* $s$ *is called smaller than* $s'$ *according to the lexicographic order if the vector of machine loads* $L(s) = (L_1(s), L_2(s), \ldots, L_m(s))$*, sorted in non-increasing order, is smaller in the lexicographic order than the vector* $L(s') = (L_1(s'), L_2(s'), \ldots, L_m(s'))$*, sorted in non-increasing order. We denote this relation by* $s \prec_L s'$*.*

The lexicographic order $\prec_L$ defines a total order on the assignments.

**Lemma 83.** *For any instance of the game, a minimal optimal schedule w.r.t. the lexicographic order is an* NE*.*

**Proof.** Let $\mathcal{A}^*$ be an optimal assignment such that no other optimal assignment $\mathcal{A}$ satisfies $\mathcal{A} \prec_L \mathcal{A}^*$. We show that $\mathcal{A}^*$ is an NE assignment. Order the machines in $\mathcal{A}^*$ according to a non-increasing order of loads. Assume by contradiction that $\mathcal{A}^*$ is not NE, that is, there is at least one job that would benefit from moving to another (lower) machine. Consider the leftmost machine with such a job, let it be $M_i$ and denote the job by $p_r$. In a slight abuse of the notation we use $p_r$ to denote both the job and its size. Job $p_r$ is selfish, and will want to move to the least-loaded machine (or one of the least-loaded machines if there are several such machines) in $\mathcal{A}^*$. Denote this machine by $M_j$, $j > i$, and the resulting assignment by $\mathcal{A}'$.

First, we observe that $L_{max}(\mathcal{A}^*) = L_{max}(\mathcal{A}')$:

If there is a unique machine with maximum load in $\mathcal{A}^*$, then $p_r$ could not be from this machine. Else, if it moves to the least loaded machine in $\mathcal{A}^*$, then $L_{min}(\mathcal{A}^*) > L_{min}(\mathcal{A}')$ if there is a single machine with load $L_{min}\mathcal{A}^*$ in $\mathcal{A}^*$, or $L_{min}(\mathcal{A}^*) = L_{min}(\mathcal{A}')$ if there are several such machines in $\mathcal{A}^*$, while $L_{max}(\mathcal{A}') < L_{max}(\mathcal{A}^*)$ as $p_r > 0$. Hence $e(\mathcal{A}^*) > e(\mathcal{A}')$, which contradicts the optimality of $\mathcal{A}^*$ with respect to the envy-ratio objective. Therefore $L_{max}(\mathcal{A}^*) = L_{max}(\mathcal{A}')$.

Obviously if there are several machines with maximum load in $\mathcal{A}$, then $L_{max}(\mathcal{A}) = L_{max}(\mathcal{A}')$, no matter where $p_r$ comes from.

As for the load of the least loaded machine in $\mathcal{A}'$ we consider two cases.

**Case 1** If there is a unique machine with minimum load in $\mathcal{A}^*$:

First, observe that $L_i(\mathcal{A}^*) - p_r > L_j(\mathcal{A}^*)$, otherwise job $p_r$ would not have migrated. Note that this migration has an effect on the loads of only two machines $M_i$ and $M_j$ in $\mathcal{A}^*$.

If machine $M_j$ with load $L_j(\mathcal{A}') = L_j(\mathcal{A}^*) + p_r$ is the lowest machine in $\mathcal{A}'$ as well, as the loads of the other machines remain unchanged, except for $L_i(\mathcal{A}')$, which is larger than $L_j(\mathcal{A}^*)$, thus $L_{min}(\mathcal{A}') > L_{min}(\mathcal{A}^*)$, and as the load of the highest machine remains the same in $\mathcal{A}'$, $e(\mathcal{A}') < e(\mathcal{A}^*)$ contradicting the optimality of assignment $\mathcal{A}^*$. If $M_j$ is not the lowest machine in $\mathcal{A}'$, there is a new lowest machine $M_t$, $t > i$ in $\mathcal{A}'$ such that $L_t(\mathcal{A}') > L_j(\mathcal{A}^*)$ (as there is no additional machine with load $L_j(\mathcal{A}^*)$ in $\mathcal{A}^*$), which implies that $e(\mathcal{A}') < e(\mathcal{A}^*)$, again contradicting the optimality of $\mathcal{A}^*$.

**Case 2** If there are at least two machines with minimum load in $\mathcal{A}^*$:

After the migration the envy ratio of $\mathcal{A}'$ remains the same as the envy ratio of $\mathcal{A}^*$ was, as there is (at least one) other machine besides $M_j$ with load $L_j(\mathcal{A}^*)$ in $\mathcal{A}^*$, and its load is not influenced by the migration.

After the migration, $L_j(\mathcal{A}') = L_j(\mathcal{A}^*) + p_r > L_j(\mathcal{A}^*)$. Also, $L_i(\mathcal{A}') < L_i(\mathcal{A}^*)$, as a job with a positive size left $M_i$.

Finally, note that $L_i(\mathcal{A}^*) > L_j(\mathcal{A}^*)$, otherwise job $p_r$ would not have migrated.

Therefore, as $\mathcal{A}^*$ and $\mathcal{A}'$ differ only in machines $M_i$ and $M_j$, by definition 82 $\mathcal{A}' \prec_L \mathcal{A}^*$, contradicting the minimality of $\mathcal{A}^*$ with respect to the lexicographic order.

We got a contradiction in all cases, thus $\mathcal{A}^*$ is an NE assignment. □

Since for any set of $n$ jobs there are finitely many possible assignments, among the assignments that are optimal with respect to this social goal there exists at least one which is minimal w.r.t. the total order $\prec_L$, and according to Lemma 83 this assignment is an NE. As we have established that there exists an optimal solution which is also a NE, we conclude that POS $= 1$. □

## 5.8 Summary and conclusions

In this chapter we have studied a non-cooperative variant of the machine covering problem for identical machines, where the selfish agents are the jobs. We considered both pure and mixed strategies of the agents. We provided various results for the POA and the POS that are the prevalent measures of the quality of the equilibria reached with uncoordinated selfish agents.

For the pure POA for $m$ identical machines, we provided nearly tight lower and upper bounds of 1.691 and 1.7, respectively. An obvious challenge would be bridging this gap. As stated previously, we believe that the actual bound is the upper bound we gave.

We had also provided values for pure POS and POA (1 and 2, respectively) for machine scheduling game on $m$ identical machines with the envy-ratio minimization objective.

# Chapter 6

# Machine Covering with selfish jobs on uniformly related machines

## 6.1 Introduction and results

In this chapter we consider a scheduling problem where the goal is maximization of the minimum load, seeing jobs as selfish agents, in the same model introduced in Chapter 5, now in the setting of uniformly related machines.

The motivations to study this problem can be found, for example, in the routing and queuing applications that were presented in Chapter 5, as in real-world applications the data rate of each of the links/queues can vary.

As we did previously for the case of identical machines, we consider the quality of the NE in the induced game on related machines via the prevalent measures of POA and POS.

In Chapter 5, we considered the problem for identical machines. We showed that the POS is equal to $1$ while the POA and show close bounds on the overall value of the POA is at least 1.691 and at most 1.7 for an arbitrary number of machines. For small numbers of machines, namely, 2,$3$ and $4$ machines, the POA is $\frac{3}{2}$, $\frac{3}{2}$ and $\frac{13}{8}$, respectively.

In contrast to these results, we show that for uniformly related machines even the POS is unbounded already for two machines with a speed ratio *larger than* 2, and the POA is unbounded for a speed ratio of *at least* 2. The same property holds for $m$ machines (where the speed ratio, denoted by $s_{max}$, or simply $s$) is defined to be the maximum speed ratio between any pair of machines). Surprisingly, we prove that the POS is equal to $\frac{3}{2}$ for two machines with the threshold speed ratio 2. We show that the POS is constant for $m$ machines of speed ratio at most 2, and the POA is $\Theta(\frac{1}{2-s})$ for $m$ machines of speed ratio $s < 2$. Finally, we focus on the case of two machines and the exact resulting values of POA. Specifically, we use a linear program to derive tight upper and lower bounds on the POA for speed ratios in the interval $(1, 2)$, and provide tight lower and upper bounds for the POS for speed ratios in the intervals $[1, 4/3]$ and $[1.78, 2]$. We give additional results for the more general problem of $m$ uniformly related machines, and in particular show cases where the POA and POS are unbounded. These results are very different from the situation for the makespan minimization social goal. For that problem, the POS is 1 for any speed combination. Chumaj and Vöcking [36] showed that the overall POA is $\Theta(\frac{\log m}{\log \log m})$ (see also [54, 79]).

Note that for identical machines, the results given in Chapter 5 are more similar to the situation for makespan minimization, where the POS is $1$ and the POA is constant [56, 99].

## 6.2 $m$ related machines

In the setting of related machines, we show that for large enough speed ratios the POA and the POS are unbounded already for two machines.

Denote the speeds of the $m$ machines by $s_1, s_2, \ldots, s_m$, where $s_i \leq s_{i+1}$ and let $s = \frac{s_m}{s_1}$. Without loss of generality, we assume that the fastest machine has speed $s \geq 1$, and the slowest machine has speed $1$.

We now characterize the situation in all cases.

**Theorem 84.** *On related machines with speed ratio $s$, $\text{POA}(s) = \infty$ for $s \geq 2$ and $\text{POS}(s) = \infty$ for $s > 2$.*

**Proof.** Consider an instance that contains $m$ identical sized jobs of size $s$. Clearly, $\text{COVER}(\text{OPT}) = 1$ for this input.

For $s > 2$, we show that any assignment where each job is assigned to a different machine is not a Nash equilibrium. In fact, in such an assignment, any job assigned to the first machine sees a load of $s$, while if it moves to the $m$-th machine, its load becomes $\frac{2s}{s} = 2 < s$. Thus, any NE assignment has a cost of $0$ and the claim follows.

For $s = 2$, consider the assignment $\mathcal{A}$ where each machine is assigned a single job, the first machine has no jobs, and the $m$-th machine has two jobs. It is not difficult to see that this is an NE. Each job assigned to the $m$-th machine will not move to the first machine, since it will have the same load. It would not move to another machine since it would be assigned there together with another job, and this machine is not faster than its current machine. A job assigned to another machine would not move to the first machine, since it is not faster. Moving to a machine which has at least one job assigned to it would result in load of at least $2$, while the current load that the job sees is at most $2$. Thus, we have presented an instance for which $\text{COVER}(\text{OPT}) = 1$ and $\text{COVER}(\mathcal{A}) = 0$, which implies the claim for $s = 2$. □

We next characterize the situation in all other cases.

We can in fact show that if we define $\varepsilon = 2 - s$, the POA grows with $1/\varepsilon$.

**Theorem 85.** *Let $\varepsilon = 2 - s$. The POA as a function of $\varepsilon$ has the order of growth $\Theta(\frac{1}{\varepsilon})$.*

**Proof.** For the upper bound, consider an instance with $\text{COVER}(\text{OPT}) = 1$, and an arbitrary NE assignment $\mathcal{A}$ for this instance. We define a weight function $g(x)$ to be applied on sizes of jobs.

$$g(x) = \begin{cases} 1 & \text{, for } \quad x \geq 1 \\ x & \text{, for } \quad 0 < x < 1 \end{cases}$$

We show that the total size of jobs in any set $I$ is at least $1$ if and only if their total weight (by $g(x)$) is at least $1$. In addition, if their total weight is strictly larger than 1, then $|I| \geq 2$.

**Claim 86.** *Let $I$ be a set of jobs. The total size of jobs in $I$ is at least $1$ if and only if their total weight is at least $1$. In addition, if their total weight is strictly larger than 1, then $|I| \geq 2$.*

**Proof.** If $I$ has a total size of at least 1, we consider two cases. If it contains a job of size at least $1$, then the total weight is at least its weight, which is $1$. Otherwise, it contains only jobs of sizes less than $1$, thus their total weight is equal to their total size. In both cases we get a total weight of at least $1$.

If $I$ has a total weight of at least 1, we consider two cases. If it contains a job of weight $1$, then the total size is at least its size, which is $1$. Otherwise, it contains only jobs of weight less than $1$, thus their total size is equal to their total weight. In both cases we get a total weight of at least $1$.

If $|I| = 1$, then the weight of jobs in $I$ is at most $1$. Thus if the total weight exceeds $1$, then $|I| > 1$. □

Let $P$ be the least loaded machine in $\mathcal{A}$, and let $G$ denote the total weight of jobs assigned to $P$ (if this set is empty then $G = 0$). Let $s_p$ denote the speed of $P$.

If $G \geq 1$, then the total size of jobs is at least $1$, thus the load of $P$ is at least $\frac{1}{s_p} \geq \frac{1}{2}$.

If $G < 1$, then since every machine of OPT has a load of at least $1$ and all speeds are at least $1$, each such machine has a total size of jobs of at least $1$ and thus weight of at least $1$. Therefore, the total weight of all jobs is at least $m$. Therefore, there exists a machine of $\mathcal{A}$ with a total weight of jobs of more than $1$, and thus at least two jobs assigned to it. Denote this machine by $Q$ and its speed by $s_q$.

Let $p$ denote the total size of jobs assigned to $P$ and let $q$ denote the total size of jobs assigned to $Q$. Since the total weight of jobs assigned to $Q$ is at least $1$, we have $q \geq 1$. Let $a$ be the size of a smallest size job assigned to $Q$ (and thus $q \geq 2a$). Since the job of size $a$ has no incentive to move, we have $\frac{p+a}{s_p} \geq \frac{q}{s_q}$ or $\frac{p}{s_p} > \frac{q}{s_q} - \frac{a}{s_p} \geq \frac{q}{2-\varepsilon} - a$, using $1 \leq s_p$ and $s_q \leq 2 - \varepsilon$.

If $a \geq \frac{1}{3}$, using $q \geq 2a$ we have $\frac{q}{2-\varepsilon} - a \geq \frac{2a-2a+a\varepsilon}{2-\varepsilon} \geq \frac{\varepsilon}{6}$. If $a < \frac{1}{3}$, we have $\frac{q}{2-\varepsilon} - a \geq \frac{1}{6}$ since $q \geq 1$. Thus the load of $P$ is $\Omega(\frac{1}{\varepsilon})$, and since $\text{COVER}(\text{OPT}) = 1$, we have a ratio of $O(\frac{1}{\varepsilon})$.

For the lower bound, consider an instance with $m$ identical jobs of size $s = 2 - \varepsilon$, and one job of size $\varepsilon$. We show that the schedule which assigns one large job to each machine, except for machine $m$ which receives two such jobs, and machine $1$ which receives the small job.

Similarly to the proof of Theorem 84 the large jobs assigned to machines $2, 3, \ldots, m-1$ have no incentive to move. Similarly, the jobs assigned to machine $m$ have no incentive to move to any machine, except for possibly the first machine. We have $\frac{2s}{s} = 2 = s + \varepsilon$, and therefore these jobs do not have an incentive to move. Finally, the small job would not move since every machine except for machine 1 is already loaded by at least $1$. For this instance, $\text{COVER}(\text{OPT}) \geq 1$, while $\text{COVER}(\mathcal{A}) = \varepsilon$. □

We next show that the POS has a finite value for any $s \leq 2$. For this purpose we use a well-known algorithm for scheduling, called LPT (see [65]). This algorithm sorts the jobs in a non-increasing order of their sizes, and greedily assigns each job to the machine which would have a smaller load (taking the speed into account) as a result of assignment of the job. In a case of a tie (a job can go to either machine), it assigns the job to the slowest machine which has the largest index among the slowest candidate machines.

It is known that for scheduling games on uniformly related machines with the same selfish goal of the players, LPT produces a pure NE schedule [57]. As this algorithm is deterministic, the value of POS is upper-bounded by its approximation ratio. We use this to derive an upper bound on the POS.

We first show that a ratio between the optimal cover and the cover of the schedule it creates (and accordingly the POS) is finite for $s \leq 2$.

**Theorem 87.** *On related machines with speed ratio $s$, $\text{POS} \leq 2$ for any $s \leq 2$.*

**Proof.** Suppose a counterexample exists, which shows that $\text{POS} > 2$ on $m$ related machines that all have speeds in $[1, 2]$. Let $z > 2$ be a ratio which can be achieved by an example, i.e., the ratio between the cover of an optimal solution, and the maximum cover achieved by every Nash equilibrium, for a specific set of jobs. By normalizing, we may assume the optimal cover of the counterexample has a value of 1. Then, the total size of the jobs in it is at least $m$. There may be an unbounded number of (normalized) counterexamples for which the cover of the best Nash equilibrium assignment is $1/z$. Let $T_z$ be the infimum total size of jobs in all normalized counterexamples with a best NE cover of $1/z$. Let $I$ be such a counterexample for which the total size of all the jobs is at most $T_z + 1/(16m)$. We will derive a contradiction.

Consider the Nash equilibrium assignment $\mathcal{A}$ which is determined by the LPT assignment of the jobs in $I$. LPT begins by sorting the machines in order of nonincreasing speed, and gives each machine a fixed index according to this sorting. Let $P$ be the least loaded machine in $\mathcal{A}$. By assumption, the load of $P$ is exactly $1/z < 1/2$, implying that the total size of the jobs assigned to $P$ is less than 1 (since $s_P \leq 2$).

We herein prove a set of claims regarding qualities of assignment $\mathcal{A}$, which will lead us to a contradiction to the existence of such $I$.

**Claim 88.** *If there are at least $m$ jobs in the input, then for $s \leq 2$, LPT assigns the $i$-th job (for $1 \leq i \leq m$) to machine $m - i + 1$.*

**Proof.** Consider a sorted list of $m$ jobs, where the size of the $i$-th job is denoted by $p_i$. We have $p_1 \geq p_2 \geq \ldots \geq p_m$. We show the claim by induction. By definition, the first job is assigned to the $m$-th machine. Assume that jobs $1, 2, \ldots, k$ ($k \leq m - 1$) have been assigned as claimed. Consider the job of size $p_k$. If the machine to which this job is assigned is chosen among the machines of indices $1, 2, \ldots, m - k + 1$, then by definition it should be assigned to the machine of index $m - k + 1$. If the job is assigned to machine $j > m - k + 1$, the resulting load would become $\frac{p_{m-j+1}+p_k}{s_j} \geq p_k$ (since $p_{m-j+1} \geq p_k$ and $s_j \leq 2$). The load resulting from assigning the job to machine $m - k + 1$ is $\frac{p_k}{s_{m-k+1}} \leq p_k$. Thus, by definition, the job is assigned to machine $m - k + 1$, as claimed. □

We show in the next claim that the set of machines with load at least 1 is fixed until the end of LPT run after the first $m$ jobs have been assigned.

**Claim 89.** *The set of machines with load at least 1 is fixed after the first $m$ jobs have been assigned. These machines do not receive further jobs.*

**Proof.** By the definition of LPT, $P$ received one of the $m$ largest jobs (Claim 88). After this and until LPT is finished, $P$ had load of less than $1/2$, so any additional single job that $P$ received after its first job could not by itself increase the load of $P$ to 1 or more. This means that no assignment of any later job $j$ increases a load of any machine above 1 once each machine has a job, because LPT would assign $j$ to $P$ instead. In addition, a machine which already has a load of 1 or more cannot receive this job. □

A job that is among the $m$ largest jobs and that on assignment by LPT makes the load of its machine 1 or more is called huge. By Claim 89, the precise size of any huge job does not affect the assignment of LPT. As long as the size satisfies the following conditions, the LPT assignment is fixed:

- the size is at least the speed of the machine that it is assigned to (because then that machine receives no further jobs),

- the size is at least the size of the next job in the LPT ordering, and
- the size is at most the size of the previous job in the LPT ordering.

However, the size may of course affect the value of the optimal cover. Regarding the last two conditions, if we have two jobs of different sizes, then making those sizes equal might lead LPT to assign them in a different order. However, we ignore this and still say that the assignment is the same (i.e., we do not distinguish between jobs that have the same size). In the remainder of our proof, we will sometimes change the size of a huge job, but always in such a way that the LPT assignment (and its cover) remain unchanged, and the value of an optimal cover would not be affected. In particular, we will not increase the size of any huge job.

We furthermore show that there is a machine with load at least $1 + 1/(4m)$.

Let $Q'$ be a machine with maximum load.

**Claim 90.** *$Q'$ has load at least $1 + 1/(4m)$.*

**Proof.** Suppose not. Then all machines have load less than $1 + 1/(4m)$, whereas $P$ has load less than 1. The total load on all the machines in $\mathcal{A}$ is then at most

$$\sum_{i=1}^{m} s_i \left(1 + \frac{1}{4m}\right) - s_P \left(1 + \frac{1}{4m}\right) + \frac{s_P}{2} < \sum_{i=1}^{m} s_i + \frac{2m}{4m} - \frac{s_P}{2} \leq \sum_{i=1}^{m} s_i.$$

On the other hand, since $\text{OPT} = 1$, the total load of all the jobs must be at least $\sum_{i=1}^{m} s_i$. This is a contradiction. □

We will return now to the proof of the main theorem.

Let $Q$ be the machine with load at least $1 + 1/(4m)$, which has the smallest index out of such machines. Machine $Q$ exists by Claim 90. Let $A$ be the set of machines which received their first job before $Q$ did, and let $A' = A \cup \{Q\}$. Let $m' = |A'|$. By Claim 89, all machines in $A'$ that have load more than 1 have only one job. In particular, $Q$ has a single, huge job $j_Q$ of size

$$x \geq \left(1 + \frac{1}{4m}\right) s_Q \geq 1 + \frac{1}{4m} > 1.$$

All machines in $A'$ have at least one job of size at least $x > 1$. All machines that are not in $A'$ have speed at most $s_Q$.

We say that the jobs on machines with load at least $1 + \frac{1}{4m}$ are *threshold jobs* (they have size at least $x$, and are huge) while the first jobs on the other machines in $A'$ are called *big* (their size is also at least $x$). Together, these are $m'$ jobs that all have size at least $x$ due to the definition of LPT. All big jobs were assigned before $j_Q$ by LPT.

Consider an optimal schedule OPT. We may assume that no threshold job $j$ is assigned to a machine of smaller index than a big job $j'$ of the same size ($j$ and $j'$ can be switched if necessary).

If OPT has any threshold job $j$ on $Q$ or on a machine that is not in $A'$ (and hence is not faster than $Q$), we claim that $j$ can be made smaller. Let $M(j)$ be the machine that $j$ is assigned to by LPT. First, note that $j$ must be one of the first $m-1$ jobs in the LPT ordering, since all machines that get a job before $M(j)$ have a job of size at least $x$ and hence a load at least $x/2 > 1/2$. Hence, none of these machines can be $P$, and $M(j) \neq P$ as well. This shows that $M(j)$ must be one of the first $m-1$ machines in the ordering used by LPT.

We now decrease $p_j$ by $1/(8m)$. As a result, LPT may assign $p_j$ to some later machine, but $p_j$ is still more than 1 and $j$ remains one of the $m-1$ largest jobs (otherwise, there are $m$ jobs of size more than 1, so both before and after the modification, $P$ receives a load more than $1/2$ which is a contradiction).

If LPT still assigns $j$ to $M(j)$, then the load of $M(j)$ remains more than 1, so $\mathcal{A}$ and $cover(\mathcal{A})$ do not change. If LPT assigns $j$ to a later machine $M'(j)$, then the machines $M'(j), \ldots, M(j)$ all have a load at least 1 after assignment of their first job, both before and after the change. This holds because they all have a job of size at least $p_j - 1/(8m)$ (by definition of LPT) and all these machines have speed at most $s_{M(j)} \leq p_j/(1+\frac{1}{4m}) < p_j - 1/(8m)$. Hence they receive no more jobs and the assignment of all later jobs and $cover(\mathcal{A})$ remain unchanged.

Hence, the value $p_j$ can be decreased by at least $1/(8m)$ without affecting $cover(\mathcal{A})$. This also does not affect the value of OPT, since OPT assigns $j$ to a machine which is not faster than $Q$. We have now found a counterexample with total size of jobs below $T_z$, a contradiction. Hence, all threshold jobs are on machines in $A$ in the optimal assignment.

If any threshold job $j$ is together with a big job on machine $M'$ in the optimal assignment, it can again be made $1/(8m)$ smaller without affecting the value of OPT, since this means the load on $M'$ is at least $2x/2 \geq 1 + \frac{1}{4m}$ on that machine (since $s_{M'} \leq 2$). (Again, the LPT assignment remains unchanged apart from possibly the order of some huge jobs, as above.)

If any big job $j$ is on a machine with higher index than $Q$ (i.e., not in $A'$) in the optimal assignment, or if $j$ is together with another big job on machine $M'$, $j$ can be switched with $j_Q$ (which is assigned to a machine in $A$ by the above) without decreasing the value of OPT (since the size of $j_Q$ is more than $s_Q$, but not more than that of $j$, and $M'$ has a load of at least $1+\frac{1}{4m}$ as above). After this, the size of $j_Q$ can again be decreased for a contradiction.

We thus find that we may assume that all big and threshold jobs are on $A'$ in the optimal assignment, one per machine. But then, $Q$ must have a big job $j'$ since it does not have a threshold job, and then $j_Q$ and $j'$ can again be switched in the optimal assignment (and, after this, $j_Q$ can be made smaller as described above). We find a contradiction to the definition of $I$ in all cases.

□

We note that unlike the POA which continuously grows from constant values to infinite values, the POS jumps from a constant value for $s = 2$ to $\infty$ for $s > 2$.

## 6.3 Two related machines

In this section we study the problem on two uniformly related machines. Recall that we denote the speed of the faster machine by $s \geq 1$, and the other machine has speed 1. We analyze the POA and the POS as a function of $s$. Thus POA$(s)$ and POS$(s)$ denote the POA and POS (respectively) of instances where the speed ratio between the machines is $s$.

The case where $s = 1$ was already analyzed in Section 1, and it was shown that POA$(1) = \frac{3}{2}$ and POS$(1) = 1$. Hence, from now on, we assume that $s > 1$. Moreover, Theorem 84 restricts us to the case $s < 2$ in the analysis of the POA as a function of $s$.

We give a complete analysis of the exact POA as a function of the speed of the faster machine, $s$, for $s < 2$.

**Theorem 91.** *For two related machines and* $1 < s < 2$,

$$\text{POA}(s) = \min\left\{\frac{s+2}{(s+1)(2-s)}, \frac{2}{s(2-s)}\right\} = \begin{cases} \frac{s+2}{(s+1)(2-s)} & \textit{for} \quad 1 < s < \sqrt{2} \\ \frac{2}{s(2-s)} & \textit{for} \quad \sqrt{2} \leq s < 2 \end{cases}$$

**Proof.** In order to generate the bounds on the value of the POA we use Linear Programming. First, we show that we can restrict our analysis for the upper bound to instances involving no more than 4 jobs. Thus, we need to consider a small number of cases. We then formulate for each one of these cases a linear program (LP) whose optimal objective function is exactly the POA for any $s$ (or for some subinterval of $s$, depends on the case). It is possible to find the tight values using the solution for the corresponding dual linear program (DLP). Since we have several cases, we choose our bound on POA for each $1 < s < 2$ to be the highest bound accepted among all the cases. The idea for this analysis is motivated by the classic paper by Graham [65], where he used a similar approach to analyze the performance of an algorithm for a variant of the makespan minimization problem for $m$ identical machines, and is often used since in the study of various scheduling problems. To the best of our knowledge, it was not applied to analyze the POA for any non-cooperative version of a scheduling problem before.

Consider an input instance $I$ and an arbitrary Nash equilibrium assignment $\mathcal{A}$ for it.

**Claim 92.** *We can assume that $I$ contains at most four jobs.*

**Proof.** If there are at least 5 jobs, then one of the machines in the optimal schedule has at least three jobs assigned to it. If a pair of jobs share a machine in both $\mathcal{A}$ and OPT (which must happen in this case), we can merge them. This is applied to any pair of jobs that behave the same in both solutions. The equilibrium properties are kept, so an alternative NE is created, which has at most four jobs, and the same ratio between the values of the optimal solution and the value of the NE solution. □

Thus we have to consider only a small number of cases. Yet, since crucial constraints in the linear programs state the fact that a given job does not have an incentive to move, it is not always possible to assume that a job, which does not exist, simply has a size of zero. Thus we need to consider the cases of one, two, three and four jobs, but some of the cases of three jobs would be seen as special cases of four jobs.

The case of one job is trivial as any schedule has a value $0$, and we defined that in this case POA $= 1$. Consider a schedule with two jobs. Let $A$ and $B$ denote both the jobs and their sizes. A schedule where both are assigned to the same machine cannot be optimal. Moreover, such a schedule cannot be an NE either. Specifically, if both jobs are assigned to $M_1$, then any job has an incentive to move. If they are both assigned to $M_2$, then the job of size $\min\{A, B\}$ would have a load of at most $\min\{A, B\} \leq \frac{A+B}{2}$ being assigned to $M_1$, while the load of $M_2$ when both jobs are assigned to it is $\frac{A+B}{s} > \frac{A+B}{2}$.

Otherwise, $\text{COVER}(\text{OPT}) > 0$ and $\text{COVER}(\mathcal{A}) > 0$. Thus each schedule has one job on each machine. We prove that for $s < 2$, POA $\leq s$. Suppose that the optimum schedule has job $A$ assigned to $M_1$ and job $B$ assigned to $M_2$. If $\mathcal{A}$ gives the same assignment we have have a ratio of $1$. Else, we consider the four possibilities: If $A \leq \frac{B}{s}$ and $B \leq \frac{A}{s}$, $\text{COVER}(\text{OPT}) = A$ and NE $= B$, we have the ratio $\frac{A}{B} \leq \frac{1}{s} < 1$ for $s < 2$. If $A \leq \frac{B}{s}$ and $B > \frac{A}{s}$, $\text{COVER}(\text{OPT}) = A$ and NE $= \frac{A}{s}$, and $\frac{A}{\frac{A}{s}} \leq s$. If $A > \frac{B}{s}$ and $B \leq \frac{A}{s}$, $\text{COVER}(\text{OPT}) = \frac{B}{s}$ and NE $= B$, and the ratio is at most $\frac{\frac{B}{s}}{B} \leq \frac{1}{s} < 1$. If $A > \frac{B}{s}$ and $B > \frac{A}{s}$, $\text{COVER}(\text{OPT}) = \frac{B}{s}$ and NE $= \frac{A}{s}$, and the ratio is at most $\frac{\frac{B}{s}}{\frac{A}{s}} < s$ for $s < 2$. We conclude that the asserted upper-bound of $s$ holds. For $s \geq \sqrt{2}$, since $\frac{2}{s(2-s)} \geq 2$, and $s < 2$, the obtained ratio $s$ does not exceed the claimed bound. For $s < \sqrt{2}$, since $\frac{s+2}{(s+1)(2-s)} \geq \frac{3}{2}$, and $s < \frac{3}{2}$, the obtained ratio $s$ again does not exceed the claimed bound.

We next consider the cases of three and four jobs. Whenever it is possible, we study the case of three jobs as a special case of the case of four jobs, with one of the jobs having size 0. There are four cases of three jobs, when each time a different job of the four has size 0. Also, we need two linear programs for every configuration, that is, one for the case where in $\mathcal{A}$ the slow machine is less loaded (which we call the *short* machine), and one for the opposite case. The linear programs do not state the fact that no job would benefit from leaving the short machine, which always holds.

We assume $\text{COVER}(\text{OPT}) = 1$ which can be achieved by scaling all instances.

So, suppose that the four jobs (and their sizes) are denoted by $A$, $B$, $C$ and $D$. Both $\mathcal{A}$ and OPT assign exactly two jobs to each machine. Suppose also that the optimum schedule has jobs $A$ and $B$ assigned to $M_1$, and jobs $C$ and $D$ assigned to $M_2$. Suppose further that $\mathcal{A}$ assigns jobs $A$ and $C$ to $M_1$ and jobs $B$ and $D$ to $M_2$. Given these NE and optimal assignments and knowing that $\text{COVER}(\text{OPT}) = 1$, we illustrate the following linear programs, where all variables are non-negative.

The constraints first state the properties of OPT (that both machines have a load of at least 1), next the fact that the jobs assigned to the more loaded machine do not have an incentive to move, and finally, the relation between the loads of the two machines in $\mathcal{A}$. We wish to minimize the minimum load, which is the inverse of the ratio between $\text{COVER}(\text{OPT}) = 1$ and $\text{COVER}(\mathcal{A})$.

Since we would like to provide an example which achieves the POA for each value of $s$, it is sufficient that we use only the primal linear programs and prove an upper bound on the POA, which is later matched by our examples.

The programs for four jobs are as follows.

(i) There are four jobs and the slow machine is short:

$\min A+C$ subject to: $A+B \geq 1$, $C+D \geq s$, $A+C+B \geq \frac{B+D}{s}$, $A+C+D \geq \frac{B+D}{s}$, $\frac{B+D}{s} \geq A+C$.

(ii) There are four jobs and the fast machine is short:

$\min \frac{B+D}{s}$ subject to: $A+B \geq 1$, $C+D \geq s$, $\frac{A+D+B}{s} \geq A+C$, $\frac{C+D+B}{s} \geq A+C$, $A+C \geq \frac{B+D}{s}$.

In the first program, it is assumed that $B > 0$ and $D > 0$, thus each one of the cases $B = 0$ and $D = 0$ need to be considered separately. In the second program, it is assumed that $A > 0$ and $C > 0$, thus each one of the cases $A = 0$ and $C = 0$ need to be considered separately. Thus we have additional four programs.

(iii) There are 3 jobs with non-zero sizes, $A = 0$, and the fast machine is short:

$\min \frac{B+D}{s}$ subject to: $B \geq 1$, $C+D \geq s$, $\frac{C+B+D}{s} \geq C$, $C \geq \frac{B+D}{s}$.

(iv) There are 3 jobs with non-zero sizes, $B = 0$, and the slow machine is short:

$\min A+C$ subject to: $A \geq 1$, $C+D \geq s$, $D+A+C \geq \frac{D}{s}$, $\frac{D}{s} \geq A+C$.

(v) There are 3 jobs with non-zero sizes, $C = 0$, and the fast machine is short:

$\min \frac{B+D}{s}$ subject to: $A+B \geq 1$, $D \geq s$, $\frac{A+B+D}{s} \geq A$, $A \geq \frac{B+D}{s}$.

(vi) There are 3 jobs with non-zero sizes, $D = 0$, and the slow machine is short:

$\min A+C$, subject to: $A+B \geq 1$, $C \geq s$, $B+A+C \geq \frac{B}{s}$ ,$\frac{B}{s} \geq A+C$.

Before we consider the first two programs, we show that the last four programs do not give a ratio larger than $s$.

Program (iii) gives a worst case ratio of at most $s$; as $B \geq 1$ and $D \geq 0$, we get that $\frac{B+D}{s} \geq \frac{1}{s} = \frac{\text{OPT}}{s}$.

Program (iv) gives a worst case ratio of at most 1; as $A \geq 1$ and $C \geq 0$, we get that $A + C \geq A \geq 1 = \text{OPT}$.

Program (v) gives a worst case ratio of at most 1; as $D \geq s$ and $B \geq 0$, we get that $\frac{B+D}{s} \geq \frac{D}{s} \geq 1 = \text{OPT}$.

Program (vi) gives a worst case ratio of at most 1; as $C \geq s$ and $A \geq 0$, we get that $A + C \geq C \geq s \geq 1 = \text{OPT}$.

We next consider the first two programs. Though it is possible to solve the programs parametrically and together with solving the corresponding parametric dual programs for all values of $1 < s < 2$. The minimum of the solutions is the POA (provided that it is not below $s$, the upper bound which we got for the case of two jobs).

For Program (i) with $s \leq \sqrt{2}$, we multiply the first two constraints by $2 - s$, and the next two constraints by $s$, and sum all of the four resulting constraints. This gives $(2 + s)(A + C) + 2(B + D) \geq 2(B + D) + (s + 1)(2 - s)$, and thus $A + C \geq \frac{(s+1)(2-s)}{s+2}$. This gives an upper bound of $\frac{s+2}{(s+1)(2-s)}$.

For Program (i) with $s \geq \sqrt{2}$, we multiply the second two constraint by $2 - s$, the third constraint by $1$, the fourth constraint by $s - 1$, the constraint $A \geq 0$ by $2 - s$, and sum all of the four resulting constraints. This gives $2(A + C) + B + D \geq B + D + s(2 - s)$, and thus $A + C \geq \frac{s(2-s)}{2}$. This gives an upper bound of $\frac{2}{s(2-s)}$.

For Program (ii) we multiply the first two constraints by $2s - 1$, and the next two constraints by $s$, and sum all of the four resulting constraints. This gives $2s(A+C)+(2s+1)(B + D) \geq 2s(A + C) + (s + 1)(2s - 1)$, and thus $\frac{B+D}{s} \geq \frac{(s+1)(2s-1)}{s(2s+1)}$. This gives an upper bound of $\frac{s(2s+1)}{(s+1)(2s-1)}$. It is easy to verify that for $1 \leq s < 2$, $\frac{s(2s+1)}{(s+1)(2s-1)} \leq \frac{s+2}{(s+1)(2-s)}$ and $\frac{s(2s+1)}{(s+1)(2s-1)} \leq \frac{2}{s(2-s)}$ hold.

We can see from this discussion that Program (i) gives an upper bound of $\frac{s+2}{(s+1)(2-s)}$ on the POA for $1 < s < \sqrt{2}$ and gives an upper bound of $\frac{2}{s(2-s)}$ on the POA for $\sqrt{2} \leq s < 2$.

We conclude that

$$\text{POA}(s) \leq \begin{cases} \frac{s+2}{(s+1)(2-s)} & \text{, for } \quad 1 < s < \sqrt{2} \\ \frac{2}{s(2-s)} & \text{, for } \quad \sqrt{2} \leq s < 2 \end{cases}.$$

This bound is tight, as can be seen from the following two feasible solutions to the corresponding LPs.

For $1 < s < \sqrt{2}$, consider the following 4 jobs of sizes: $A = \frac{2-\mathcal{A}^2}{s+2}$, $B = \frac{s(s+1)}{s+2}$, $C = \frac{s}{s+2}$ and $D = \frac{s(s+1)}{s+2}$. The optimal solution has a balanced schedule of value 1. In the NE, the load of $M_1$ is $\frac{2+s-\mathcal{A}^2}{s+2}$, and the load of $M_2$ is $\frac{2(s+1)}{s+2}$. Moving a job of size $\frac{s(s+1)}{s+2}$ from $M_2$ to $M_1$ would result in a load of $\frac{2(s+1)}{s+2}$, thus this is indeed a NE. Its value is $\frac{(s+1)(2-s)}{s+2}$.

For $\sqrt{2} \leq s < 2$, consider the following 3 jobs of sizes (where the job of size $A$ is absent): $B = \frac{\mathcal{A}^2}{2}$, $C = \frac{(2-s)s}{2}$ and $D = \frac{\mathcal{A}^2}{2}$. The optimal solution has a load of 1 on $M_2$ and a load of $\frac{\mathcal{A}^2}{2} \geq 1$ on $M_1$. In the NE, the load of $M_1$ is $\frac{2s-\mathcal{A}^2}{2}$, and the load of $M_2$ is $s$. Moving a job of size $\frac{\mathcal{A}^2}{2}$ from $M_2$ to $M_1$ would result in a load of $s$, thus this is indeed a NE. Its value is $\frac{(2-s)s}{2}$. □

We can see that as $s$ approaches 2, POA(s) tends to infinity, as implied by Theorem 85.

**Theorem 93.** *For two related machines and $s \leq 2$, POS $\leq 2 - 1/s$. For $s \leq 3/2$, POS $\leq 1/(2s - s^2)$. These bounds are tight for $s \in [1, 4/3]$ and $s \in [\frac{1}{4}(3 + \sqrt{17}), 2]$. (We have $\frac{1}{4}(3 + \sqrt{17}) \approx 1.78078$).*

**Proof.** Consider an instance $I$ that shows a POS of at least $2 - 1/s$ for some $s \in [1, 2]$. For this instance, the optimal solution OPT is not a Nash equilibrium. Hence, there must be at least one job in OPT that could improve its delay by moving.

Suppose there is such a job $j$ that is alone on its machine in the optimal assignment. Then it must be on the slow machine, else it could not improve. Normalize all the job sizes so that $j$ has size 1. The total size $q$ of all other jobs must be less than 1, else $(1 + q)/s \geq 2/s \geq 1$, contradicting that job $j$ can improve. Consider what happens if we switch the contents of the two machines. The cover is $\min(1/s, q)$, whereas in the optimal assignment, it was $\min(q/s, 1)$. Since $q < 1$, we have $\min(q/s, 1) = q/s$. However, the cover of the new schedule is either $1/s > q/s$ or $q > q/s$, contradicting that the original schedule is optimal. This shows that this situation cannot happen.

Therefore, any job that can improve its delay in the optimal assignment is sharing its machine with at least one other job. We consider the smallest job that can improve and denote it by $j$. We now normalize such that this job has size 1. Denote the delay that $j$ experiences by $D$, and the delay on the other machine by $D'$. If $j$ is on the fast machine, then $D'$ must be less than $D - 1 < D - 1/s$. But then the current assignment does not give an optimal cover, since the cover is at most $D' < D - 1/s$ and $\min(D - 1/s, D' + 1)$ can be achieved by moving $j$.

We conclude that $j$ must be on the slow machine. Then $D \geq 2$. Since $j$ can improve by moving, we have

$$D' + 1/s < D.$$

We also have

$$D' \geq D - 1,$$

else the cover of the optimal assignment could be improved by moving $j$. Hence, after $j$ moves, no other job can improve its delay by moving to the fast machine, since the delay on the slow machine is now only $D - 1 < D' + 1/s$. The cover of the new assignment is hence $\min(D - 1, D' + 1/s) = D - 1$, whereas the cover of the optimal assignment was $\min(D, D') = D' < D - 1/s$, where the equality holds because $j$ can improve. This gives

us a ratio of

$$\frac{D'}{D-1} < \frac{D-1/s}{D-1} \leq \frac{2-\frac{1}{s}}{1} = 2 - \frac{1}{s}. \tag{6.1}$$

We have now reached a Nash equilibrium unless there are jobs on the fast machine which can improve. Any such job must have size strictly smaller than $1$, since $j$ preferred the fast machine.

We let such jobs move one by one, starting with the largest. Note that any job $q$ which moves towards the slow machine (and thereby improves its delay) *strictly improves* the cover. The only case where this is not immediately clear is where the minimum load is achieved on the fast machine after $q$ moves. Denote the loads before $q$ moves by $D_1$ on the fast machine and $D_2$ on the slow machine. Then $D_2 + q < D_1$, the old cover was $D_2$ and the new cover is $D_1 - q/s > D_2$.

We claim that we end up in a Nash equilibrium. Suppose not. Then some job $r$ can improve by moving back to the fast machine. Since all the jobs that moved to the slow machine have size strictly less than $1$, this must be one of the jobs that moved in our process (after $j$ moved). It cannot be the last job that moved. Consider the first time that $r$ can improve. This happens just after another job $r'$ moves to the slow machine which is not larger than $r$. But then, if $r'$ prefers the slow machine at this point, $r$ must do so as well, a contradiction. Since we showed that the cover only improves after job $j$ moves, this process ends in a Nash equilibrium with a cover of at least $2 - 1/s$ due to (6.1).

We now improve this upper bound for $s \leq 3/2$. Consider the schedule $\text{OPT}^-$ that we get by switching the jobs on the fast and the slow machines. The new loads are $D's$ and $D/s$.

The optimal cover is $\min(D', D) = D'$. The cover of $\text{OPT}^-$ is $\min(D/s, D's)$. For $s \in [1, 3/2]$, we have $D's \geq (D-1)s > D/s$ since $D \geq 1/(s-1)$. For $s \in [3/2, 2]$, this holds because $D \geq 2$. Therefore, $\text{COVER}(\text{OPT}^-) = D/s$. This cannot be more than $\text{COVER}(\text{OPT}) = D' < D - 1/s$. We conclude that $D/s < D - 1/s$ and hence $D > 1/(s-1)$. This is a stronger condition than $D \geq 2$ for $s \leq 3/2$. Instead of (6.1) we now get

$$\text{POS} \leq \frac{D-1/s}{D-1} < \frac{\frac{1}{s-1} - \frac{1}{s}}{\frac{1}{s-1} - 1} = \frac{s-(s-1)}{s-s(s-1)} = \frac{1}{2s-s^2} \quad \text{for } s \leq 3/2.$$

Since $D/s < D's$, the only job that might be able to improve in $\text{OPT}^-$ is a job which is now on the slow machine. Such a job currently experiences a delay of $D's$, and after moving, has a delay of at most $D/s + D'$ (the bound is reached if there is a single job on the slow machine).

We now present lower bounds for the subintervals $s \in [1, 4/3]$ and $s \in [1.78, 2]$. Let $s_0 = \frac{1}{4}(3+\sqrt{17})$. Consider some value $s_0 \leq s \leq 2$. Let $M_1$ be the slow machine and $M_2$ be the fast machine. Consider a list of $3$ jobs, having sizes $2s-1$, $1+\varepsilon$ (for some small $\varepsilon > 0$) and 1. The jobs are listed in a non-increasing order of the sizes, as $2s-1 > 1+\varepsilon$ for any $s > 1+\frac{\varepsilon}{2}$, which is the case here. The LPT algorithm applied to this list of jobs first assigns the job of size $2s-1$ to $M_2$, then assigns the job of size $1+\varepsilon$ to $M_1$, and as $2+\varepsilon > \frac{2s-1+1}{s} = 2$, assigns the job of size 1 to $M_2$. This schedule is an NE(as any schedule produced by the LPT rule), and has a value of $1+\varepsilon$. Obviously, LPT produces an NEschedule, but there may exist other schedules that are NEfor this list of jobs. We claim that for $s \geq s_0$ this is the best NEschedule, that is, this is a schedule with the largest cover among all possible NEschedules for this instance. A different schedule, that assigns a job of size 1 to $M_1$ and the jobs of sizes $2s-1$ and $1+\varepsilon$ to $M_2$, is also an NE, but has a smaller cover (of 1). Clearly, no schedule that has all three jobs assigned to one of the machines and has a cover of value 0 is stable. Now consider a schedule that has the job of size $2s-1$ assigned to $M_1$ and the two jobs of sizes $1+\varepsilon$ and 1 assigned to $M_2$. As $\frac{2+\varepsilon}{s} > 2s-1$ for any $s > 1.28$, it has a cover of $2s-1$, which is larger than the cover of the schedule produced by the LPT for $s$ in the considered interval. This is not a NEschedule for $s \geq s_0$ though, as for these values of $s$ we find $2s-1 > \frac{2s-1+2+\varepsilon}{s}$ holds, and the job of size $2s-1$ will benefit from moving to $M_2$. The optimal schedule assigns the two jobs of sizes $1+\varepsilon$ and 1 to $M_1$, and the job of size $2s-1$ to $M_2$, and has a cover of $\frac{2s-1}{s} = 2-\frac{1}{s}$ and ($2-\frac{1}{s} < 2$, for positive $s$). It is not stable, though, as the job of size 1, for example, will benefit from moving to $M_2$. We conclude that $\text{POS}(s) \geq 2-\frac{1}{s}$ for $s \in [s_0, 2]$ (letting $\varepsilon$ tend to 0).

Now consider the case $1 \leq s \leq 4/3$. We now use an instance with four jobs, two of size $2-s+\varepsilon$ and two of size $s-1$ (note that $2(s-1) < 2-s+\varepsilon$). The optimal cover is $(1+\varepsilon)/s$ (two jobs on each machine). Consider the possible allocations of the jobs. If there is a machine with one job, then the cover is at most $2-s+\varepsilon$ for a bound that tends to $1/(2s-s^2)$ as desired. If the two small jobs are on one machine, the cover is at most $2(s-1)$, which is even less. If each machine has two different jobs, we do not have a Nash equilibrium, because the small job on the slow machine can improve by moving. Its current delay is $1+\varepsilon$, and the delay on the fast machine would be only $(2(s-1)+2-s+\varepsilon)/s = 1+\varepsilon/s$. Hence in all cases, either the cover is at most $2-s+\varepsilon$ or the schedule is not a Nash equilibrium.

□

It is interesting to note that while the POS is equal to $\frac{3}{2}$ for $s=2$, it becomes infinite already for $s = 2+\varepsilon$ for arbitrary small positive $\varepsilon$.

## 6.4 Summary and conclusions

In this chapter we studied the Prices of Anarchy and Stability for the Machine Covering problem on related machines, as a function of the parameter the maximum speed ratio $s_{max}$ between any two machines. We show that on $m$ related machines, the Price of Stability is unbounded $s_{max} > 2$, and the Price of Anarchy is unbounded for $s_{max} \geq 2$. In an interesting manner, the behavior of these measures for $s_{max}$ close to the threshold speed ratio of 2 is very different; We show that while the Price of Anarchy tends to grow to infinity as $s_{max}$ tends to $2$, the Price of Stability is constant and has a value of at most $2$ for any $s_{max} \leq 2$.

# Chapter 7

# Conclusion and open problems

This chapter presents the conclusion of this thesis, coming back on the main issues. Our contributions are also mentioned with the main results we established. Finally, we describe the limits of this thesis throughout the suggestion of interesting further works.

## 7.1 Context

This thesis continues the line of research that studies the effect of selfishness and lack of coordination on functionality of distributed systems by considering combinatorial problems that are appropriate to describe such settings, using a game theoretic approach.

The objective of this thesis is to contribute to the existing body of work by studying various combinatorial problems that model routing and networking problems. In particular, we consider the Bin Packing problem in both classic and parametric versions, the Job Scheduling and the Machine Covering problems in different machine models with various objective functions.

We address various aspects concerning different economic solution concepts in the induced games, such as existence, quality, recognition, computation etc..

This is achieved by applying combination of both game-theoretic tools that are tailored to analyze consequences of strategic interactions and combinatorial techniques that are commonly used for these problems.

The quality measures featured include the well known regular and Strong Prices of Anarchy and Stability, as well as the newly introduced (weak and strict) Pareto Prices of Anarchy and Stability.

## 7.2 Contribution

First, we considered the (Pareto) efficiency of NE schedules in the Job Scheduling game with makespan minimization objective and established tight bounds on the (weak and strict) Pareto Prices of Anarchy and Stability in the identical, uniformly related and unrelated machines models. We gave a complete classification of recognition complexity of weakly and strictly Pareto schedules. In the cases where it is possible, we provided efficient algorithms to recognize such schedules. We also considered the effect of preserving deviations on such schedules.

Next, we considered the classic and the parametric versions of the Bin Packing problem and provided nearly tight bounds on the regular Prices of Anarchy for any value of the parameter $\alpha \geq 1$. As previously stated, we believe that their exact values match the respective lower bounds. We also gave tight bounds on the Strong Prices of Anarchy and Stability (which have identical value in the considered game), provided a characterization of SNE packings in this game and gave an (exponential time) algorithm that calculates such packings. We showed that computing an SNE packing is, in fact, NP-hard.

Finally, we considered the quality of NE schedules in the Job Scheduling game, with respect to the following fair optimality criteria: minimization of the maximal load over the machines in the schedule (which gives rise to the Machine Covering game) and minimization of the envy-ratio (that is the ratio between loads of the most and the least loaded machines in the schedule). In the model of identical machines, we established nearly tight bounds on the Price of Anarchy and tight bounds on the Price of Stability for the former fairness criterion, and tight bounds on these measures for the latter. For the first fairness criterion we have also considered mixed NE schedules, and showed that the mixed Price of Anarchy grows exponentially in the number of machines.

In the model of uniformly related machines, we considered the Price of Anarchy towards the first fairness criterion as a function of the maximum speed ratio between any two machines in the schedule. We showed that for ratios greater than 2 already the Price of Stability may be arbitrarily unbounded, and the Price of Anarchy may be arbitrarily unbounded starting at 2. We considered the behavior of these measures for remaining ratios, and finally proceeded to analyze these measures for schedules on two machines.

## 7.3 Further works

This thesis leaves some open issues which we list here, chapter by chapter.

In Chapter 2 we gave a full classification of the complexity of recognition of weak and strict Pareto efficient NE schedules in the Job Scheduling game. It can be interesting to consider the complexity of computation of such schedules, as well. In the cases of identical machines and weak/strict Pareto NE (where, as we proved, any NE schedule is strictly Pareto efficient) and also for uniformly-related machines and weak Pareto NE (where any NE schedule is weakly Pareto efficient, as proved in [11]) this question becomes equivalent to the question of complexity of computation of a regular (pure) NE schedule that can be answered easily as it was shown that any run of the (polynomial time) LPT scheduling algorithm on uniformly-related machines produces a pure NE schedule [57]. The complexity of computing strict Pareto NE in the model of uniformly-related machines as well as of weak and strict Pareto NE schedules in the model of unrelated machines remains open.

In Chapters 3 and 4 we gave nearly tight bounds on the Price of Anarchy in the classic and in the parametric Bin Packing games. Finding the exact values is an open problem. As previously stated, we believe that the values equal to the lower bound that we provided.

In Chapter 5 we gave nearly tight bounds on the Price of Anarchy in the Machine Covering game in identical machines model. Bridging this gap is an obvious challenge. As previously stated, we believe that the value equals to the upper bound that we provided. Another interesting direction is to consider the Strong Prices of Anarchy and Stability in this game.

In Chapter 6 (among other things) we considered the Price of Stability as a function of the maximum speed ratio $s$ between machines in the schedule for a setting of two machines. Whereas for $s \in [1, 4/3]$ and $s \in [1.78, 2]$ the given bounds are tight, determining its exact value for $s \in (\frac{4}{3}, 1.78)$ is left for further study.

Other than resolving these issues, there are other possible directions for further study.

The first option is to study similar matters for the restricted variants of these combinatorial problems that can be used to describe real life scenarios which the simple models neglect to consider. They are all natural extensions of basic problems from a theoretical point of view, and their behavior gives useful hints on the significance of previously known results. For example, if we talk about the bin packing problem, the possible variants to be considered include the bin packing with variable sized bins, which can be used to describe a parallel link network where the links are of various capacities, and the bin packing with cardinality constrains, where in addition to the capacity constraint a cardinality constraint is imposed on each link, which demands that the number of users that may utilize the link at the same time is limited. This latter cardinality restriction is often important in order to maintain transmission quality over the link due to ‘noise’ considerations.

Another direction is to define different cost sharing schemes and analyze the resulting models (for the classic problems and additional variants) with comparison to the previously studied model. For example, in this thesis we considered the bin packing game where the cost of each bin in a packing is proportionally shared among the items it contains according to their sizes. However, we can consider a more general cost scheme and charge the cost of the bin to the items according to some more complex function of their sizes, that, for instance, can charge more cost from a ‘very large’ item that occupies more than half a bin. Also, the activation cost of a bin that is charged to the items sharing it in our model equals bin’s capacity. We can consider a more general problem with a non-uniform cost structure, where the activation cost of a bin depends on a number of bins used in a packing.

It can be also interesting to further explore similar issues to those considered in this thesis for fair NE schedules (fair according to various fairness criteria), which is very important from an economic point of view. This is especially relevant to modern computer networks and communication networks, where time, bandwidth and other expensive resources are scarce. A very well known fairness notion that can be considered is, for example, is the min-max fairness. It is defined as follows:

**Definition 94.** *A configuration $x$ is called min-max fair if moving to another configuration $y$, for every player $i \in N$, if $c_i(y) < c_i(x)$ then there either exists a player $j \in N \setminus i$ such that $c_j(x) \geq c_i(x)$ and $c_j(y) > c_j(x)$ or a player $j \in N \setminus i$ such that $c_j(x) < c_i(x)$ and $c_j(y) \geq c_i(x)$.*

That is, min-max fairness requires that there is no improvement at the cost of someone who pays already higher cost while an improvement that increases the cost of a player with smaller original cost (but not by more than the original cost of an improving player) is allowed. Every min-max fair configuration is also strictly Pareto (and therefore also weakly Pareto) optimal.

The existence of such min-max fair NE schedule in the Job Scheduling game was proved in [67]. For a scheduling game with the makespan minimization objective the min-max-fair Price of Anarchy (and Stability) is 1, as it easy to verify that any schedule which is not optimal with respect to makespan is not min-max-fair. However, for a scheduling game with the covering maximization objective this is not necessarily so, and this problem requires further study.

Obviously, any other relevant packing, scheduling and covering problem can also be considered for study.

We believe that this research has both theoretical as well as practical importance. From a theoretical point of view, the motivation to study these optimization core problems using the game-theoretic approach is clear. Combination of concepts and tools drawn from different application domains already had and will produce elegant results and proofs.

From a practical point of view, many of these problems simulate actual current systems and proposed systems. The analysis of equilibria with various efficiency and fairness properties can lead to conclusions on whether a system can survive even without a centralized protocol. These results will be of a great help for system designers when they plan a multi-user multi-component systems with no central control, that have to be economically and computationally efficient despite the inability of the parties involved to communicate.